\def\vsk{\vskip.1in \noindent}
\hsize3.5in
\font\title=cmssdc10 scaled 1200
\magnification=\magstep1
\baselineskip 22 true pt \parskip=0pt plus3pt
\hsize 5.5 true in \hoffset .125 true in
\vsize 8.5 true in \voffset .1 true in

 \def\Schrod{Schr\"{o}dinger }
\def\pmb#1{\setbox0=\hbox{#1}
 \kern-.025em\copy0\kern-\wd0
\kern.05em\copy0\kern-\wd0
 \kern-.025em\raise.0433em\box0 }

\def\ket#1{|#1\rangle}
\def\doubleket#1{|#1\rangle\rangle}
\def\ref#1{$^{\bf #1}$}
\def\bra#1{\langle #1|}
\def\doublebra#1{\langle\langle #1|}

\def\kvac{\ket{0}}
\pageno=1
\def\bvac{\bra{0}}

\def\eq#1{\eqno{(#1)}}
\centerline{{\title Quantum Surveying: 
 How Entangled Pairs Act as}}
\centerline{{\title Measuring Rods on Manifolds
of Generalized
Coherent States.}}
\vskip.2in
\centerline{Daniel I. Fivel}
\centerline{Department of Physics,
University of Maryland}
\centerline{College Park, Md.}
\vsk
\centerline{May 2, 2002}
\vskip.2in
\centerline{\bf Abstract}
\vskip.2in
\noindent
Generalized coherent states arise from  reference states by the
action of locally compact transformation groups and thereby form
manifolds on which there is an invariant
measure.  It is shown that this implies the existence of canonically
associated Bell states that serve as measuring rods by relating the
metric geometry of the manifold to the observed EPR correlations. It
is further shown that these correlations can be accounted for by a
hidden variable theory which is non-local but invariant under the
stability group of the reference state.
\vskip.4in

Quantum dynamics provides the mathematical machinery for  computing 
the orbit in Hilbert space ${\cal H}$ of an initially given state
vector. But to experimentally identify   the state vector $s^\prime$ in
${\cal H}$ into which an initial state
$s$ 
has evolved with arbitrary precision  knowing only that it lies in a
neighborhood
$N$ of s, we must be able to choose for any $\epsilon > 0$ 
a {\it finite} set of states $s_1,s_2,\cdots,s_{J(\epsilon)}$ in $N$
such that 
$|s^\prime - s_j| < \epsilon$ for some $j$. This implies that the
space of states must be a {\it locally compact} subset of ${\cal H}$,
i.e.\   every point has a neighborhood with compact closure\ref{1}. 
\vskip.1in
While the
infinite dimensional Hilbert space 
${\cal H}$ required for particle dynamics
 is not a locally compact space, the groups of transformations
such as the Poincar\'{e} and Weyl-Heisenberg groups by which we relate
particle detectors to one another
 are locally
compact groups, and so then is the space of states that arise from a
reference state by their action. In fact, all of the relevant groups
are Lie groups\ref{2} and hence the states
lie on finite dimensional manifolds in ${\cal H}$.
\vskip.1in 
By restricting the set of allowed
states to such manifolds
we have the required  locally compact space, {\it but it will not
be a linearly closed space}, i.e.\  the superposition
principle will not hold.
Since the Dirac-von Neumann interpretation of measurement which
identifies observable properties with eigenvalues of self-adjoint
operators is not implementable without the superposition principle, 
the restriction to locally compact spaces requires an alternative
interpretation of the measurement process. Since the so-called
measurement problem of quantum mechanics arises from the Dirac-von
Neumann interpretation\ref{3}, we have an  additional motivation for
seeking such an alternative.
\vskip.1in
Because the set of allowed states is a manifold, we are led to
interpret  quantum measurement as a means of determining the
metric geometry of that manifold. 
There is a useful analogy which
clarifies the relationship of this interpretation with that of
Dirac-von Neumann. When Gauss undertook the survey of Hanover he
pointed out that the properties of a surface that can be determined by
``small, flat bugs" that live on its surface are of intrinsic interest,
and these do not require the assignment of
Cartesian coordinates to its points relative to a reference point in
space\ref{4}.  The Dirac-von Neumann measurement
paradigm which seeks to characterize a state by its membership in
eigenspaces of observables is analogous to the latter, while the
interpretation we shall describe below corresponds to the Gaussian
approach. In view of the analogy I shall call this {\it quantum
surveying}. 
\vskip.1in
The restriction to locally compact groups has a profound
implication: For each such group G there will be an invariant
measure\ref{5} which allows us to integrate over the group and thereby,
as we shall see, construct pair states canonically related to G which
will serve as measuring rods for quantum surveying. To this end we first
introduce suitable kinematics for the manifold of states.
\vskip.1in
Let U be an irreducible, unitary representation on ${\cal H}$ of a Lie
group G, and let 
$\ket{0}\in {\cal H}$ be a reference unit vector. We shall use $g$ to
indicate both a group element and its representation in U. The
allowed states will be the images of
$\ket{0}$ by a transformation $g \in G$. To label a state by the $g$
which produces it is ambiguous because two elements $g_1,g_2$ can
produce  equivalent states, i.e.\ states that differ by a phase factor.
This means that $h = g_1^{-1}g_2$ has $\ket{0}$ as an eigenstate. Such
elements form a subgroup  $G_o$ called the stability subgroup of the
reference state. If we select one element $g$ from each coset $gG_o$
and define $\ket{g} = g\ket{0}$ we shall have a one-one
correspondence between states and labels. The set of states so labeled
is referred to as a set of {\it generalized coherent states}\ref{6}
 and can be identified with the coset space $F =
G/G_o$. It is a homogeneous space, i.e.\ every pair of states is
related by a transformation in the group.
\vskip.1in
As noted above, the assumptions made about G imply that
F is a manifold upon which there exists an invariant
measure $d\mu$  by which we can integrate over F. The invariance of
the measure implies that the operator 
$${\bf I} \equiv \int_F d\mu \ket{g}\bra{g}\eq{1}$$
 commutes with every $g$. Since U is irreducible, Schur's
Lemma informs us that we can take {\bf I} to be the unit
operator by suitably scaling
$d\mu$.
 Observe
that
$$Tr({\bf I}) = \int_F d\mu \equiv V_F\eq{2}$$
is the ``volume" of the space of states and is finite if and only if
$G$ is compact. 
\vskip.1in
It 
follows from (1) that although F is itself not linearly closed, its
linear closure is all of ${\cal H}$. The
states of F are not mutually orthogonal, and in general F is
infinitely over-complete, i.e.\ one can delete an infinite subset, and
the linear closure of the remaining states is still ${\cal H}$.
\vskip.1in
We next use the representation of {\bf I}  to
construct a pair state with remarkable properties: Consider the object
$$\doubleket{B} \equiv C \int_F d\mu \ket{g}\otimes \bra{g},\eq{3}$$
where C is a normalization constant.
We can interpret this as a pair state in the following way: Let
$\tau $ be any {\it anti}-unitary operator and observe that the map
$$\tau\ket{g} \to \bra{g}\eq{4}$$
is unitary. Thus one can think of $\bra{g}$ as the $\tau$-reversal of
$\ket{g}$, i.e.\ we might just as well have written the state as
$$\doubleket{B} \equiv C\int_F d\mu \ket{g}\otimes \ket{g^\tau},\hbox{
with }
\ket{g^{\tau}} \equiv \tau\ket{g}.\eq{5}$$
In (3) the representation space  is the tensor product of
${\cal H}$ with its dual, whereas in (5) it is the tensor
product of ${\cal H}$ with itself. We shall use the form (3) which has
a more transparent structure. Let us compute the probability in
state
$\doubleket{B}$ that one member is found in state
 $\ket{g_1}$ and its
partner in state
$\bra{g_2}$, i.e.\ that the pair is found in the state
$$\doubleket{g_1,g_2} \equiv \ket{g_1}\otimes \bra{g_2}.\eq{6}$$ This
is computed using (1) to be
$$p(g_1,g_2) =
|\doublebra{g_1,g_2}B\rangle\rangle|^2
= |C|^2|\bra{g_1}g_2\rangle|^2.\eq{7}$$ It follows that there is equal
probability $|C|^2$ for finding one member in any state if nothing is
given about its partner, but that the conditional probability for
finding one member in state
$\ket{g_1}$ if its partner is found in the state $\bra{g_2}$ is
$$p(g_1|g_2) = |\bra{g_1}g_2\rangle|^2.\eq{8}$$
 When the measure is scaled to make {\bf I} the unit
operator we find that
$$\doublebra{B}B\rangle\rangle = |C|^2 V_F, \eq{9}$$
which  means that $\doubleket{B}$ can be normalized by choosing
$C = V_F^{-1/2}$ if and
only if $V_F$ is finite, i.e.\ G is compact. Since we are interested in
groups that may be locally compact but not compact we shall have to
make sense of
$\doubleket{B}$ in general through a limiting process. This creates no
problems because $C$
disappears in computing the conditional probability (8). 
\vskip.1in
We see from (8) that $\doubleket{B}$ exhibits perfect EPR correlation,
i.e.\ one concludes with certainty that one member of the pair will
be in state $\ket{g}$ if its partner is in state $\bra{g}$. If U is
is the spin-1/2 representation of $SU_2$ and $\tau$ is the time-reversal
operator, it is shown in the Appendix that $\doubleket{B}$ is the
familiar Bohm-Ahronov singlet. We shall refer to $\doubleket{B}$ as the
{\it generalized Bell state canonically associated with the manifold F}.
\vskip.1in
Let us now see that measurement of $p(g_1|g_2)$ using
$\doubleket{B}$ reveals the metric structure of the manifold F:
Because
$\bra{g_1}g_2\rangle$ is a scalar-product, it follows that
$$d(g_1,g_2) \equiv \sqrt{1 - |\bra{g_1}g_2\rangle|^2}\eq{10}$$ is
a metric on F, i.e.\ it is non-negative, symmetric, obeys the
triangle inequality, and vanishes if and only if $g_1 \approx g_2$
where $``\approx"$ means membership in the same coset of $G_o$.
Since
$$\bra{g_1}g_2\rangle = \bra{0}g_1^\dagger g_2\ket{0} = 
\bra{0}g_1^{-1} g_2\ket{0},\eq{11}$$ 
we have
$$p(g_1|g_2) = |\bra{0}g\ket{0}|^2 \equiv p(g),\;\, g =
g_1^{-1}g_2.\eq{12}$$
 Observe that this is invariant under the
substitution
$$g \to g_o g g_o^\prime,\;\, g_o,g_o^\prime \in G_o.\eq{13}$$
Thus $p(g)$ is constant on the double cosets $G_o\backslash g/G_o$.
 The distance $d(g_1,g_2)$ can be written
$$d(g_1,g_2) = d(g) \equiv \sqrt{1 - p(g)},\;\; g =
g_1^{-1}g_2 \eq{14}$$ and is therefore a function of the double coset
to which the relation
$g$ between $g_1$ and $g_2$ belongs. The double cosets partition G just
as left and right cosets do\ref{7}. We shall refer to the double cosets
as {\it coherence relations} between the two states and refer to
$d(g)$ as the ``diameter" of the coherence relation g. 
\vskip.1in
The function $d(g)$ has remarkable properties which follow from the
metric properties of $d(g_1,g_2)$:
$$ 0 \leq d(g) \leq 1, \eq{15a}$$
where $d(g) = 0$ if and only if $g$ is the identity coherence relation;
$$d(g) = d(g^{-1});\eq{15b}$$
$$d(g) + d(h) \geq d(gh).\eq{15c}$$
\vskip.1in
Now observe that each $g_o \in G_o$ determines an automorphism $g \in G
\to g^\prime = g_ogg_o^{-1} \in G$ which leaves F invariant since
$G_o \to G_o$. Since  
$d(g_1^\prime,g_2^\prime) = d(g_1,g_2)$ this is an {\it
isometry} of F regarded as a manifold.
 The one-parameter subgroups $t \in R \to
g_o(t)$ of
$G_o$ define dynamical processes that take $F$ into itself, i.e.\
$$\ket{g} \to \ket{g(t)}, \;\, g(t) = g_o(t)g
g_o(t)^{-1}.\eq{16}$$ Since G is a Lie group, we will be able to
write $$g_o(t) = e^{-itH}\eq{17}$$
with some Hermitian operaltor H and so obtain a \Schrod equation
in the Heisenberg picture:
$$ dg(t)/dt = - i[H,g(t)].\eq{18}$$
{\it Thus the linear dynamics is preserved even
though the space of allowed states is not linearly closed.}
\vskip.1in
Now let us calculate the diameter of the coherence relation between
nearby points on an orbit generated by H. We find for small $\delta t$:
$$d_g(\delta t) \equiv d(g(- \delta t/2),g(\delta t/2)) =
d(g^{-1}e^{-iH\delta t}g) = \delta t \Delta_g(H), \eq{19}$$
where
$$\Delta_g(H) \equiv (\bra{g}H^2\ket{g} -
\bra{g}H\ket{g}^2)^{1/2}\eq{20}$$ is the dispersion of H in the state
$\ket{g}$. Thus we arrive at the useful conclusion that {\it the
dispersion of the generators of the one-parameter subgroups of $G_o$
determine the local differential geometry of the manifold, and this is
expressed by the diameters of coherence relations in the neighborhood
of the identity. }
\vskip.1in
Bell's EPR Theorem informs us that
it is not possible to construct a local hidden variable theory that
will reproduce the function $p(g_1|g_2)$.
 Such a theory 
would provide a map from states $\ket{g}$ to sets $\Lambda(g)$ with
measure
$\mu$ such that 
$$p(g_1|g_2) = \mu(\Lambda(g_1)\cap\Lambda(g_2)).\eq{21}$$
In fact the incompatibility between the right and left sides
of (21) is known to be a consequence of the difference between the
metric structures implied by them\ref{8}. A new possibility
emerges, however, from our restriction of allowed states to F which
makes
$p(g_1|g_2)$  a function only of the combination $g = g_1^{-1}g_2$.
For let us suppose that in each run of a correlation experiment
with $\doubleket{B}$  a random element $h$ of F serving as a hidden
variable is generated in such a way that the probability of an $h$ with
$d(h) < r$ is $r^2$. Suppose further that a correlation between
detectors for
$\ket{g_1}$ and $\bra{g_2}$ is observed when the diameter $d(g)$ of the
relation
$g = g_1^{-1}g_2$ between them  is smaller than that of h and otherwise
not. Then the probability of a correlation will be $1 - d(g)^2 = p(g)$,
i.e.\ we will reproduce the quantum mechanical result. Since $d(g)$ is
invariant when
$g_1$ and $g_2$ evolve under the dynamical transformation (17), this
type of hidden variable theory preserves the $G_o$
symmetry of the theory. In particular in relativistic theories where
$G_o$ includes the Poincar\'{e} group the non-locality does not destroy
the covariance of the theory. Because the hidden variable h is
non-local there is no violation of Bell's Theorem.
\vskip.1in
We now illustrate the general ideas above by an important example,
namely the quantum surveying of the electromagnetic field of a laser.
We present the analysis for a single mode laser and then note its
generalization to any number of modes. 
\vskip.1in
The relevant group G is the Weyl-Heisenberg (WH) group. This group has,
according to the Stone-von Neumann theorem, a unique
unitary representation (up to equivalence) known as the Fock
representation obtained as follows: Let  
$a,a^{\dagger}$ be the Bose operators 
with the commutation rules $[a,a^\dagger] = I$.
The group elements are then  $$U(\theta,\lambda) =
e^{i\theta}u(\lambda),\qquad u(\lambda) = e^{\lambda\cdot a^\dagger -
\lambda^*\cdot a}. \eq{22}$$

Here $\lambda$ ranges over the finite complex plane and $0 \leq \theta
< 2\pi$. The composition law is obtained from
$$u(\lambda)u(\mu) = e^{i\theta}u(\lambda + \mu),\qquad \theta =
Im(\lambda^*\cdot \mu).\eq{23}$$ 
Taking the reference state $\ket{0}$ to  be the Fock vacuum,
i.e.\ the state annihilated by $a$, the stability subgroup is $G_o =
U(\theta,0),\; 0 \leq \theta < 2\pi$. Hence the manifold
$F = G/G_o$ of coherent states is in one-one correspondence with the
set $\lambda$, i.e.\ with the complex plane. The area element
$d^2\lambda$ is an invariant measure on F.  We may thus take F to be 
the set of states of the form
$$\ket{\lambda} \equiv u(\lambda)\ket{0} =
e^{-|\lambda|^2/2}e^{\lambda\cdot
a^\dagger}\ket{0}\eq{24}$$ which we recognize as the familiar
Glauber\ref{9} coherent states of optics, and we  
 then have
$$p(\lambda|\mu) = |\bra{\lambda}\mu\rangle|^2 =  e^{-|\lambda -
\mu|^2} = p(\lambda - \mu), \eq{25}$$
and (1) becomes:
$$I = \pi^{-1}
\int d^2\lambda
\ket{\lambda}\bra{\lambda}.\eq{26}$$
To construct a generalized Bell state we observe that (3)
gives
$$\doubleket{B} = C\pi^{-1}\int d^2\lambda
\ket{\lambda}\otimes\bra{\lambda}  = C\int d^2\lambda
e^{-|\lambda|^2}(e^{\lambda
a^\dagger}\otimes e^{\lambda^* a})(\ket{0}\otimes \bra{0}) 
= $$ $$C (e^{a^\dagger \otimes a})(\ket{0}\otimes \bra{0}).\eq{27}$$
As noted above the fact that the WH group has infinite volume (i.e.\
is not compact) means that $\doubleket{B}$ is not normalizable.
However, we can make sense of it as follows. For $0 \leq r < 1$ one
verifies that the norm of the state
$$\doubleket{B,r} \equiv (1 - r^2)^{1/2} e^{r(a^\dagger \otimes
a)}(\ket{0}\otimes\bra{0})\eq{28}$$
is unity. Although the state becomes improper as $r \to 1$, the
singular normalization factor will, as noted above, disappear when
conditional probabilities are computed. We may thus perform our
surveying by studying correlations in the state
$\doubleket{B,r}$, and compute the metric properties from the limit
of the correlations as $r \to 1$ at which they become perfect EPR
correlations.
\vskip.1in
It is instructive to observe that $\doubleket{B,r}$ can be regarded as
a ``twisted" form of the two particle vacuum $\ket{0}\otimes\bra{0}$.
For observe that the operator
$$a_r \equiv (1 - r^2)^{-1/2}(a - ra^\dagger)\eq{29}$$
satisfies
$$[a_r,a_r^\dagger] = 1,\;\; (a_r \otimes I)\doubleket{B,r} =
0 = (I \otimes a_r^\dagger)\doubleket{B,r}.\eq{30}$$
Thus the states $\doubleket{B,r}$ for $0 \leq r < 1$ are normalizable
two-particle states in which there is a correlation that approaches
perfect EPR correlation as $r \to 1$. The fact that the states
$\ket{B,r}$ are twisted vacuua removes some of the mystery from the
non-locality of  EPR correlations.
\vskip.1in
Observe that the unitary operator 
$$V(t) \equiv e^{-itH},\;\; H = \omega a^\dagger a \eq{31}$$
has the properties
$$V(t)aV^{-1}(t) = e^{-i\omega t}a, \;\; V(t)\ket{0} = \ket{0}.\eq{32}$$
Hence we can enlarge the WH group by adjoining $V(t)$.
This simply enlarges the stability subgroup of $\ket{0}$
so that it now contains the one parameter subgroup $g_o(t) = V(t)$. 
The dispersion of $H$ in the state $\ket{\lambda}$ is readily
computed to be
$$\Delta_\lambda(H) = \omega |\lambda|.\eq{33}$$
Hence the diameter of the coherence relation between two nearby points
on an orbit generated by $V(t)$ passing through $\ket{\lambda}$ will be
$$d_\lambda(\delta t) = |\lambda|\omega\delta t =
\sqrt{N_\lambda}\omega\delta t,\eq{34}$$  where $N_\lambda =
|\lambda|^2$ is the mean photon number in the state
$\lambda$.
\vskip.1in
Let us now see how the non-local hidden variable theory described
above works for this example. Suppose that on each run of an
experiment a random element labeled by $\lambda$ of the Weyl-Heisenberg
group is generated such that the probability of choosing $\lambda$ in
an area
$d^2\lambda$ centered on $\lambda$ is $\rho(\lambda)d^2\lambda$ with
$$\rho(\lambda) = \pi^{-1}e^{-|\lambda|^2}.\eq{35}$$
Then since $d(\lambda) = \sqrt{1 - e^{-|\lambda|^2}}$, the probability
for $d(\lambda) \geq d(\lambda_o)$ is the probability for $|\lambda|
\geq |\lambda|_o$, which is
$$\pi^{-1}\int_{|\lambda|\geq |\lambda_o|}d^2\lambda e^{-|\lambda|^2} =
e^{-|\lambda_o|^2} = 1 - d(\lambda_o)^2.\eq{36}$$
Thus the probability for $d(\lambda) < r$ is $r^2$. Thus if a
correlation betwen
\ $\ket{\lambda_1}$
and $\ket{\lambda_2}$ is recorded whenever $|\lambda_1 - \lambda_2| \leq
|\lambda|$ we will reproduce the quantum mechanical prediction that
the probability of correlation is $e^{-|\lambda_1 - \lambda_2|^2}$.
Thus the required distribution for the non-local hidden variable is the
Maxwellian distribution (35).
\vskip.1in
Let us now see how the results for a single mode field generalize to an
n-mode field with bose operators $a = (a_1,\cdots,a_n)$ satisfying
$[a_i,a_j^\dagger] = \delta_{ij}I,\; [a_i, a_j] = 0$. The group
G is now the direct product of n copies of the Weyl-Heisenberg group
and the coherent states are still described by (24) if we interpret
 $\lambda$ is an n-component complex vector with 
$|\lambda|^2 = |\lambda_1|^2 + \cdots + |\lambda_n|^2$, and
$a = (a_1,\cdots,a_n)$. The reference state $\ket{0}$ is the n-mode Fock
vacuum. We can now describe an enormous group of dynamical processes
by adjoining the group of transformations:
$$V_H(t) = e^{-it{\cal H}},\; {\cal H} = a^\dagger\cdot H \cdot
a\eq{37}$$ in which $H$ ranges over the set of all $n \otimes n$
Hermitian matrices.
$V_H(t)$ is unitary (for real t) because
$$a^\dagger\cdot H \cdot a = \sum_{j,k = 1}^n a^{\dagger}_j H_{jk}
a_k\eq{38}$$ is a Hermitian operator. Since it is quadratic in the
bose operators one readily verifies that
$$ V_H(t)\ket{\lambda} = \ket{\lambda(t)} ,\; \lambda(t) =
e^{-iHt}\lambda.\eq{39}$$
Thus the unitary evolution of the state vector in $F$ is described by
that of an n-component vector
$\lambda(t)$   which  evolves
under a unitary transformation $e^{-iHt}$. 
\vskip.1in
Let us now summarize our results: Any quantum mechanical system
that can be characterized by transformations $g$  belonging to a
locally compact group G of a reference state $\ket{0}$ can be
identified with a manifold F of generalized coherent states. There will
be a canonically associated generalized Bell state $\doubleket{B}$ 
which serves to determine the metric geometry of the manifold through
observable EPR correlations. Dynamical transformations are identified
with elements of the stability subgroup $G_o$ of the reference state,
and such transformations take the states of F into one another. Thus
the dynamics remains linear even though F is not a linear space. There
is a canonically associated hidden variable interpretation of the
stochastic behavior which, albeit non-local, preserves the symmetry of
the theory under $G_o$. In particular this will be the case in a
covariant theory in which $G_o$ includes the Poincar\'{e} group.
\vskip.2in
\centerline{Appendix}
\vskip.2in
 Let us examine
$\doubleket{B}$ when $G$ is $SU_2$. 
Choosing $\ket{0}$
as the north-pole on a sphere, a state which is stabilized by a $U_1$
subgroup, we see that $F$ can be identified with the points of the
sphere (the Poincar\'{e} sphere). The invariant measure is the solid
angle
$d\Omega$ and we find with suitable C:
$$\doubleket{B} = C \int d\Omega (\cos\theta \;\;
e^{-i\phi}\sin\theta)\otimes\left(\matrix{\cos\theta \cr
e^{i\phi}
\sin\theta}\right)$$
$$ = 
2^{-1/2}\left((1,0)\otimes\left(\matrix{1
\cr 0}\right) + (0,1)\otimes\left(\matrix{0 \cr
1}\right)\right) = 
2^{-1/2}\left(\ket{\uparrow}\otimes \bra{\uparrow}
+ \ket{\downarrow}\otimes \bra{\downarrow}\right). \eq{A1}$$ 
Choosing $\tau$
to be the time-reversal operator for which
$\ket{\uparrow ^\tau} = \ket{\downarrow}$ and $\ket{\downarrow ^\tau}
= -\ket{\uparrow}$, the mapping $\bra{x}\to \ket{x^\tau}$ gives
$$\doubleket{B} = 2^{-1/2}\left(\ket{\uparrow}\ket{\downarrow} -
\ket{\downarrow}\ket{\uparrow}\right),\eq{A2}$$
which is recognized as the Bohm-Aharonov singlet.
\vskip.2in
\centerline{References}
\vskip.2in
\noindent
1. Folland, G., {\it A Course in Abstract Harmonic Analysis}, CRC
Press, Boca Raton 1995 Chap. 2.
\vskip.1in
\noindent
2. It is known that connected, simple, locally compact groups are
necessarily Lie Groups. See Mackey, G., {\it Unitary Group
Representations in Physics, etc.}, Benjamin/Cummings, Reading. p.87.
\vskip.1in
\noindent
3. Bub, J., {\it Interpreting the Quantum World}, Cambridge University
Press, 1997.
\vskip.1in 
\noindent
4. Gauss, K.F., {\it Disquisitiones generales circa superficies
curvas.},
\vskip.1in
\noindent
5. Folland, G., {\it op. cit.}
\vskip.1in
\noindent
6. Perelomov, A., {\it Generalized Coherent States and Their
Applications}, Springer-Verlag, Berlin 1986. Chap. 2.
\vskip.1in
\noindent
7. Hall, M., {\it The Theory of Groups}, Macmillan, 1959, p. 14.
\vskip.1in
\noindent
8. Fivel, D., 
Phys. Rev. Lett, {\bf 67}, 285 (1991).
\vskip.1in
\noindent
9. Glauber, R. Phys. Rev, {\bf, 131} 2766 (1963). 
 
\end

\end

s
\end
Then 
$$\doublebra{r}r\rangle\rangle =

\end

\end
 a map
 $$F \to F$ that preserves the metric $d(g_1,g_2)$.  
$u
\in U$ such that 
$\ket{0}$ is an eigenstate. This transformation leaves the
diameter of coherence relations invariant. Clearly $G_o$ is a subgroup
of $S$.
 Recognizing that{\it it is
$d(g)$ that is measured in EPR correlation experiments,} we can now
construct a non-local hidden variable model for EPR correlations which
is nonetheless manifestly covariant:
\vskip.1in
 Imagine that in each run of an
experiment a random $h$ is generated such that the probability
for $d(h) \leq r$ is $r^2$. Suppose that when we test for a correlation
between $g_1,g_2$ we get a coincidence when the relation $g =
g_1^{-1}g_2$ has a smaller diameter than h and otherwise not. Then
the probability of a coincidence will be the probability for
$d(h) \geq d(g)$ which is $1 - d(g)^2 = p(g)$. We thus reproduce
the quantum mechanical result. There is no violation of Bell's Theorem
because the model is non-local. Since the hidden variable is
constructed from the diameter of coherence relations which are
invariant under the symmetry group, this hidden variable theory is
manifestly
  covariant with respect to S. 
\vskip.1in
To illustrate the theory developed above let us examine the important
special case wherein G is the group associated with multimode lasers.
\end
of 
\end

 G such that if $S_r$ is the subset of $F$ for which 
$$\int_{F{d(g) \leq r} d\mu = r^2.$$
Thus if 

 Suppose that in each run of an
experiment in which we test for a correlation between pairs $g_1,g_2$,
a random group element $h$ is generated in such a way that
the probability that $d(h)$ is smaller than $r$ is $r^2$

\vskip.1in

\end
 
$\ket{g} \equiv g\ket{0}$ for $g \in G$. Let $G_o$ be the stability
subgroup for $\ket{0}$, i.e.\ the subgroup of $G$ for which $\ket{0}$
is an eigenvector. Then  $\ket{g_1}$ and $\ket{g_2}$ differ by a
phase factor if and only if $g_1^{-1}g_2 \in G_o$, i if.e.\ $g_1$ and
$g_2$ belong to the same left coset $gG_o$. Thus the distinct images of
$\ket{0}$ under $G$ are in one-one correspondence with the set $F =
G/G_o$ of left cosets of $G_o$. The set of states $F$ obtained by
selecting one $\ket{g}$ for each $gG_o$ is referred to as a set of 
{\it generalized coherent states}. When $G$ is a Lie group this will be
the manifold of allowed states. 
\vskip.1in 
We can now introduce an invariant measure $d\mu$ on $F$ and consider
the operator:
$$I \equiv \int_F d\mu \ket{g}\bra{g}.\eq{1}$$
One readily checks that  
$$I g = gI.\eq{2}$$
If G is simple, there is no invariant subspace under $g$ and it
follows from Schur's Lemma that by suitably scaling the measure we can
make I the unit operator. It follows that the generalized coherent
states are complete. They are not mutually orthogonal and in general
form an infinitely over complete set. Note, however, that as observed
earlier they do not form a linear space. 
\vskip.1in 
Now consider the object 
$$\doubleket{B} \equiv C \int_F d\mu \ket{g}\otimes \bra{g},\eq{3}$$
where C is a normalization constant.
We can interpret this as a pair state in the following way: Let
$T$ be any anti-unitary operator and observe that the map
$$T\ket{g} \to \bra{g}\eq{4}$$
is unitary. Thus one can think of $\bra{g}$ as the T-reversal of
$\ket{g}$, i.e.\ write
$$\doubleket{B} \equiv C\int_F d\mu \ket{g}\otimes \ket{g^T},\hbox{
with }
\ket{g^T} \equiv T\ket{g}.\eq{5}$$
To obtain some insight into this structure let us examine
$\doubleket{B}$ when $G$ is $SU_2$. 
Choosing $\ket{0}$
as the north-pole on a sphere which is stabilized by a $U_1$ group we
see that $F$ can be identified with the points of the sphere (the
Poincar\'{e} sphere). The invariant measure is the solid angle
$d\Omega$ and we find with suitable C:
$$\doubleket{B} = C \int d\Omega (\cos\theta \;\;
e^{-i\phi}\sin\theta)\otimes\left(\matrix{\cos\theta \cr
e^{i\phi}
\sin\theta}\right)$$
$$ = 
2^{-1/2}\left((1,0)\otimes\left(\matrix{1
\cr 0}\right) + (0,1)\otimes\left(\matrix{0 \cr
1}\right)\right).\eq{6}$$
Choosing T to be the time-reversal operator, the spin-states
$\ket{\uparrow},\ket{\downarrow}$ are related by
$T\ket{\uparrow} = \ket{\downarrow},\; T\ket{\downarrow} =
-\ket{\uparrow}$ so that we can write
$$\doubleket{B} = 2^{-1/2}\left(\ket{\uparrow}\ket{\downarrow} -
\ket{\downarrow}\ket{\uparrow}\right),\eq{7}$$
which is recognized as the Bohm-Aharonov singlet.
\vskip.1in

\end
\end

$$\doubleket{B} = \int d\Omega 
\left(\cos(\theta/2), e^{-i\phi}\sin(\theta/2)\right)\otimes \lft(
\matrix{\cos(\theta/2) \cr e^{i\phi}\sin(\theta/2)}\right)
\right\}$$
\end

\end 

 evolves
into an identifyable state means that  we can identify a state to within
$\epsilon$ by a finite number of measurements. Among the states in some
neighborhood of s we can  pinpoint to within
$\epsilon$ a state into which $s$ evolves. The state into which it
evolves by a finite number of measurements.
ate
Argue that the space of states must be locally compact to make sense
of measurement.

 Every state has a neighborhood which is compact. Can
cover neighborhood with a finite $\epsilon$ grid for any $\epsilon$.
Can
 Quantum dynamics supplies an operator for the evolution of a state
vector and gives the conditional
probability $p(b|a)$ that a system in initial state $a$
will be in a final state $b$. Bell's theorem (ref Bell) demonstrates
that the form of the function $p$ is incompatible with an underlying
local, hidden variable theory. More precisely it is not possible to
identify states
$x$ with sets $\Lambda_x$ on which there is a measure $\mu$ such that
$$p(b|a) = \mu(\Lambda_{b}\cap\Lambda_{a}).\eq{1}$$
 
The incompatibility of the quantum mechanical $p(b|a)$ with the right
side of (1) reveals the essential difference between the space of
quantum states and the space of classical states in the following way
(ref Fivel): If $f(a,b)$ is a symmetric function then
 $$d(a,b) \equiv \sup_z|f(a,z) - f(b,z)|, \eq{2}$$
is a metric. Evaluating it for 
 $f(x,y) = p(x|y)$ and for $f(x,y) =
\mu(\Lambda_{x}\cap\Lambda_{y})$ (see Appendix) we find that
$$d(x,y) = (1 - f(x,y))^\nu, \eq{3}$$
where the exponent is $\nu = 1/2$ and $\nu = 1$  respectively. 
\vskip.1in
Now if $d$ saturates then no power $d^\nu, \, \nu > 1$ is a metric,
for one verifies that the triangle inequality will fail (see
Appendix). Bell inequality violations are manifestations of that
failure. Thus experiments that test for Bell inequality violations
are essentially probes of the metric geometry of the space of states.
\end 
 Suppose
now that we can measure
$f$ and from it get a metric
$d$.  Now one readily checks that if $d$ saturates it will violate the
triangle inequality if it is replaced by $d^\eta,\, \eta > 1$.
In  particular since the quantum mechanical metric
corresponds to $\nu = 1/2$ its square must violate the $\nu = 1$
triangle inequality.
Thus there must be
$x,y,z$ such that there is a violation of the inequality
$$(1 - p(x,z)) + (1 - p(z,y)) \geq (1 - p(x,y),$$
which is the Bell inequality:
$$p(x,z) + p(z,y) \leq 1 + p(x,y)$$
Thus the essential difference between a quantum mechanical
system lies in the 
 metrical structure of the space of states.
\end

 inequality for
$\nu = 1$ and
$f(x In particular
 Suppose
further that the triange inequality saturates to any degree of
accuracy. Then one readily finds that $d^\nu$ is not a metric for $\nu
> 0$. Thus there is a   Suppose further that we can find $x,y,z$ such
that such that
$d(x,z) + d(z,y) \approx d(x,y)$ to any degree of accuracy (sat

The fact that the quantum mechanical metric) $\nu = 1/2$ rather
than $\nu = 1$ can 
If the quantum mechanical $p(x|y)$ obeyed (1), the triangle inequality
for the corresponding metric (3) with $\nu = 1$ implies an inequality
for
$p$ (Bell's inequality) which is violated for the quantum mechanical
$p$.  For a pair of causally  related event pairs (time-like or
light-like separated) one can construct a local hidden variable theory
in which the hidden variable relaxes during the evolution of one event
to the other. For a pair of space-like separated events, however, this
is not possible, and by experimentally demonstrating the violation for
such pairs Aspect et al demonstrated the impossibility of accounting
for the correlation with a local hidden variable.
\vskip.1in
The question then arises as to whether it is possible to account for
the correlations by a non-local hidden variable in a manifestly
covariant way. We shall show that this is indeed the case.
\vskip.1in
We consider experiments of the Aspect type in which coincidences
are observed between detections of the two members of a pair
in a maximally entangled state $B$ referred to as a Bell state. If
$S$ is the set of all detectors, the probability  that  
 a member is detected by  $x \in S$ is independent of $x$, but
the conditional probability $p(x|y)$ for coincidence of detection of
one member by $x$ and the other by $y$ depends on a relationship
between $x$ and $y$. In particular there is a one-one map $x \to
x^\prime$ such that $p(x|x^\prime) = 1$. Quantum theory assigns unit
ray
$e^{i\theta}\ket{x}$ to the detectors such that
$$p(x|y^\prime) = |\bra{x}y\rangle|^2.\eq{4}$$
\end

\end

$$d(x,y) = \sqrt(1 - p(x|y))$$
distinguishes it from classical mechanics, it is interesting to
ask whether

\end

 In classical dynamics
knowledge of the state at one time determines that at the other.
Quantum dynamics is modeled on this in the sense that \Schrod 
dynamics determines the evolution of the state vector, and the theory
predicts the 
 
In the Dirac-von Neumann interpretation of quantum mechanics the
task of measurement is to determine the values of ``observables"
identified with Hermitian operators in a Hilbert space ${\cal H}$. A
state has a determinate value of an observable if and only if it is
an eigenstate of the observable. Otherwise the outcomes are
distributed, and from this distribution one obtains the projection of
the state vector on the various eigen-spaces. Thus the goal
of measuring observables is to locate a state in ${\cal H}$. 
\vskip.1in

Suppose we know that a state $\psi$ 
 lies in a sphere $S_r(\psi_o)$ of radius $r$ about some state
$\psi_o$, i.e.\
$|\psi - \psi_o| \leq r$. Suppose we wish to improve our knowledge by
covering $S_r$ with spheres of smaller radius centered on states
$\psi_j,\; j = 1,\cdots,N_\epsilon$ such that for an 
$\psi$ is determined within $\epsilon$ by its projection on this set
or, what amounts to the same thing that $\psi$ lies in a sphere
$S_\epsilon(\psi_j)$ for some $j$.

must be recognized, however, that some constraint on the location must
be imposed {\it ab initio} if one is to make sense of the measuring
process. To see this suppose we try to locate an unknown state $\psi$
to within $\epsilon$ (in the sense of the Hilbert space metric) by
determining the probability
$p_j(\psi)$ that it is found in the states $\phi_j, \; j=
1,2,\cdots,N_\epsilon$. It is {\bf not} the case that for any $\epsilon
> 0$ we can choose a set
$S_\epsilon$ of states with {\it finite} $N_\epsilon$  such
that
$\psi$ lies within
$\epsilon$  of the
$\phi_j$ for which
$p_j$ is largest.
\vskip.1in
 The reason is that
Hilbert space is not locally compact, i.e.\ one can hop from one
basis vector to another without approaching a limit even though one
remains within a bounded set. 
\vskip.1in
While in this sense Hilbert space is too large an arena for quantum
mechanics, any finite dimensional subspace is too small because,
 such spaces cannot support unitary representations of
non-compact groups such as the Galilee and Poincar\'{e} groups needed
to do physics in the real world.
\vskip.1in
These groups are locally compact, however, and  if we restrict
the allowed states to a set ${\cal F}$ that arises from a refefence
state $\ket{e}$ by the action of such a group the set 
$S_\epsilon$ will exist. But while this set is  a manifold in ${\cal
H}$, it is not a linear subspace. Hence the notion of observables
represented by linear operators will not be implementable, and we shall
have to find a replacement for Dirac-von Neumann interpretation of
measurement.
\vskip.1in
 In this paper I will describe how this can be done. We will see that
the quantum mechanical correlations we detect are simply related to
the metric geometry of the manifold ${\cal F}$. Thus we may adopt a
point of view for quantum measurement analagous to the point of view
introduced by Gauss for the surveying of surfaces. The idea is
to examine the surface from the point of view of an observer who can
move around on it rather than embedding the surface in a higher
dimensional vector space and attaching coordinate vectors to its
points. Analogously we do not attempt to locate the state vector in
Hilbert space but rather to relate the intrinsic geometry of the
manifold ${\cal F}$ to correlations between pairs of detectors for
the states belonging to it.
\end

shall thus produce a form of quantum mechanics in which the symmetry
group of the theory is not the full unitary group
$U$ acting on ${\cal H}$ but rather is a locally compact
subgroup $G$ of $U$.
\vskip.1in
We begin with kinematics. For each $g \in G$
define the image
$$\ket{g} = g\ket{e} \eq{1}$$
of the reference unit vector under transformation by
$g$. Let $G_o$ be the stability subgroup of $\ket{e}$, i.e.\ the
subgroup of $G$ for which $\ket{e}$ is an eigenvector. One
verifies that $g_1,g_2$ produce the same physical state if and only if
$g_1^{-1}g_2 \in G_o$. We obtain a set of distinct images
of $\ket{e}$ by selecting one element $g$ from each coset $gG_o$. The
set of distinct images is thus in one-one correspondence with the
coset space ${\cal F} = G/G_o$. This is a homogeneous space, i.e.\
it enjoys a transitivity property, and is referred to as a set of
{\it generalized coherent states}. The coherent states of optics
(Glauber states) are a particular case (see below) in which G is the
Weyl-Heisenberg group, and $\ket{e}$ is the Fock vacuum. The space
${\cal F}$ will now be the arena for quantum mechanics instead of the
the Hilbert sphere ${\cal S}$, i.e.\ the set of all unit rays in
${\cal H}$.
\vskip.1in
Now observe that the restriction to ${\cal F}$ implies that amplitudes
and probabilities have a group-theoretic significance. Thus
$$\bra{g_1}g_2\rangle = \bra{e}g\ket{e} ,\eq{2a}$$
$$p(g_1,g_2) = |\bra{g_1}g_2\rangle|^2 = |\bra{e}g\ket{e}|^2 \equiv
p(g),\; g = g_1^{-1}g_2.\eq{2b}$$
This is unchanged if $g \to g_o g g_o^\prime,\;\; g_o,g_o^\prime \in
G_o$, i.e.\
 $p(g)$ 
{\it is a function on the double cosets}
$G_o \backslash G/G_o$.
The double cosets partition G just as left and
right cosets do. We shall refer to them as {\it coherence
relations}. Thus the quantum mechanical probability is
a function on the coherence relations between the elements of ${\cal
F}$.
\end

 If $\ket{g_2} = e^{i\theta}ket{g_1}$, i.e.\ $g_1$ and $g_2$ produce
the same physical state, then 
$$g\ket{e} = e^{i\theta}\ket{e},\; \; g = g_1^{-1}g_2,\eq{1}$$
which means that $g$ belongs to the so-called {\it stability subgroup}
$G_o$ of G, i.e.\ the subgroup for which the reference state is an
eigenstate.
\end
in which $\ket{e}$ is a reference unit vector in ${\cal H}$. 
$U$
\end

Quantum mechanics is concerned with correlations between events.
  The events are possible states of a system localized 
in space-time that may be space-like, light-like, or time-like to
one another. EPR correlations belong to the first class, and
correlations between dynamically evolving states belong to the second
or third. 
A state is represented by a unit ray $e^{i\theta}\ket{s}$ in a
Hilbert space ${\cal H}$. The correlation $p(s_1,s_2)$ between events
associated with states $e^{i\theta_1}\ket{s_1},e^{i\theta_2}\ket{s_2}$
is given by
$$p(s_1,s_2) = |\bra{s_1}s_2\rangle|^2.\eq{1}$$
If $U$ is the group of unitary transformations of ${\cal H}$,
$p$ is invariant when the two
states are transformed by the same element of U and when the
 the elements are interchanged.
\end
 
relativistic covariance require that the states transform among
themselves under a unitary representation of the Poincar\'{e} group.

 correlations are invariant under the group $U$ of unitary
transformations of ${\cal H}$ together with complex conjugation.

The systems are characterized by the way in which they are prepared.
This information is given by characterizing systems as images
of a reference system under a group G. Thus correlations between
systems 

 Correlations between two systems
are invariant when the same  
g is applied to both. Correlations are then
functions of $g_1^{-1}g_2$. Measuring devices are functions on $g$ to
R.  Any probability on relations must satisfy 
$$\int p(g^{-1}h)d\mu(h) = 1 \forall g \in G$$
Must have invariant measure for this. In qm
$$sup_x|p(gx) - p(x)| = \sqrt}{1 - p(g)}.$$

$$sup_x|d^2(gx) - d^2(x)| = 2d(g)$$
$$sup_x(|d(gx) - d(x)||d(gx) + d(x)|)
\leq d(g)\sup_x|d(gx) + d(x)| \to 2d(g)\;\; g \to e$$

\end
\vskip.1in
Let us give precise statement of the task to be undertaken: We shall
construct a form of quantum mechanics in which the space of allowed
states is a e Hilbert sphere ${\cal S}$, i.e.\ the set of all unit rays
of an infinite dimensional, complex vector space ${\cl H}$. The theory
is formulated so that it enjoys a symmetry under the group U of unitary
transformations of ${\cal H}$. The group U is not locally  compact. Let
G be a locally compact subgroup of U. Let $\ket{0}$ denote a reference
unit vector of

${\cal H}$, and let ${\cal F}$ be the set of distinct unit rays obtained
by the action of elements $g$ of G on $\ket{0}$. We shall construct a
form of quantum theory in which ${\cal F}$ replaces ${\cal S}$ as the
set of allowed states, and 
\  
\end

We shall further assume that
G is a Lie group so that F is a manifold  with a locally Euclidean
metric, and finally we assume that FD

Our goal is to construct a form of quantum mechanics
in which F replaces S as the set of allowed states.

 We shall use the term
$G-restricted$ quantum mechanics (GQM) to mean a quantum theory in
which the set $F$ of allowed states are characterized by  

T
Our goal will be to construct a restricted form of quantum mechanics
which has a locally compact symmetry

\end

\vskip.1in

 associated with
Galilean or Poincar\'{e} invariant  systems are only locally compact
and  compact as well as locally compact it cannot accommodate Galilean
or Poincare
 {\it locally
compact}, however, and the answer to our question is ``yes" {\it if we
restrict the set of allowed states to the set of images of a given
state under the action of such groups}.
\vskip.1in
But if we restrict the set of allowed states in this way we lose the
superposition principle, for while these sets form a manifold in Hilbert
space, {\it they are not linearly closed}. It is then not possible to
implement the notion of ``observable", and the Dirac-von Neumann
interpretation of measurement must be replaced.

\end
\vskip.1in
The 
\end

\centerline{Presented to CHPS February 7, 2002}
\vskip.5in
\centerline{Outline of talk:}
\vnoind
1. Motivation for investigating maolds of generalized
coherent states. The relationship of EPR correlations to the metric.
\vnoind
2. Why measurability demands the existence of canonically
associated Bell states.
\vnoind
3. Why linear dynamics is possible without a
 superposition principle.
\vnoind
4. The mechanism for peaceful coexistence and the nature
of quantum non-locality.
\vnoind
5. Conclusions
\vfill\eject
\pageno=2
Quantum mechanical analogue: 
\vnoind
Outer approach - Dirac-von Neumann.
\vnoind
States represented by
$$\Psi = \alpha_1 \psi_1 + \cdots \alpha_N\psi_N,$$
in eigen-basis of an observable. Measurement seeks to
obtain the eigen-manifold to which $\Psi$ belongs. This
is the eigenvalue-eigenstate link.
\vnoind
Inner approach (to be described): Show that
accessible states form a manifold (not necessarily a linear
subspace!) in Hilbert space and that detectable correlations determine
``distances".
\vnoind
Measuring rod: Generalized Bell states produced, for example,
by spontaneous parametric down conversion (SPDC).
\vfill
\eject

\centerline{Labeling the set of accessible states.}
\vnoind
Assume that all accessible states  are 

\noindent obtained
from a reference state $\ket{0}$ by 

\noindent unitary transformations $g$
belonging to

\noindent a group G. 

$$\ket{g} = g\ket{0} .$$
\vskip.2in
\vnoind
If $g_1\ket{0} = e^{i\theta}g_2\ket{0}$ then
$g = g_2^{-1}g_1$ satisfies
$$ g\kvac = e^{i\theta}\kvac $$
so $g$ is in subgroup $G_o$ of G with  $\ket{0}$ as an eigenstate.
\vnoind
$G_o$ is called the {\it stability subgroup} of the reference
state.
\vnoind
 $g \in G_o$ means
$g_1 = g_2 g$, so $g_1,g_2$ belong to same left coset
of $G_o$ 
\vskip.2in
\noindent
 Select one element $g$ 
from each coset to form the set
 $$F =
G/G_o$$ 
 of state labels.
\vnoind
 F
is a homogeneous space (transitivity) and referred to
as set of
{\bf generalized coherent states}
 \vnoind Concept due to  A.
Perelomov generalizes familiar Glauber states of optics (see
below). 
\vnoind In cases of interest G
is a Lie group and F is a manifold.
\vnoind
States of F not mutually orthogonal and in cases of
interest is infinitely over-complete.
\vnoind However,
{\bf it will not be linearly closed}, i.e.\ linear combinations
are not in general coherent.
\vnoind Thus {\it superposition principle
doesn't
 hold .} 
\vnoind
 Dirac-von Neumann
measurement 
paradigm based on identifying properties with
eigenvalues of observables is not implementable. (It would be
the equivalent of trying to attach coordinates to points of a
surface from a platform in space.)
\vnoind
 Measurement problem goes away
{\bf but} must

\noindent (1) {\it redefine
 measurement} and (2)
{\it verify that linear dynamics takes F into itself.}
\vfill
\eject
\vskip.3in
\vnoind
The group structure of G is imparted to

\noindent quantum mechanical
correlations:
$$p(g_1,g_2) = |\bra{g_1}g_2\rangle|^2$$
as follows:
$$
\bra{g_1}g_2\rangle = \bra{0}g_1^\dagger g_2\ket{0} = 
\bra{0}g\ket{0},\; g = g_1^{-1}g_2.$$
Observe that $p$ depends on $g_1,g_2$ through $g$, and we write
$$p(g) = p(g_1,g_2). $$
It is the same for all $g_1,g_2$ such that $g$ belongs
to the {\it double coset} $G_o\backslash g /G_o$.
\vnoind 
 Double cosets
partition a group in the same way that left and right cosets do.
\vnoind
{\bf Thus each double coset defines a type of relation that we
shall refer to as a {\it coherence relation}. It is about these
relations that we learn from experiment.}
\vfill
\eject
\vskip.2in
\noindent
Because $\bra{g_1}g_2\rangle$ is a scalar product
$$d(g_1,g_2)  \equiv \sqrt{1 - |\bra{g_1}g_2\rangle|^2}$$
defines a {\it metric} on $F$, i.e.\ it is symmetric and obeys
a triangle inequality. 
\vskip.1in
Thus
$$d(g) \equiv d(g_1,g_2) = \sqrt{1 - p(g)}$$
can be regarded as the {\it size} of the coherence relation
defined by $g$ or the distance of the coherence relation $g$
from the identity relation.
\vskip.1in
\noindent
From the metric properties of $d(g_1,g_2)$ there follows:
$$d(g) \geq 0  \eq{i}$$
 with equality if and only if 
 g  is
in the identity double coset.
$$d(g) = d(g^{-1}).\eq{ii}$$
$$d(gh) \leq d(g) + d(h).\eq{iii}$$
\vfill
\eject
\noindent
The set of $h$ such that $d(g^{-1}h)< \epsilon$ is referred to as
an $\epsilon$-sphere about $g$. 
\vskip.1in
\noindent The $\epsilon$ spheres define a system of neighborhoods
by which $G$ becomes a topological

\noindent
 group. One verifies that
$$|d(h) - d(g)| \leq d(g^{-1}h)$$
so that $d(g)$ is continuous in this topology.
\vnoind
Use this to construct a measurement

\noindent paradigm:
\vnoind
Suppose that given any $\epsilon > 0$ there is a {\it finite} set
$g_1,\cdots, g_n$ such that every $g$ lies within an $\epsilon$
sphere about $g_j$ for some $j$. We can then say that $g$ has the
property $g_j$ to within $\epsilon$. It is then easy to show that
\vskip.1in
\vnoind
(1) Closed spheres about the identity 

\noindent $\{h: d(h) \leq r < 1\}$
are compact.
\vskip.1in
\vnoind
(2) F is {\it locally compact}, i.e.\ every point has a
neighborhood whose closure is compact.
\vskip.1in
\vnoind
Important consequence:
  {\it 
Invariant measure $d\mu$ on F exists by which we can
integrate over it F.}
\vfill
\eject
\vskip.2in
\centerline{EXAMPLES}
\vskip.2in
\vnoind
 Single mode laser states. The group G is the Weyl-Heisenberg
group defined by:
$$u(\lambda,\theta) = e^{i\theta}U(\lambda),$$
where $\lambda$ is a complex number and 
$$U(\lambda)U(\mu) = e^{i\phi}U(\lambda + \mu),\;\; \theta =
Im(\lambda^*\mu)$$
By Stone-von Neumann theorem there is 

\noindent only one irreducible representation up to 

\noindent equivalence
obtained by setting
$$U(\lambda) = e^{\lambda a^\dagger - \lambda^* a}$$
in which $a,a^\dagger$ are bose operators $[a,a^\dagger] = 1$ 
and the reference state $\kvac$ is the Fock vacuum for which
$a\kvac = 0$:
$$\ket{\lambda} = U(\lambda)\kvac = e^{-|\lambda|^2/2}e^{\lambda
a^\dagger}\kvac.$$
$$p(\lambda) = e^{-|\lambda|^2}.$$
\vskip.1in
\vnoind
The stability subgroup $G_o$ of $\kvac$ is the group of phase
multiplications $e^{i\theta}I$, so that $F = G/G_o$ is just the
set of states $\ket{\lambda}$ as $\lambda$ runs over the complex
berplane. Thus the Weyl-Heisenberg coherent states are the familiar
Glaustates for a single mode. 
\vskip.1in
\vnoind
One can immediately generalize to the n-mode Weyl-Heisenberg group
by simply replacing $\lambda$ by an $n$-component complex vector,
$$\lambda\cdot a \equiv \lambda_1 a_1 + \cdots + \lambda_n a_n ,$$
and
$$|\lambda|^2 = |\lambda_1|^2 + \cdots + |\lambda_n|^2 .$$
Observe that the n-component complex

\noindent space is locally compact and
the ``sphere"
$$d(\lambda) = (1 - e^{-|\lambda|^2})^{1/2} \leq r < 1$$
is bounded and closed and hence compact as required for the
implementation of our measurement paradigm.
\vfill
\eject
\vskip.2in
\noindent
How linear dynamics is implemented without a superposition
principle:
\vskip.2in
\noindent
Let $H_{ij}$ be any hermitian matrix and let
$${\bf H} = \sum _{ij}H_{ij}a_i^\dagger a_j .$$
\noindent
Then if
$$V(t) = e^{-it{\bf H}}$$
one observes that
$$V(t)U(\lambda)\kvac = 
V(t)U(\lambda)V^{-1}(t)\kvac  
$$ $$ = U(\lambda^\prime)\kvac,\;\;
\lambda^\prime = e^{-itH}\lambda.$$ Thus the 
 dynamical 
 transformations must be {\it
automorphisms of G} 
 that leave the reference state invariant. 
In relativistic quantum mechanics this will be satisfied by
having a Poincar\'{e} invariant vacuum.
\vfill
\eject
\centerline{States that act as ``measuring rods"}
\vskip.1in
\noindent
Measuring devices determine relations between pairs of states.
We shall describe one member of the pair by vectors in the Hilbert
space ${\cal H}$ and the other by vectors in the dual space
$\overline{{\cal H}}$. Pair states will thus be linear combinations
of factored states:
$$\doubleket{g_1,g_2} = \ket{g_1}\otimes\bra{g_2}.$$
$$\doublebra{g_1,g_2} = \bra{g_1}\otimes\ket{g_2}.$$Observe that if T
is any {\it anti-unitary} operator, the map
$$\bra{g} \to T\ket{g} \equiv \ket{g^T}$$
is {\it unitary}. Thus one can also write pair states
using only kets as $$\doubleket{g_1,g_2} = 
\ket{g_1}\otimes\ket{g_2^T}.$$
\vfill
\eject
\noindent
 Now use
existence of invariant measure $d\mu$ 
to construct a remarkable class of states which generalize
the familiar Bell states. 
Consider pair states of the form:
$$\doubleket{B} \equiv C \int_F d\mu \ket{g}\otimes \bra{g}.$$
which will be called {\it generalized Bell states}. Here C is a
normalization constant (see below).
\vnoind
The amplitude in $\doubleket{B}$ for the state
$$\doubleket{g_1,g_2} \equiv \ket{g_1}\otimes \bra{g_2}.$$
is found to be
\vfill
\eject
$$\doublebra{g_1,g_2}B\rangle\rangle
 =  
C\int_F d\mu \bra{g_1}g\rangle\bra{g}g_2\rangle = $$
$$
C\bra{g_1}I\ket{g_2},$$
$$I = \int_F d\mu \ket{g}\bra{g}.$$
From invariance of the measure one sees 
that 
$$gI = Ig$$
for all g. If the reresentation of G is irreducible it follows
from Schur's Lemma that by suitably scaling the measure we can
make I the unit operator. 
\vnoind
Note that this proves the completeness of the generalized coherent
states. They are not mutually orthogonal and it turns out that in
general they are infinitely over-complete, i.e.\ one can delete an
infinite subset without losing completeness.
\vfill
\eject
\noindent
Hence the probability in $\doubleket{B}$ for one member
to be in $\ket{g_1}$ {\it and} the other in $\bra{g_2}$ is 
$$p(g_1;g_2) =|C|^2 |\bra{g_1}g_2\rangle|^2.$$
\vnoind
It follows that in state $\doubleket{B}$ there is equal
likelihood for one member to be in any state $\ket{g_1}$ if nothing
is specified about its partner, but the conditional probability
for the ket-particle to be in $\ket{g_1}$ if its partner is in
$\bra{g_2}$ will be
 $$p(g_1,g_2) = |\bra{g_1}g_2\rangle|^2 = p(g), \; g = g_1^{-1}g_2.$$
\noindent
Thus the state $\doubleket{B}$ exhibits perfect EPR correlation.
and this correlation gives the metric
 $$d(g) = \sqrt{1 - p(g)}.	$$ {\it Thus EPR correlations  act as
the measuring rods for the ``little flat bugs" on the manifold of
generalized  coherent states.
}
\vfill
\eject
One finds that
$$\doublebra{B}B\rangle\rangle = \int_F d\mu$$
which means that $B$ is normalizable if and only if $F$ is
compact.
\vnoind In the cases where F is only locally compact we can
still integrate over $g$ on the
 compact spheres $r < 1$ and then pass to the
limit
$r
\to 1$. This creates no problems because $C$ disappears when we
compute the conditional probability.
\vfill
\eject
\centerline{EXAMPLE - 1 - Bohm-Aharonov singlet.}
\vnoind
$F = U_2/U_1$ coherent states:
$$\doubleket{\beta} =$$ $$ C \int d\Omega (\cos\theta \;\;
e^{-i\phi}\sin\theta)\otimes\left(\matrix{\cos\theta \cr
e^{i\phi}
\sin\theta}\right)$$
$$ = 
2^{-1/2}\left((1,0)\otimes\left(\matrix{1
\cr 0}\right) + (0,1)\otimes\left(\matrix{0 \cr
1}\right)\right)$$
The anti-unitary time-reversal 
 transformation T gives
$$\ket{\uparrow^T} =
\ket{\downarrow},\qquad \ket{\downarrow^T}=  - \ket{\downarrow}$$
so this can also be written:
$$2^{-1/2}(\ket{\uparrow}\otimes\ket{\downarrow} -
\ket{\downarrow}\otimes
\ket{\uparrow})$$
which is the Bohm-Aharonov singlet.

\vfill
\eject
\centerline{EXAMPLE - 2: Weyl-Heisenberg coherent states.}
\vnoind
$$\doubleket{B} = $$
$$C\int d^2\lambda \ket{\lambda}\otimes \bra{\lambda} =$$
$$
C\int d^2\lambda e^{-|\lambda|^2} e^{\lambda
a^\dagger}\ket{0}\otimes \bra{0}e^{\lambda^* a},$$
where $\ket{0}$ is the Fock vacuum.
 Because G is locally compact
but not compact we must perform the limiting process described
above with the result
$$\doubleket{B} = \lim_{\xi \to 1} \doubleket{B,\xi},$$
$$\doubleket{B,\xi} \equiv \sqrt{1 - |\xi|^2}e^{\xi
a_1^\dagger \otimes a_2^\dagger}\ket{0}\otimes\ket{0}.$$
Here we have introduced the creation operators $a_1,a_2$ for a
state and its T-reversed state to put $\doubleket{B}$
into the more familiar form where both particles are described by
kets.
\vfill
\eject

The family of states 
$$\doubleket{B,\xi}
\equiv \sqrt{1 - |\xi|^2}e^{\xi
a_1^\dagger \otimes a_2^\dagger}\ket{0}\otimes\ket{0}$$
are generated from a vacuum $\doubleket{B,0}$ by the group
$SL_2(R)$, which describes spontaneous, parametric
down conversion. The parameter $\tanh^{-1}(|\xi|)$ is related to
the expectation value of the number of pairs produced. The
probability for one member to give 

\noindent photo-current
$\lambda$ given that its partner gives photocurrent $\mu^*$ turns
out to be
$$p_\xi(\lambda|\mu^*) = e^{-|\lambda - \xi \mu|}$$
yielding the metric for $\xi \to 1$
$$d(\lambda - \mu) = (1 - e^{-|\lambda - \mu|^2})^{1/2}.$$
\vfill
\eject
\pageno=20
\centerline{Some insight into the non-locality:}
\vskip.1in
\noindent
$$a_1\ket{0} = 0 = a_2\ket{0}$$
where $\ket{0}$ is the Fock vacuum.
\vskip.1in
\noindent
$a_1,a_2$  have commutation rules
$$[a_1,a_1^\dagger] = 1 = [a_2,a_2^\dagger]$$
$$[a_1,a_2] = [a_1,a_2^\dagger] = 0$$
Make the transformations
$$a_1 \to A_1 = { {a_1 - \xi a_2^\dagger}\over {(1 - |\xi|^2)^{1/2}}}
 $$
$$a_2 \to A_2 = {{a_2 - \xi a_1^\dagger}\over {(1 - |\xi|^2)^{1/2}}}$$
\vskip.1in
\noindent
All of the commutation rules of the $A$'s are the same as the
corresponding ones of the $a$'s, and 
$$A_1\doubleket{\xi} = 0 = A_2\doubleket{\xi}.$$
Hence $\doubleket{B,\xi}$ can be thought of as a 

\noindent ``twisted" vacuum.
That something like the vacuum is non-local does not seem
implausible.
\vfill
\eject

\centerline{How the correlation is transmitted}:

$$\doubleket{\xi} \propto e^{\xi a_1^\dagger
a_2^\dagger}\doubleket{0} = \sum_n{{\xi^n}\over {n!}}a_1^{\dagger
n}a_2^{\dagger n}\doubleket{0}.$$
\vskip.2in
\noindent
This says that  the number of $a_1$ photons is always exactly the
same as the number of $a_2$ photons.
\vfill
\eject
\centerline{ Peaceful coexistence}
\vnoind
Consider a coherent state with
$$\lambda\cdot a = \int_L \,d\Omega \;\lambda(k)\cdot a(k)$$
the integration being over the forward light cone (invariant measure
$d\Omega$). 
\vnoind The correlation between a state with $\lambda$ and a state
with $\mu$ will depend on the Poincar\'e invariant distance
$|\lambda -
\mu|$ where
$$|\lambda|^2 = \int_L \, d\Omega|\lambda(k)|^2.$$
 The computation by which we
arrived at this depended on the commutation rules of the $a's$ and
$a^\dagger$'s and the fact that
$a$'s annihilate the vacuum $\ket{0}$.
\vfill
\eject
\noindent
Under space-time translations and Lorentz transformations
the operators $a,a^\dagger$ are

\noindent  transformed by unitary operators
$$a \to VaV^{-1}$$  so the commutation rules are
unchanged. 
\vnoind

\noindent Since the operators $V$ are exponentials of quadratic
forms
$a(k)^\dagger a(k^\prime)$ the Fock vacuum is invariant.
\vnoind
It follows
that {\it the metric $|\lambda - \mu|$ {\it is a 

\noindent Poincar\'{e}
invariant.}
\vnoind
{\it Thus, although the EPR effect is non-local, it is  manifestly
covariant and hence there can be no super-luminal signalling.}}
\vfill
\eject
\centerline{Metric formulation of quantum cryptography:}
\vskip.1in
\noindent
Suppose Alice modifies $\lambda$ by the phase transformation
$$\lambda \to e^{i\phi}\lambda ,$$
and  that Bob is in possesion of the correlation detector.
Bob finds that the correlation changes because
$$|e^{i\phi}\lambda - \mu| \neq |\lambda - \mu|.$$
He discovers that he can restore the orignal correlation on by
the transformation $\mu \to e^{i\phi}\mu$ because
$$|e^{i\phi}\lambda - e^{i\phi} \mu| = |\lambda - \mu|.$$
Hence he can deduce what Alice did.
\vskip.in
\noindent
But the eavesdropper EVE can detect nothing since 
EVE can only discern a change in $|\lambda|$, and this
is unaffected by what Alice did.
\vskip.1in
\noindent
Observe that these conclusions are also

\noindent reached 
for $|\lambda - \xi \mu|$ so that even a weak SPDC signal
can be used for this.
\vfill
\eject
\centerline{Canonical Hidden-variable models}
\vnoind
Suppose that in each run of an experiment using the generalized Bell
state a random element $h$ of the group G is generated with a
distribution $\rho(h)$ with respect to the invariant measure on G.
Suppose that a pair of detectors in relation $g$ give a conincidence
if and only if $d(g) \leq d(h)$. If $\rho$ is so chosen that
the measure of F for which $d(h) \leq r$ is $r^2$, then the probability
of a coincidence is $1 - d(g)^2 = p(g)$, i.e.\ we reproduce the
quantum mechanical prediction.
\vnoind
Observe that this does not violate no-go theorems because the hidden
variable  depends on the distance $d$ between members of a correlated
pair rather than on the states independently.
\vnoind
This type of model was noted long ago by Bell. What we see here is that
it has a natural generalization to all kinds of coherent states and for
relativistic theories is  acceptably covariant. For $SU_2$ coherent
states $\rho$ is constant and for Weyl-Heisenberg groups it is
Maxwellian. It would be interesting to know the general form for all
Lie groups.

\vfill
\eject
 
\centerline{CONCLUSION}
\vskip.1in
\noindent
The basic idea of this investigation was to reinterpret
the notion of quantum measurement as a process for 
discovering the metric structure of the manifold of accessible
states rather than a process for discovering properties of
individual states. Within this limited scope there is no need
for wave-function collapse to complete the measurement process.
\vnoind
The generalized Bell states serve as measuring rods for determining
the metric, and, although they are necessarily non-local
objects, they carry out their task in a Poincar\'{e}
invariant manner. 
\vnoind
It is possible to construct simple non-local hidden variable
models for this kind of measurement process without violating
Bell or other no-go theorems.
\vnoind
 Thus there is neither a peaceful coexistence problem nor
a measurement problem within
this framework.
\end

. With this limited purpose there is 

that
asking for the coefficients of a state relative to a basis is not
the right question if one wants to understand the manifold of
states to which we {\it actually have access}. Rather we regard
the purpose of measurement as the discovery of the metric
properties of that manifold 
\vskip.1in 
\noindent
Metrical measurements on that manifold  
performed using generalized EPR states do not

can account in principle
for the stochastic behavior by a non-local hidden variable. Thus
no collapse process is needed to complete the measurement, and there
is no measurement problem.
\end
\centerline{Conclusion}
\vskip.1in
\noindent The measurement problem arises from Dirac-von Neumann
interpretation of measurement.
\vskip.1in
\noindent
 In this interpretation the task of
measurement is to determine values of observables represented by
Hermitian opererators. Each such operator has a set of
orthogonal eigenspaces, and the measurement is supposed to find
out in which of these the state is located.
\vskip.1in
\noindent
 However, the measurement process involves an
entanglement between the state and the measuring device, and this
state is not in general in any one of the eigenspaces. It is then
supposed to {\it collapse} into one of the constituents with a
probability that
\end

\noindent indicates the projection of the state on the
associated eigenspace. 
\vfill
\eject
\noindent
Thus the linear deterministic \Schrod \nl evolution of the state
described by the \nl Hamiltonian must, at some unknown time and in some
unknown way change into the non-linear, stochastic collapse process.
\vskip.1in
\noindent
The failure of the Dirac-von Neumann interpretation to account for
this is called ``the measurement problem" by philosophers and the
\Schrod cat problem by \nl ordinary people.
\vfill
\eject 
\noindent
\centerline{NO MEASUREMENT PROBLEM }
\centerline{FOR DISTANCE MEASUREMENTS}
\vskip.1in
\noindent
The basic idea of our investigation was that asking for
the coefficients of a state relative to a basis is not
the right question if one wants to understand the manifold of
states to which we {\it actually have access}.
\vskip.1in 
\noindent
By restricting to metrical measurements on that manifold we
can account in principle
for the stochastic behavior by a non-local hidden variable. Thus
no collapse process is needed to complete the measurement, and there
is no measurement problem.
\vfill
\end
\noindent the
Lorentz group in 2 + 1 dimensions. For small $\xi$ these are
the states produced by

\noindent spontaneous, parametric down-conversion.

\end
\end
\vskip.1in

\end

 2. 
\vskip.1in

\end

\vskip.1in

\end

\end

Weyl-Heisenberg coherent states - Spontaneous parametric down
conversion  Construction off

\vskip.1in
2.

In the  conventional formulation of quantum mechanics
any ray in the Hilbert space ${\cal H}$ can represent a
state. The symmetry group of the theory is the full
unitary group ${\cal U}$ together with complex
conjugation. Since ${\cal H}$ is not even locally
compact there is no procedure by which an arbitrary
state can be located to a given degree of accuracy by
its relation to some sufficiently large but finite set
of states. 
\vskip.1in
While ${\cal H}$ is in this sense too large an arena for quantum
mechanics,  any finite dimensional subspace of ${\cal H}$ is too
small. For such spaces cannot accommodate unitary representations
of non-compact groups such as the  space-time translations and Lorentz
transformations of  the Poincar\'{e} group or the
creation-annihilation operations of the Weyl-Heisenberg group.
Although these groups are non-compact they are locally compact,
metric spaces. This 
suggests the possibility of constructing a form of quantum
mechanics which makes predictions identical to those of
conventional quantum mechanics, but in which {\it the only allowed
states are those that are accessible from a reference state by
transformations belonging to a locally  compact subgroup} ${\cal
G}$ of ${\cal U}$.
\vskip.1in
The set ${\cal F}$ of states obtained in this way will be a locally
compact subset of ${\cal H}$. It will be complete (see below), in fact
infinitely overcomplete, but {\it will not be a subspace of }${\cal
H}$, i.e.\  {\it  it will not be linearly closed}. Thus we will not
have a superposition principle, and it will be incumbent upon us to
show that with a sufficiently large ${\cal G}$ we can describe observed
dynamical processes, interference phenomena, and entanglement {\it
within} ${\cal
 F}$ itself. 
\vskip.1in
The groups ${\cal G}$ that occur in physics are
manifolds, so there will  be a natural metric structure on ${\cal
F}$. We shall then find  that experimentally observed correlations
between pairs of states depend on the distance between them in this
metric. Thus the observation of  correlations between pairs of states
becomes a means of probing the metric structure, an activity we shall
call ``quantum surveying".  
\vskip.1in
Since ${\cal F}$ is not a linear space the Dirac-von Neumann
notion of an ``observable" is no longer meaningful, and quantum
surveying will take the place of the conventional
interpretation of measurement. There is a close analogy here with
the
 surveying of an ordinary surface. When Gauss undertook the
survey of Hanover in 1818 he pointed out that ``the properties of a
surface that are of greatest interest to geometers are those that can
be observed by little flat bugs who live upon it."  The flat bugs do
not assign Cartesian coordinates to the points relative to an origin in
space, but rather set up devices upon the surface with which to
measure  distances between points to which they have access.
The conventional interpretation of quantum measurement
using observables views it as a means of determining subspaces --- the
eigenmanifolds of the observables --- to which the state belongs,
which supplies information about the coordinates of the state vector
relative to the eigenbasis. In contrast quantum surveying does
not attempt to say anything about individual states but only about {\it
metric relations} between them that are manifest in the observed
correlations. We will see that generalized Bell states in which such
correlations are observed become the ``measuring rods" for quantum
surveying.
\vskip.1in
Let $\ket{0}$ denote a reference state in ${\cal H}$ and $\ket{g} =
g\ket{0}$ be its image under $g \in {\cal G}$. This is consistent
with the use of $e$ as the identity of ${\cal G}$. If $g_1,g_2$
produce the same image, i.e. if $\ket{g_1}$ and $\ket{g_2}$ differ
only  by a phase factor, then $\ket{0}$ is an eigenstate of $g =
g_1^{-1}g_2$. The subgroup
${\cal G}_o$ of ${\cal G}$ for which $\ket{0}$ is an eigenstate is called the {\it
stability subgroup} of $\ket{0}$. Evidently we then obtain the set of
distinct  images by selecting one element $g$ from each coset
$g{\cal G}_o$. Thus the  distinct images are in one-one
correspondence with the points of the coset space ${\cal F} = {\cal
G}/{\cal G}_o$ which is referred to as a {\it system of generalized
coherent states} (Perelomov). It is a homogeneous (transitive) space
and will replace the Hilbert sphere (the set of unit rays in ${\cal
H}$). In the following we shall use $g$ to indicate  the state
associated with
$g{\cal G}_o$ when no confusion will result.
\vskip.1in
When we restrict the set of allowed states to ${\cal F}$, the group
structure is imparted to quantum mechanical amplitudes in the
following way:
$$\bra{g_1}g_2\rangle = \bra{0}g_1^\dagger g_2\ket{0} =
\bra{0}g_1^{-1}g_2\ket{0} = \bra{0}g\ket{0},\;\, g =
g_1^{-1}g_2.\eq{1}$$ If $g_o,g_o^\prime \in {\cal G}_o$ then
$$\bra{0}g_ogg_o^\prime\ket{0} =
e^{i\theta}\bra{0}g\ket{0},\eq{2}$$
so that the quantum mechanical conditional probability function 
$$p(g_1|g_2) = |\bra{g_1}g_2\rangle|^2 = |\bra{0}g\ket{0}|^2
\equiv p(g)\eq{3}$$ {\it is constant on double cosets} ${\cal G}_o g
{\cal G}_o$. Thus
$p(g)$ is a function on the partition of ${\cal G}$ defined by the set
 of double cosets
$${\cal S} \equiv {\cal G}_o\backslash {\cal G}/{\cal G}_o$$. We
shall say that $g_1,g_2$ {\it are in the coherence relation } $g$
if
$g_1^{-1}g_2$ belongs to the double coset
${\cal G}_o g {\cal G}_o$. Evidently the experimental determintion of
$p(g_1|g_2)$ can only provide information about the coherence relation
to which
$g_1,g_2$  belongs.
\vskip.1in
Now observe that a scalar procuct $\bra{a}b\rangle$
determines a metric $d(a,b)= (1 - |\bra{a}b\rangle|^2)^{1/2}$
so that we can interpret the experimental determination of $p(g)$
as the means of determining the metric:
$$d(g_1,g_2) = (1 - p(g_1^{-1}g_2))^{1/2} \equiv d(g).\eq{4}$$
We can interpret $d(g)$ as the ``diameter" of the coherence relation
$g$ or as its distance from the identity relation $e$. 
\vskip.1in
The function $d(g)$ has remarkable properties deriving from 
the metric properties of $d(g_1,g_2)$ combined with its dependence
on the group-theoretic combination $g_1^{-1}g_2$. Thus
one readily deduces:
$$d(g) \geq 0. \eq{5a}$$
$$d(g) = d(g^{-1}),\eq{5b}$$ 
$$d(g) + d(h) \geq d(gh).\eq{5c}$$
Observe that one can deduce from (5) that
the subset of ${\cal G}$ for which $d(g) = 0$ form a subgroup of
${\cal G}$ which is the stability subgroup ${\cal G}_o$ of the
reference state. 
\vskip.1in
We now make the crucial assumption that the space ${\cal S}$ is
locally compact and connected in the topology defined by $d$, i.e.\
taking as neighborhoods the ``spheres" for $r \in (0,1)$
$N_r(g) = \{h \in {\cal G}| d(g^{-1}h) < r \}$. The following
theorem is proved in the appendix:
\vskip.1in
{\it Theorem}: $N_r(e)$ is compact for all $r < 1$. 
\vskip.1in
This theorem can be exploited to construct a measurement paradignm.
For given any $r < 1$ and any $\epsilon > 0$  we can select a {\it
finite} set of elements $g_1,\cdots,g_n$ such that every point in
$N_r(e)$ lies within $\epsilon$ of $g_j$ for some $j$. The
assignment of a $j$ value to $g$ in this way shall be called
a $g_j$-measurement. Our next task is to explain how this may be
carried out.
\vskip.1in
In each experiment we assume that there are two kinds of detectors
for coherent states which we will call $R$ and $L$ detectors
represented by $\ket{g}$,  $\bra{g^\prime}$ with $g,g^\prime \in
{\cal F}$. Now consider the pair state 
$$\doubleket{B} = \int_{\cal F}d\mu \ket{g}\otimes\bra{g},\eq{6}$$
where $d\mu$ is an invariant measure on ${\cal G}$ {\it the
existence of which is insured by local compactness}. 
The amplitude for $ \doubleket{g_1,g_2}
\equiv \ket{g_1}\otimes\bra{g_2}$ in 
$\doubleket{B}$ is seen to be
$$\doublebra{g_1,g_2}B\rangle\rangle = \bra{g_1}I\ket{g_2} \eq{7}$$
where
$$I = \int_{\cal F}d\mu \ket{g}\bra{g}.\eq{8}$$
From the invariance of the measure one readily checks that $I$
with $h$ for all $h \in {\cal G}$. From the assumed irreducibility
of the representation and Schur's Lemma it follows that by
suitably scaling the measure $I$ can be taken to be the unit
operator. With this choice it follows that the probability for
detection of $\ket{g_1}$ and $\bra{g_2}$ in the state
$\doubleket{B}$ is $p(g_1|g_2)$ given by (3). 
\vskip.1in
Now observe that if $T$ is an {\it anti-unitary} transformation,
the map
$$T\ket{g} \to \bra{g} \eq{9}$$
is {\it unitary}. Thus one may think of $L$ states as simply
$T$-reversed $R$ states.
\vskip.1in
If ${\cal G}$ is $SU_2$ and the two dimensional representation is
selected one will find that $\doubleket{B}$ is the Bohm-Aharonov
singlet, with $T$ being time-reversal. In this case ${\cal F}$ is
the Poincar\'{e} sphere and $d(g)$ is the chordal metric.
\vskip.1in
Let us now see how all of this works out for an important
non-compact group namely the Weyl-Heisenberg group. The coherent
states for this group are the familiar Glauber states of optics.
The elements of the Weyl-Heisenberg group are labeled by a complex
number $\lambda$ and a phase $\theta$ and are represented in terms
of Bose operators
$a,a^\dagger$ by $e^{i\theta}u(\lambda)$ where
$$u(\lambda) = e^{\lambda a^\dagger - \lambda^* a}, \;
[a,a^\dagger] = 1. \eq{10}$$
The states are 
$$\ket{\lambda} = u(\lambda)\ket{0},\eq{11}$$
The composition law is obtained from
$$u(\lambda)u(\mu) = e^{i\theta}u(\lambda + \mu), \;\; \theta =
Im(\lambda^{*}\mu).\eq{12}$$
One then obtains
$$p(\lambda) = e^{-|\lambda|^2} \eq{13}$$
so that 
$$d(\lambda) = \left(1 - e^{-|\lambda|^2}\right)^{1/2}.\eq{14}$$
Observe that the theorem above in this case is just the statement
that $|\lambda|$ is finite for $d(\lambda) < 1.$
\vskip.1in
Let us construct the state $\doubleket{B}$ for this group. Observe
that
$$ \doublebra{B}B\rangle\rangle = \int_{\cal F}d\mu.$$
Thus the state will be normalizable if and only if ${\cal G}$ is
compact. For the Weyl-Heisenberg group.

This example can be generalized to multi-mode bose systems by
simply replacing $\lambda$ by an N-component complex vector with
$|\lambda|$ being the norm.

\end
1. Surveying: - Relations vs Coordinates.
\vskip.1in
\indent
Gauss' point of view and its quantum mechanical analogue. Manifolds

\hskip.2in of generalized coherent states.

Metric structure of coherence relations.

Determination of states in finite terms --- local compactness.

Example: Multimode laser states.

\vskip.1in
\noindent
2. Generalized Bell states as ``measuring rods" for coherence
relations
\vskip.1in
\indent
Example (i):
(a) $SU_2$ coherent states ---  Bohm-Aharonov entanglement.

\indent Example (ii): Weyl-Heisenberg coherent states
--- entanglement 

\indent
\hskip.2in produced by
 spontaneous parametric down conversion.
\vskip.1in
\noindent
3. Implications for the measurement problem and EPR.
\vskip.1in
\indent
Abandoning the Dirac-von Neumann paradigm --- linear dynamics

\indent  \hskip.2in without the superposition principle.

\indent Vacuum-like non-locality of generalized Bell states. 
\vfill
\eject
\centerline{ABSTRACT}

It is possible to construct a form of quantum mechanics in which the symmetry
group
G is  locally compact  in contrast to the full unitary group which is not. The
set M of states
thereby obtained is a manifold of generalized coherent states. Because M is
 not a linear  space we are compelled to
 give up the superposition principle. However  when  the group G is large enough
it
can accommodate all interference phenomena that we actually observe.
Without the superposition principle  "observables" in the
Dirac von Neumann sense do not exist. However, we can regard measurement as a
means
by which the metric geometry of the manifold of coherent states is
experimentally probed by
its manifestation in EPR correlations. Generalized Bell states then become the
mesuring rods
for what may be termed "quantum surveying".
\vfill
\eject
\centerline{1. Surveying: Relations vs Coordinates}
\vskip.1in

\end

By local compactness there is for each $g$ an $r$ in $(0,1)$
such that the closure of $N_r(g)$ is compact.

 Now let us use this
to show that 

To prove that $d(g)$ is continuous at $g$ in $N_r(e)$ for all $r
\in (0,1)$ must show that for any $\epsilon > 0$ there is a
$\delta$ such that $|d(g) - d(h)| < \epsilon$ for $h \in
N_\delta(g)$. From (5) it follows that $|d(g) - d(h)| \leq
d(g^{-1}h)$ so that it suffices to choose $\delta = \epsilon$

\end

 Let us next exploit the
assumed local compactness of
${\cal G}$ to construct a measurement paradigm. Local compactness
means that for each $g$ there is a neighborhood whose closure is
compact. In particular if $g = e$ suppose the subset $S_r$ of $g$
for which
$d(g)
\leq r < 1$ is compact for any $r$. Fixing $r$, we can, for any
$\epsilon > 0$, cover $S_r$ with a finite number of neighborhoods 
$s_1,\cdots, s_N$ and choose a point $g_j$ in each $s_j$ such that
all points $g$ in $s_j$ are within $\epsilon$ of $g_j$, i.e.\
$d(gg_j^{-1}) \leq \epsilon$. For an arbitrary $g$ it follows that
a determination of the N quantities $d(gg_j^{-1})$ allows us
to decide in which $s_j$ the $g$ will be found except for a set of
measure zero. One then may regard $g_j$ as an $\epsilon$ 
approximant to $g$. What quantum surveying does is, for each
desired degree of approximation, provide a finite set of tests
by means of which such an approximation can be found.
\vskip.1in
Our next task is to describe an experimentally implementable
``measuring rod" for $d(g)$. 
\end

 find a finite set of points $g_1,g_2,\cdots g_N$
(with N depending on $\epsilon$) in $S_r$ 

The set of elements $g$ for
which $d(g) \leq d_o <1$ is  
\end

In all cases of interest in physics the group ${\cal G}$ will be a
Lie group. The points of the manifold a ssociated with ${\cal G}$
will be labelled by a finite set of parameters associated with
one parameter subgroups. Suppose 
${\cal G}$ contains the
Poincar\'{e} group. 
$$u(\Lambda,x)a_k u^{-1}(\Lambda,x) = e^{ik\cdot x}a_{\Lambda k}.$$
\end

 each element $g$ of ${\cal G}$ will have
(among others)

\end

Since it is $p(g)$ that determines the
correlations we observe, it follows that 

 (\end

Each experiment E will involve a pair of states labelled $(g_1,g_2)$
which are identified with a pair of detectors in the
laboratory.  Each run of an experiment produces a one or a zero
by comparing $g_1^{-1}g_2$ with a random h

\end

$\ket{0}$ by some element $g$ of a locally
compact subgroup
${\cal G}$ of ${\cal U}$. We then write $\ket{g} = g\ket{0}$.
We may consistently use $e$ for the identity of the group. 
\vskip.1in
Observe that if $g_1,g_2$ produce the saeme state when acting on
$\ket{0}$ then $g_1^{-1}g_2$ is an element of the subgroup ${\cal G}_o$ of
${\cal G}$ that {\it stabilizes } $\ket{0}$. It follows that the distinct
images of $\ket{0}$ are in one-one correspondence with the cosets
$gG_o$. Thus we obtain an unambiguous labelling by selecting one
element from each coset. Viewed as a set of states the coset space
$F = G/G_o$ is called a set of ``generalized coherent states"
(Perelomov). We write $g \approx g^\prime$ to indicate that
$g,g^\prime$ belong to the same coset. 
\vskip.1in
 
$$d(g_1,g_2) \geq 0,\eq{1a}$$
$$d(g_1,g_2) = 0 \hbox{ iff } g_1 \approx g_2,\eq{1b}$$
$$d(g_1,g_2) = d(g_2,g_1),\eq{1c}$$
$$d(g_1,g_2) + d(g_2,g_3) \geq d(g_1,g_3).\eq{1d}$$ 
By G-symmery we have
$$d(gg_1,gg_2) = d(g_1,g_2)$$
whence
$$d(g_1,g_2) = d(e,g_1^{-1}g_2) = d(e,g) \equiv d(g),\; g =
g_1^{-1}g_2.\eq{2}$$
From the metric properties the function $d(g)$ satisfies:
$$d(g) \geq 0,\eq{3a}$$
$$d(g) = 0 \hbox{ iff } g \approx e,\hbox{ i.e.\ } g \in G_o,\eq{3b}$$
$$d(g) = d(g^{-1}),\eq{3c}$$
$$d(g) + d(h) \geq d(gh). \eq{3a}$$
Using this we deduce that $d(g)$ is constant on double cosets of ${\cal G}_o$,
i.e.\
$$d(g_ogg_o^\prime) = d(g),\quad g_o,g_o^\prime \in G_o.\eq{4}$$
\vskip.1in
Let us now imagine a scenario by which the function $d(g)$ might be
manifest experimentally. Suppose that in each run of an experiment
a pair of events labelled $(g_1,g_2)$ occur. Since G is locally
compact there is an invariant measure on G which induces such a
measure $d\mu$ on $F$. Then the probability $dP$ for a pair of events
in a neighborhood $d\mu(g_1)d\mu(g_2)$ will be given by
$$dP = p(g_1,g_2)d\mu(g_1)d\mu(g_2),$$
with some distribution function $p$. If G-invariance is respcted we
will have $p(g_1,g_2) = p(gg_1,gg_2)$ whence from the invariance of
the measure we find that the probability of an event with given 
$g = g_1^{-1}g_2$ is
$$dP = M p(g)d\mu(g)$$
where
 $$ p(g) = p(e,g_1^{-1}g_2),\; \; M = \int_F d\mu(g).$$
In other words each $g_1$ and $g_2$ is equally likely but
the probabilty of the {\it relation} $g$ is $p(g)d\mu(g)$.
 
\end

\vskip.1in
 The space
$S$ acts like a self contained universe in that it is no longer
necessary to separate the observer from the system observed. Indeed
because $S$ is not a linear space the Dirac-von Neumann notion of an
observable is meaningless. 
\vskip.1in

 Since G is locally compact there is
an invariant measure $d\mu$. What will ``happen" in S is this: The
probability of a pair of states being in
$d\mu(g_1)d\mu(g_2)$ will be $p(g_1,g_2)d\mu(g_1)d\mu(g_2)$ such that
$\int p(g_1,g_2)d\mu(g_2)$ is independent of $g_1$, and the
similar integral over $d\mu(g_1)$ is independent of $g_2$. In other
words the theory will tell us nothing about the probability of $g_1$
and $g_2$ independently, but will tell us something about the
probability of a relation between them. To see what that relation is
observe that the invariance of the theory under ${\cal G}$ means that
$$p(gg_1,gg_2) = p(g_1,g_2)$$
which means that 
$$p(g_1,g_2) = p(e,g) \equiv p(g),\; g = g_1^{-1}g_2.$$
Thus the theory provides us with a function $p(g)$ on ${\cal G}$ which gives
the probability that the {\it relation} $g$ ``happens" without
specifying which particular pair $(g_1,g_2)$ is the particular
{\it manifestation} of it.
\vskip.1in
All of the structure of the quantum theory on G emerges from properties
of the function $p(g)$.

Let us first observe that the group theoretic
structure precludes the possibility of deriving $p(g)$ from a Boolean
hidden variable substructure. To see this suppose there were an
assignment of sets $\Lambda(g)$ and a measure $\Gamma$ on the
sets such that 
$$p(g_1,g_2)\, {\buildrel \rm ? \over = }\,
\Gamma(\Lambda(g_1)\cap\Lambda(g_2)).$$ By the symmetry above
putting $g_1 = g_2$ implies that
$\Gamma(\Lambda(g_1))$ is independent of $g_1$ making the theory
trivial. 
\vskip.1in
Because of the local compactness 
$$d(g_1,g_2) \equiv \sup_h|p(g_1,h) - p(g_2,h)|$$
exists. 
\end

\end

\end
\vskip.1in

 characterized by metric relations to other states and the relations
themselves determine the correlations we observe

 The subgroup ${\cal G}_o$ of ${\cal G}$ that leaves $\ket{0}$
invariant (i.e.\ multiplies it by a phase) is called the {\it stability
subgroup of} $\ket{0}$. The set of distinct states is in one-one
correspondence with the cosets $F = G/G_o$, i.e.\ we can label the
states by choosing one element from each coset $gG_o$. The set of
states so labeled is called a {\it system of generalized coherent
states}.
\vskip.1in
Suppose then that we wish to reformulate quantum mechanics on $F$
instead of ${\cal H}$. The problem we immediately confront is that
$F$ {\it is not a subspace of }${\cal H}$. Thus even though it will
turn out that the coherent states are infinitely overcomplete, linear
combinations of coherent states are not in general coherent states.
Thus to formulate quantum mechanics on F we must abandon the
superposition principle.
\vskip.1in
To justify this we will show that if ${\cal G}$ is big enough the
interference effects we observe can be accommodated within it, i.e.\
without forming superpositions. 
\vskip.1in
Without linear closure the supply of operators available to describe
observables will be drastically reduced, in general to the one
dimensional projectors onto the allowed states. Thus the Dirac-von
Neumann characterization of measurement using observables becomes
meaningless and we must find a replacement.
\vskip.1in

\vskip.1in
What the Dirac-von Neumann measurement scheme seeks to accomplish is to
find out what subpaces of Hilbert space a given state vector belongs to.
This is analogous to characterizing the points of a surface by
cartesian coordinates relative to some basis. When Gauss was asked to
survey Hanover in 1818 he \d that this is not the best
characterization. Rather ``the properties of a
surface most worthy of study by geometers are those that can be
measured by little flat bugs who live on it." In other words it is
the metrical relations that are important. 
\vskip.1in
Thus since the DvN scheme is not available to us once we have replaced
${\cal H}$ with $F$, we can make a virtue of necessity and redefine
measurement as the process of extracting the metrical relations of the
space of states through experimentally obtained correlations. I call
this ``quantum surveying". 

 "quantum surveying". Since all measurements will take place within
$F$ so that there is no necessity to separate the observer from the
measuring devices, we will see that $F$  will act like a
``self-contained" quantum universe. This is not a topological property I
use the term ``self-contained" rather than ``closed", because it is not
a topological term.

Replacing the Dirac-von Neumann notion of measurent with quantum
surveying is analogous to r
\end
of

 which means that
we shall not be able to approximate it to any given
accuracy by means of 

From Wigner's theorem the symmetry group ${\cal U}$ of
quantum mechanics is the full unitary group $U$ together
with complex conjugation, so that $U$ is the component
of the identity. Since
$U$ is not locally compact it is clear that it has
elements that could under no stretch of the imagination
be implemented in a laboratory. One is motivated,
therefore, to consider the possibility of constructing a
form of quantum mechanics in which the symmetry group is
a locally compact subgroup ${\cal G}$ of $U$ together with
complex conjugation. If ${\cal G}$ is large enough to contain
the space-time translations plus Lorentz
transformations, i.e.\ the Poincar\'{e} group, I will
call the version of quantum mechanics we obtain a {\it
quantum universe}. 
\vskip.1in
To construct a quantum universe we start with a 
reference state $\ket{0}$ in the Hilbert space ${\cal H}$
and consider the set of states $\ket{g} = g\ket{0}$ as
$g$ ranges over ${\cal G}$. Even if ${\cal G}$ is such that
this set of states is complete or even overcomplete, it
will not be a linearly closed set, i.e.\ in general
linear combinations of such states are not in the set.
Thus we must give up the superposition principle. 
\vskip.1in
The superposition principle is a pillar of quantum
mechanics. Thus to be interesting physically we must show
that despite the absence of linear closure, a quantum
universe  can account for interference effects
that we know how to implement in the laboratory. 
\vskip.1in
Once we
have done this we have an opportunity to take a fresh
look at the measurement problem which arises fs iom th
superposition principle. The Dirac-von Neumann (DvN)
interpretation of measurement regards it as a process for
determining the value of observables which are
identified with linear (Hermitiana) operators. The
absence of linear closure in a quantum universe
forecloses this interpretation. 
\vskip.1in
We shall replace it with the following: In a quantum
universe we shall have two kinds of objects called
polarizers and analyzers
each of which can be labeled by an element g of G that
describes how it is obtained from a reference object.z
The polarizer and analyzer states labeled by the same g
are called ``duals" and are supposed to be related by
an anti-unitary map $\ket{g} \to \bra{g}$. Note that
given one such correspondence we can generate all others
namely $\ket{g} \to \bra{g}h$ for any $h\in G$. We may
take this into account by agreeing that $h$ is absorbed
into our understanding of the $\ket{},\bra{}$
relationship.

Since G is locally compact there is an invariant
measure $d\mu$ and we can define the notion of
probability using this measure. We then construct a
function
$\rho(g_1,g_2) \geq 0$ such that
$\rho(g_1,g_2)d\mu(g_1)d\mu(g_2)$ is the probability
that the 

 The theory  is supposed to produce a 

 There is a function
$f(g_1,g_2)$ which is invariant when its members are
subjected to

${\Psi}
\in {\cal H}\otimes {\overline{\cal H}}$ e
 which is invariant under the
subgroup $G^2$ of $G\otimes G.$
\end

 Considser the following
task: Decide whether two arbitrarily given unit vectors
in a Hilbert space
${\cal H}$ are a distance apart smaller than some number
$0 <
\epsilon < 1$ .  Suppose for example we choose a
basis and determine the coefficients of the vectors
relative to the first N basis elements. If it happens
that these are equal for the two vectors we will learn
nothing about the actual distance between them from the
measurements. In fact there is no finite set of
measurements which will provide this kind of information
for an arbitrary pair of vectors. 
\vskip.1in
The impossibility of making such decisions in a finite
number of steps shows that ${\cal H}$ is much too large
a space in which to formulate quantum mechanics. The
cause of the above limitation on our ability to
learn what we need to learn in a finite number of steps
is the fact that it is not {\it locally
compact}.
 On the
other hand the finite dimensional vector spaces ${\cal
H}_N$   are not only locally compact but also
compact. Hence they are too small to accommodate the
states we produce by turning on photocurrents, applying
Lorentz transformations, or any of the other non-compact
groups that we need to account for what we observe.
 But if we need an infinite number of linearly
independent states to characterize the system, and {\it
also} demand that our space be a linear space, we appear
to be stuck with
${\cal H}$.
\vskip.1in 
The only way out is to give up the requirement that the
space be a linear space, i.e.\ to give up the
superposition principle. The task addressed in this
paper is that of showing that there
 are locally compact spaces
large enough to accommodate all of the
states that we can produce in the laboratory (including
states produced by  interference) without requiring that
the space be linear.
\vskip.1in
The symmetry group of ${\cal H}$ is the group $U$ of 
unitary transformations of any dimension. The space we
seek will have a symmetry group G which is a locally
compact subgroup of $U$. The set of allowed states will
then be the images of some reference state under the
action of elements $g$ of ${\cal G}$. We can then use $g$ as a
label for the state produced by it, but must keep in
mind the possiblity that distinct elements $g,g^\prime$
have the same effect on the reference state and in such
case write $g \approx g^\prime$. If $e$ is the identity
of ${\cal G}$ the elements satisfying $g \approx e$ form a
subgroup ${\cal G}_o$ of G called the {\it stability subgroup}
of the reference state. One observes that if $g \approx
g^\prime$, then the two elements belong to the same
coset $gG_o$. Thus the set of distinct states are in
one-one correspondence with the coset space $F = G/G_o$
obtained by selecting one representative from each coset.
This set of states is referred to as {\it system of
generalized coherent states}.
\vskip.1in
We will see that $F$ contains a complete set of states,
but linear combinations of its elements are not in
general elements of $F$. Thus F is a manifold not a
subspace in ${\cal H}$. We can think of it as a surface
and imagine methods of studying its properties by
analogy with surveying.
\vskip.1in
In 1818 Gauss was asked to survey Hanover and made the
profound discovery that ``the properties of a surface
most worthty of the attention of geometers are those
that can be obtained by little flat bugs living upon
it." In other words it is the metric properties that
are important. Thus even if one could set up a platform
in space from which to send out light beams and assign
cartesian coordinates to points on the surface relative
to some basis, this information would have to be
manipulated into metric terms to reveal the structure of
the surface.
\vskip.1in
The quantum mechanical analogue of attaching cartesian
coordinates is the use of observables to find out
projections of a state on a basis. But there is no
observable other than that associated with the unit
operator for which all the points of
${\cal F}$ will be eigenvectors. Thus the Dirac-von
Neumann measurement paradigm based on observables will
not be available, and we shall have to replace it with a
measurement paradigm that can be used by little flat
bugs who carry detectors from one point to another to
investigate the metric structure of F.
\vskip.1in
Once we have demonstrated that the the intrinsic
geometry of ${\cal F}$ can be determined by its resident
little flat bugs, it will be worthy of being called a
closed quantum universe if G is large enough to include
the space-time transformations and such others that are
needed to characterize internal symmetries of
the particles one
 supposes to inhabit it.
\vskip.1in
The fact that G is both the symmetry group of the
theory and the means by which the states are labelled
puts a powerful constraint on the form of the metric.
Thus suppose the distance between elements labelled
$g_1,g_2$ is given by a metric $d(g_1,g_2)$. Symmetry
under G means
$$d(gg_1,gg_2) = d(g_1,g_2),\; \forall
g,g_1,g_2.\eq{1a}$$
Hence
$$d(g_1,g_2) = d(e,g) \equiv d(g) , \;\; g =
g_1^{-1}g_2.\eq{1b}$$
Thus the important fact emerges that what one measures
is something about the {\it relation} $g_1^{-1}g_2$
between states which may  be the same for other pairs.
\vskip.1in
Let us see what the fact that $d(g_1,g_2)$ is a metric
tells us about $d(g)$.
$$d(g_1,g_2) \geq 0 \Longrightarrow d(g) \geq 0
.\eq{2a},$$
$$d(g_1,g_2) = 0 \Longleftrightarrow g_1 \approx g_2
\hbox{ implies }d(g) = 0 \Longleftrightarrow
g \in G_o.\eq{2b}$$
$$ d(g_1,g_2) = d(g_2,g_1) \hbox{ implies } d(g) =
d(g^{-1}).\eq{3}$$
And finally from the triangle inequality
$$d(g_1,g_2) + d(g_2,g_3) \geq d(g_1,g_3)\eq{4a}$$
we have
$$d(g) + d(h) \geq d(gh).\eq{4b}$$
An easy corollary is 
$$|d(gh) - d(g)|\leq d(h),\;\; |d(hg) - d(g)|\leq
d(h),\eq{5}$$
from which there follows
$$d(g) = d(g_o g g_o^\prime), \;\; g_o,g_o^\prime \in
G_o. \eq{6}$$

The double cosets $G_o\backslash g/G_o$ partition
a group just as left and right cosets do, and  we see
that the function $d(g)$ being constant on double cosets
is a function on this partition of G. We shall say
that two $g_1,g_2$ are in {\it coherence relation} $g$ if
$g_1^{-1}g_2$ lies in the double coset generated by $g$.
When there can be no confusion we shall use $g$ to
denote the double coset to which it belongs.
\vskip.1in
From its metric structure we can identify $d(g)$ with a
``size" of the coherence relation $g$. We shall call it
the ``diameter" of $g$. Now suppose there exists a
distribution on G such that the probability of finding
two randomly chosen $g_1,g_2$ in the coherence relation
$g$ is determined by $d(g)$. Then the little flat bugs
need only have access to a generator of random $g_1,g_2$
with this distribution to determine the metric.
\vskip.1in
I will now show that for every system of generalized
coherent states there exists a device that acts as such
a generator, namely a type of generalized Bell state.
\end

 F We see that what
the little flat bugs determine  these it is the coherence
relations that are what the little flat bugs on F will be
determining from measurements. From its metric
properties we can think of
$d(g)$ as the ``diameter" of the coherence relation
generated by $g$.
\vskip.1in
Our next task is to see what sort of
 measuring device
 for determining $d(g)$ must be made available to the
little flat bugs to carry out their task and it is here
that we must take into account the statistical aspect
of quantum mechanics. Each measurement  gives
an unpredictable outcome but on average the result us
is $d(g)$. Thus $g_1,g_2$ can be selected randomly and
one discovers a correlation $d(g)$ depending only the
coherence relation between them.
\end

.function
$d(g)$ is constant on {\it double cosets} $G_o\backslash
g/G_o$. 
\vskip.1in

\end

 symmetry group of the theory
about the metric on
$F$e 
\end

 If quantum mechanics can be
confined to $F$ we must find a measurement scheme that
does not depend on linearity. That means that we cannot
employ the Dirac-von Neumann (DvN) paradigm in which we
learn about a state with devices that detect eigenvalues
of observables. 
\end

e 

 a consequence of the fact that because of the
infinite dimensionality, Hilbert space is not locally
compact.
 According to the Dirac-von
Neumann (DvN) paradigm we obtae in information about the
states of a system from devices that register
eigenvalues of observbles. If the system is not in an
eigenstate the result of the measurement is collapse of
the the state vector into one of the  eigenstates with a
probability given by the squared modulus of the
corresponding expansion coeffient. The failure of the
theory to explain how the linear, deterministic \Schrod
evolution  changes into non-linear, stochastic collapse
is called the ``measurement problem" of quantum
mechanics. 
\vskip.1in
The  ``dynamic reduction" school attacks the problem
by replacing the
\Schrod equation with a stochastic, non-linear equation
incorporating some form of intrinsic noise. The
so called FAPP  approach
regards the observed system as intrinsically open so that
complex interactions with the environment makes
phase information unavailable ``for all practical
purposes".
\vskip.1in
In this paper I am going to take a fresh look at the
measurement problem by re-examining the DvN measurement
paradigm for quantum mechanics in the light of an
analogy with the problem of surveying a surface.
\vskip.1in
When Gauss was asked to survey Hanover in 1818 he argued
that ``the properties of a surface most worthy of study
by geometers are those that can be determined by little
flat bugs who live on it". In other words it is the
{\it metric structure of the surface that is
significant}. Hence even if one had a platform in space
from which to send out light beams and assign
cartesian coordinates to points on the surface relative
to some arbitrarily selected basis, the data would have
to be converted to metrical form to reveal the essential
structure.
\vskip.1in
The DvN paradigm is the quantum mechanical equivalent of
surveying by assigning cartesian coordinates. The
problem is that there is only one observable
(represented by the unit operator) for which every state
is an eigenstate and hence no way to experimentally
assign coordinates to all the states. Thus 
information gotten from observables will not elicit the
geometric structure of the quantum mechanical equivalent
of a surface, i.e.\ a manifold in Hilbert space. To
determine this structure we must look at such manifolds
from the perspective of ``little flat bugs" who live on
them.
\vskip.1in
 What little flat bugs can do is to move their
instruments around on the surface. Thus {\it the various
points to which the instruments are carried can be
labelled by the procedure used to get there from a
reference point}. Assuming that all points can be gotten
to from one another, there will be a group G describing
the transformations and we may then attach to each point
the set of elements of ${\cal G}$ by which it is obtained.
\vskip.1in
 If two group elements $g,g^\prime$ have the same effect
on the reference point we write $g \approx g^\prime$. The
set ${\cal G}_o$ of elements for which $g \approx e$ is a
subgroup of G called the ``stability subgroup" of the
reference point. Then the points for which $g \approx
g^\prime$ belong to the same coset $gG_o$, and we thus
obtain a one-one correspondence between labels and
points by selecting one element from each coset. Thus
the set of states is labelled by the points of the coset
space $F = G/G_o$. This set of states is referred to as
a ``system of generalized coherent states" (Perelomov).
\vskip.1in
It is a fundamental tenet of quantum mechanics that the
set of quantum states of the universe lie in a Hilbert
space ${\cal H}$. The corresponding manifold would be
the Hilbert sphere, the set of all unit
rays in that infinite dimensional complex vector space.
The theory is then invariant under the 
group ${\cal U}$ of all unitary transformations
regardless of dimension. This is a huge group which is
not even locally compact. 
\end

 Although this is bounded, it is
not even locally compact, i.e.\ infinite sequences do not
necessarily have convergent subsequences.

Having placed instruments at various points on the
surface the little flat bugs must now determine the
metric by finding out how the distance between points
depends on the labels attached to them. Thus if two
points are labelled by $g_1,g_2 \in G$ there will be some
experimental procedure for determining the metric
function $d(g_1,g_2)$
\end

\Schrod evolution

By a closed quantum universe I mean a subset ${\cal S}$
of the Hilbert space ${\cal H}$ which is large enough
to include states that we can in principle produce but no
others, and . Thus the states in ${\cal S}$ will be
obtainable from some reference state by a locally
compact group
${\cal G}$ of transformations
\vskip.1in

 While finite dimensional complex
vector spaces are compact and hence too small to
accommodate all physical phenomena, infinite dimensional
Hilbert space is not even locally compact and hence is
so large that it accommodates objects that cannot be
realized in any conceivable laboratory. We are therefore
motivated to look for an appropriately sized ``arena"
for  quantum phenomena that is only as large as it has
to be to describe what we can in principle measure. To 

${\cal H}$  is much too large a space to be
the arena
 for  phenomena we actually
observe in laboratories. It is not even 
 locally compact, and this makes it possible to find
 grotesque objects in it. Its symmetry group ${\cal U}$, the
 group of unitary transformations of unrestricted
dimension, is correspondingly large, containing
transformations that are inconceivable to realize.
Hence the form of kinematics gives no
essential guidance to the form of dynamics other than
dictating that it be implemented by unitary
transformations.
\vskip.1in
If we grant that ${\cal U}$ should be replaced by some
locally compact subgroup ${\cal G}$ as the symmetry group of
the theory, we must confront the loss of the
superposition principle. For in general not all states
will transform among themselves under ${\cal G}$. A set of
states that does so will not form a subspace
but rather a manifold  in
${\cal H}$. We can think of it as
a surface ${\cal F}$ in ${\cal H}$ generated by the
action of transformations $g$ belonging to ${\cal G}$ on a
reference state associated with the identity element
$e$. We leave open the possibility that distinct
elements $g,g^\prime$ may have the same effect on the
reference state and thus are equivalent labels for the
same state. In such case we write
$g \approx g^\prime$. The set of elements equivalent to
the identity are a subgroup ${\cal G}_o$ called the ``stability
subgroup" of the reference state. The distinct states
are then in one-one correspondence with set of
cosets ${\cal F} = G/G_o$. In other words we
can obtain
${\cal F}$
 by selecting one element from each coset
$gG_o$. The set of states ${\cal F}$ is referred to as
a {\it system of generalized coherent states}.
\vskip.1in
  We will see that even though 
${\cal F}$ is not  linearly closed we can accommodate
observed dynmical processes by making ${\cal G}$
sufficiently large but still locally compact.
However, since it is not a linear space it will not be
possible to explore its properties using the Dirac-von
Neumann paradigm. Thus we cannot investigate it
 by looking for eigenvalues of suitable
observables. We are in effect in the same position
Gauss was in when in 1818 he was asked to survey Hanover.
He realized that
``the properties of a surface most worthy
of study by geometers are those that can  be obtained by little
flat bugs living upon it." In other words it is the metric
properties of ${\cal F}$ we seek, and hence we must
describe the ``measuring rods" by which metrical
properties are to be discovered.
\vskip.1in
Suppose then that there is a metric $d(g_1,g_2)$ on
${\cal F}$ which can be experimentally determined and
has ${\cal G}$ as a symmetry group. It thus has the properties
that for arbitrary elements of ${\cal G}$
$$d(gg_1,gg_2) = d(g_1,g_2), \eq{1a}$$
$$d(g_1,g_2) \geq 0, \eq{1b}$$
$$d(g_1,g_2) = 0 \Longrightarrow g_1 \approx
g_2,\eq{1c}$$
$$d(g_1,g_2) = d(g_2,g_1),\eq{1d}$$
$$d(g_1,g_2) + d(g_2,g_3) \geq d(g_1,g_3).\eq{1e}$$
From (1a) we can write
$$d(g_1,g_2) = d(e,g_1^{-1}g_2)\equiv d(g),\; g =
g_1^{-1}g_2.\eq{2a}$$ 
It then follows from the properties of $d(g_1,g_2)$
that the function $d(g)$ has the properties:
$$d(g) \geq 0,\eq{2b}$$
$$d(g) = 0 \Longrightarrow g \approx e, \hbox{ i.e.\ } g
\in G_o,\eq{2c}$$
$$d(g) = d(g^{-1}),\eq{2d}$$
$$d(g) + d(h) \geq d(gh).\eq{2e}$$
An easy deduction from these properties is
$$|d(gh) - d(g)| \leq d(h),\; |d(hg) - d(g)| \leq
d(h),\eq{2f}$$
whence 
$$d(g_o g g_o^\prime) = d(g),\;\;\forall g_o,g_o^\prime
\in G_o. \eq{2g}$$
Thus $d(g)$ {\it is constant on double cosets}
$G_o\backslash G/G_o.$
We shall say that two elements $g_1,g_2$ are in
{\it coherence relation} $g$ if $g_1^{-1}g_2$ belongs to
the double coset $g$. The function $d(g)$ can be
regarded as the ``diameter" of the coherence relation
$g$, and it is only this that the little flat bugs can
learn from experiments.
\vskip.1in
Our task now is to see what kind of experiments will
determine $d(g)$.
\end

determination of metric
properties 

\end

 In this paper I will construct quantum
``universes", i.e.\ systems in which the kinematic and
dynamic processes that we actually observe can be
accommodated

 We shall therefore construct a
quantum universe in which the symmetry group ${\cal G}$ is a
{\it locally compact} subgroup of
${\cal U}$ but which is still large enough to include
space-time transformations and transformations produced
dynamically, e.g.\ by turning on photocurrents.

to accommodate
required symmetries, e.g.\ Poincar\'{e} invariance,
and dynamical transformations that are observed.
\vskip.1in

It is useful to think of ${\cal G}$ as a surface

Let us see whether we can construct a quantum universe
with this group replacing
${\cal U}$ as the symmetry group of the theory.
\end

It is natural therefore to look for  replacements for
${\cal H}$ which are small enough to be locally compact
but large enough

to permit the operation of
such non-compact groups of transformations as 
space-time transformations and the Weyl-Heisenberg
groups which are needed to define a closed universe. Let
us suppose

transform under the non-compact groups nature employs, e.g.\
Lorentz transformations, and the Weyl-Heisenberg group (see
below).

 allownot large enough to accommodate all accessible
states e.g.\ those obtainable from any given state by Lorentz
transformations or the states of arbitrarily large photon number
 obtainable from the vacuum state by turning on a photocurrent.
\vskip.1in
 The purpose of this paper is to describe a subset ${\cal S}$
of ${\cal H}_\infty$ that includes the accessible states and no
others. It will be a manifold {\it not a subspace} that we can
think of as a surface in ${\cal H}_\infty$.  Because the
states on ${\cal S}$ are accessible we can investigate it in
the ideal way recommened by Gauss when he undertook the survey
of Hanover in 1818: ``The properties of a surface most worthy
of study by geometers are those that can  be obtained by little
flat bugs living upon it." In other words it is the metric
properties of ${\cal S}$ we seek.
\vskip.1in
Because ${\cal S}$ is not a linear space the investigation of
metrical properties has a very different character than the
measurement of ``observables" in the Dirac-von Neumann
interpretation of quantum mechanics. Information from detecting
eigenvalues of observables relates to 
the coefficients in the expansion of
the state vector in an eigenbasis of the observable. Were we to
insist on this form of measurement for $S$ we would be doing
the eqivalent of surveying Hanover by setting up a platform
in space and sending out laser beams to assign coordinate
vectors to points on the ground  relative to some
arbitrary basis with origin on the platform.
\vskip.1in
 If Gauss had 
imagined such a possibility he might not have regretted lacking
the technology to implement it. For it is 
what he himself acting as a little flat bug armed with a
measureing rod could observe that he recognized to be more
mathematically significant. But had the technology been
available he would have no doubt taken advantage of it to
simplify his labor and then converted the results to metrical
ones.
\vskip.1in
When we wish to study a manifold ${\cal S}$ of quantum states,
however, the Dirac-von Neumann observables do  not provide a
satisfactory method for determining its geometry for the simple
reason that even though the states of ${\cal S}$ are complete
they will not be mutually orthogonal. Hence there will be no
observable for which these points are eigenstates much less
eigenstates with different eigenvalues by which to distinguish
them.
\vskip.1in
Thus in a quantum theory based on ${\cal S}$ the Gaussian
``flat bug" measurements will replace observables both for
mathematic gral and practical reasons.
My first task will be to characterize ${\cal S}$ for quantum
mechanics, and I will show that its points are generalized
coherent states. My next task will be to describe the  ``rods"
that will be needed by the flat bugs for investigating the
metrical geometry. I will show that this role is played by
maximally entangled pairs such as down-converted photons. We
will see that the form of these pairs is dictated by the
structure of ${\cal S}$.
\vskip.1in
When quantum mechanics is formulated on ${\cal H}_\infty $
we know from Wigner's Theorem that the physics is invariant
under the infinite dimensional unitary group $U_\infty$ and
a duality relation (complex conjugation). To avoid
grostesqueries we require that
${\cal S}$  be locally compact. To be considered accessible the
states of ${\cal S}$ are supposed to be producible by the
action of some implementable transformation on a reference
state. The set of these transformations will then form a
{\it locally compact subgroup G of} $U_\infty$ which we may use
to label the states of
${\cal S}$. 
\vskip.1in
The physics is now to be related to a metric $d(g_1,g_2)$ with 
$g_1,g_2 \in G$.
In order for the physics to be invariant under ${\cal G}$ we
require that $d$ satisfy
$$d(gg_1,gg_2) = d(g_1,g_2),\forall g,g_1,g_2 \in G.\eq{1a}$$
If $e$ is the identity of ${\cal G}$ it follows that
$$d(g_1,g_2) = d(e,g_1^{-1}g_2) \equiv d(g),\; g =
g_1^{-1}g_2.\eq{1b}$$
From the symmetry property of the metric
$$d(g_1,g_2) = d(g_2,g_1)\eq{2a}$$
we obtain
$$d(g) = d(g^{-1}).\eq{2b}$$
From the triangle inequality of the metric 
 $$d(g_1,g_3) \leq d(g_1,g_2) +
d(g_2,g_3),\eq{3a}$$
and noting that $(g_1^{-1}g_3) = (g_1^{-1}g_2)(g_2^{-1}g_2),$
we have
$$d(gh) \leq d(g) + d(h).\eq{3b}$$
Replacing $g$ with $gh^{-1}$ and then replacing $h$ with
$h^{-1}$ one deduces 
$$|d(gh) - d(g)| \leq d(h),\; |d(hg) - d(g)| \leq d(h).\eq{4}$$
Now let ${\cal G}_o$ be the subset of ${\cal G}$ for which $d(g) = 0$.
It follows that 
$$d(gh) = d(hg) = d(g),\; \forall h \in G_o.\eq{5}$$
It follows that ${\cal G}_o$ is a subgroup and that $d(g)$ is
constant on the double cosets $G_o\backslash G/G_o$, i.e.\ on
elements $g_ogg_o^\prime$ as $g_o,g_o^\prime$ range over ${\cal G}_o$.
\vskip.1in
The requirement that a metric be non-negative and vanish only
when its arguments are identical means that we must identify
states labelled $g_1$ and $g_2$ whenever $d(g_1^{-1}g_2) =
0$, i.e.\ when $g_1^{-1}g_2 \in G_o$. Thus $g_1$ and $g_2$
are identified if they belong to the same coset of ${\cal G}_o$. Hence
the distinct states are in one-one correspondence with the
points of the coset space $F = G/G_o$. This is a homogeneous
space and the corresponding states will be called, following
Perelomov (ref), a set of generalized coherent states.
\vskip.1in
It may be observed that if $d(g)$ has all of the properties
above, so also will $\alpha(g) \equiv d(g)^\nu$ where $0 \leq
\nu \leq 1$. To see this observe that from
\end
$$\alpha(gh)^{1/\nu) \leq \alpha(g)^{1/\nu} + \alpha(h)^{1/\nu)
\leq (\alpha(g) + \al
\geq
\alpha(gh)
\leq 1$. For 
\end

If $d(g) = d(h) = 0$ it follows from  (4) that $d(gh)
= 0$. Hence the subset ${\cal G}_o$  of ${\cal G}$ on which $d(g)$
vanishes is a subgroup of ${\cal G}$. Moreover if $g_o,g_o^\prime \in
G_o$ then
$$d(g) = d(g_og g_o^\prime) with
$gh^{-1}$ we have
$$d(gh^{-1} )\geq d(g) - d(h) \Longrightarrow d(gh) \geq d(g) -
d(h)$$
whence $$|d(gh) - d(g)| \leq d(h), \hbox{ and likewise },\;
|d(hg) - d(g)| \leq d(h)$$
It follows that $d(g)$ is constant on {\it double cosets}
$G_o\backslash G/G_o$, i.e.\ on elements of the form $g_o g
g_o^\prime$ with $g_o,g_o^\prime \in G_o$.

 the transformations that
leave the physics invariant will be a locally compact subgroup
${\cal G}$ of
$U_\infty$.

\end
\vskip.1inf
Thus for quantum mechanics 

  approach for theod 
simple reason that the states of ${\cal S}$
will not be mutually orthogonal even though they may be complete.

\end

 Restr
a restriction will mean that arbitrary linear combinations of
allowed states will not be allowed. Thus we will have to
reformulate the theory without benefit of the superposition
principle, and hence will have to verify that observable
interference effects can still be accounted for.

\vskip.1in
As we know from Wigner's theorem the transformations that leave
quantum mechanical predictions invariant are unitary or
anti-unitary, and it is the unitary ones that connect
continuously to the identity. The large size of Hilbert space
is manifest in the fact that the infinite dimensional unitary
group $U_\infty$ is not locally compact. The transformations that
we can implement in the laboratory, however, will necessarily
belong to some locally compact subgroup G of $U_\infty$. Thus we
shall try to  reformulate quantum mechanics so that the only
allowed states are accessibe from   one another by
transformations belonging to such a group. These states can be
thought of as a ``surface" in Hilbert space. 
\end

 not a locally compact space, i.e.\ bounded will 
sequences need not converge. On the other hand finite
dimensional complex vector spaces are not

In the DvN interpretation of quantum mechanics the state vector
is the compendium of all information we can acquire about the
state by measuring the value that ``observables" have in the
state. When the state is not an eigenstate of the observable
the measurement results in ``collapse" of the state into one of
the eigenstates with a probability determined by the
squared modulus of the coerricient of the state vector in an
eigenbasis. Thus measurements can be regarded as attempts to
obtain information about the coefficients in various bases.
\vskip.1in

\vskip.2in
\centerline{{\bf Abstract}}
\vskip.1in
Quantum surveying is an experimental procedure for determining
the structure of a manifold S of quantum states defined by a
group G of transformations that relate them to one another. The
inner geometry (metrical structure) of S is manifest
 through EPR correlations observed when the two
members of a certain type of entangled pair are incident on
detectors for states lying on  S. The form of these
pairs is dictated by the structure of G. Surveying S differs
from measuring properties of the individual states lying
on it in that the former, being non-invasive, avoids
the necessity of wave function collapse for its completion.

\vskip.3in
 When
K. F. Gauss carried out his survey of Hanover in 1818 he realized
that the intrinsic geometry of a surface can be determined  by
``little flat bugs  who live on it". Thus  it is not
necessary and may not even be useful to
represent the points on
the surface by coordinates relative to a basis as one
is inclined to do when  thinking of the surface as something
embedded in a three dimensional Cartesian  space. 
\vskip.1in
An analagous situation arises in quantum kinematics.
According to Dirac and von Neumann the properties of states
are associated with observables. Each basis determines a family
of observables for which the basis vectors are eigenvectors. 
For all but a set of measure zero these observables have
non-degenerate eigenvalues, and the state can then be said to
possess the property indicated by the eigenvalue  if and only
if all but one of its coordinates is zero, i.e.\ if the state
vector is itself one of the basis vectors. This is referred to
as the ``eigenvalue-eigenstate link". 
\vskip.1in
The Dirac-von Neumann interpretation of measurement treats all
of Hilbert space as accessible. In practice, however, that is not
the case. The states we can examine in the laboratory are ones
that nature gives us or that we can obtain by transforming a
given state. The transformations can take many forms such as
the rotation of Stern-Gerlach magnets or the turning on of
photocurrents. But in all cases the accessible states are
obtained by implementing a group G of transformations that act
on a fiducial state.
\vskip.1in 
We can think of a set $S$ of  accessible
states as a kind of ``surface" in Hilbert space and are then
motivated to examine its experimental investigation
from the Gaussian point of view. Thus, instead of asking about
eigenvalues of observables which is tantamount to asking about
coordinates relative to a basis, we ask only what 
 ``little flat
bugs" can find out about $S$ by moving around on it with
a suitable instrument for measuring the equivalent of distances
and angles. We will see that such instruments
act like the 
crystals that emit entangled pairs by spontaneous parametric down
conversion (SPDC). {\it The metric properties of $S$ are 
extracted from the   EPR
correlations observed between detectors associated with the
states of S.}
\vskip.1in
Based on the analogy I will refer to this type of 
measurement as quantum mechanical ``surveying". The
correlations may be observed in space-like separated events
so that we are insured that the interactions between one member
of the pair and a detector do not distort the interaction
between the other member of the pair with the other detector.
Thus the ``mysterious" non-locality of EPR correlations is 
exploited in surveying to guarantee that the measurement
is effectively non-invasive.
Unlike
 Dirac-von Neumann measurements which require that the
entanglement of the state with a measuring device  be followed
by its ``collapse"  into an eigenstate of the observable,
 no such process is needed for surveying. 
  Thus by restricting to quantum mechanical surveying, the
counter-intuitive behavior of EPR pairs becomes a way of
avoiding the measurement problem.
\vskip.1in
Let us begin by describing the way in which accessible states
are to be labelled. We start with a fiducial state denoted $x_o$
that we know how to prepare and consider the set of states $x$
we can get to from it by applying a transformation $g$ belonging
to a group
${\cal G}$. If $g_1 x_o = g_2 x_o$ then $(g_1^{-1}g_2) x_o = x_o$, i.e.\
$g_1^{-1}g_2$ belongs to the subgroup ${\cal G}_o$ of ${\cal G}$ that keeps
$x_o$ fixed. This is called the {\it stability subgrooup} of
$x_o$. Choosing one element from each coset $gG_o$, the set of
images of $x_o$ are distinct, and we refer to this as a set
of {\it generalized coherent states}. Thus given any group ${\cal G}$
and any subgroup we can identify such a set of states with the
coset space $F = G/G_o$. 
\vskip.1in
Two familiar examples will illustrate the idea:
\vskip.1in
\noindent
(1) Let G be the set of $2\times 2$ unitary matrices. It
contains a subgroup ${\cal G}_o$ of diagonal matrices. Every element
of ${\cal G}$ can be written in the form:
$$g = \left(\matrix{\cos(\theta/2) & e^{i\phi}\sin(\theta/2)\cr
-e^{-i\phi}\sin(\theta/2) & \cos(\theta/2)}\right)
\left(\matrix{e^{i\psi_1} & 0\cr 0 & e^{i\psi_2}}\right).$$
The second factor runs over ${\cal G}_o$, so selecting one element from
each coset means that $ F = G/G_o$ can be identified with the
first factor with $0 \leq \theta \leq \pi$, $0 \leq
\phi < 2\pi$. Thus the images of a reference state $(1,0)$
under 
$(1,0) \to (1,0)g$ as $g$ runs over the first factor will be
distinct points on a sphere known as the Poincar\'{e} sphere.
Spin-1/2 states and polarized light are represented in this way.
\vskip.1in
\noindent
(2) Let G be the so-called Weyl-Heisenberg group. The elements
$g$ depend on a complex number $\lambda$ and a phase $0 \leq
\phi < 2\pi$ with the composition law 
$$g(\lambda_1,\phi_1)g(\lambda_2,\phi_2) = g(\phi,\lambda)$$
$$\lambda = \lambda_1 + \lambda_2,\;\; \phi = \phi_1 + \phi_2 +
Im(\lambda_1^*\lambda_2)\; \hbox{ mod }2\pi.$$
There is a subgroup ${\cal G}_o$ with elements of the form $g(0,\phi)$
and every element can be written in the form
$g(\lambda,0)g(0,\phi)$. The set of elements $F = G/G_o$ can
thus be identified with the set $g(\lambda,0)$ as $\lambda$
runs over the complex plane. The set states that are usuallype
referred to as coherent states in optics (Glauber states) are
of this type. The parameter $\lambda$ gives the amplitude and
phase of the photocurrent that produces the state. Multimode
Glauber states can be analagously be identified with a
generalization of the Weyl-Heisenberg group by the simple
expedient of replacing each $\lambda$ with an $N$-component
complex vector. 
\vskip.1in
The groups in the above examples are topological groups, i.e.\
one knows what it means to say that two elements are ``near"
each other. They are a particularly nice kind of topological
group known as ``locally compact", which means that bounded
infinite sequences have limits. The first example is compact
which means that no elements get too far away from the others.
The important thing about being locally compact is that there
is an invariant measure $d\mu$ on the group which alows one to
perform integrations over the group. All of the groups that
occur in physical applications are locally compact, so we will
freely make use of the existence of an invariant measure.
\end

\end

 represented by the images of
$(1,0)$ which  e
 As long as distinct
transformations produce different effects on $\kvac$ we can usee
the transformations  as labels for the accessible states. 
\vskip.1in
We shall assume that the
transformations form a group ${\cal G}$. This simply means that they
can be inverted and combined. If two different elements
$g_1,g_2$ have the same effect on $\kvac$ then $g_1^{-1}g_2$
leaves the reference state invariant. The subgroup ${\cal G}_o$ of ${\cal G}$
for which $\kvac$ is an eigenvector is called the {\it stability
subgroup} of $\kvac$. We can then achieve a unique labeling of
the accessible states by selecting one element from each coset
$gG_o$. The set of cosets $F = G/G_o$ partitions ${\cal G}$, and we may
then identify the set of accessible states with $F$. This is
referred to as a ``system of generalized coherent states"
(Perelomov).
\vskip.1in
The reader will find it useful to keep two examples in mind.
(1) The set of states of polarization of light or of a spin-1/2
particle. We write

$$\kvac = \left(\matrix{1 \cr 0}\right)$$
and

 and have little 
measuring rods". This idea has metaphysical implications because
it recognizes the importance of experimental ``accessibility".
There is an analogy to the notion of accessibility in quantum
mechanics that suggests an interesting way to reformulate
quantum kinematics.
\vskip.1in
We shall use the term ``detector" to mean a device that
recognizes a specific quantum state. Thus detectors can be
labelled by unit rays $\ket{g}$ in a Hilbert space 
${\cal H}$. Suppose that we only have access to detectors that
we can produce by transforming a reference detector in some
prescribed way. Let us make the assumption that these
transformations form a group ${\cal G}$. If $e$ is the identity of the
group we label the reference detector $\kvac$ and denote by
$\ket{g} = g\kvac$ its image under the transformation g. We shall
abuse notation by using $g$ to mean both an element of the
abstract group G and its representation in ${\cal H}$ by a
unitary operator $u(g)$. 
\vskip.1in
One readily checks that if $g,h$ both
produce identical detectors, then $s = g^{-1} h$ has $\kvac$ as
an eigenvector, i.e\ multiplies it by a phase factor
$e^{i\theta}$. The subgroup ${\cal G}_o$ of ${\cal G}$ consisting of elements
$s$ with this property is called the {\it stability subgroup} of
$\kvac$. We thus obtain a unique labelling of the accessible
detectors by selecting one element $f$ from each coset $gG_o$
and writing
$$\ket{f} = f\kvac, \; f \in F  = G/G_o.\eq{1}$
The set of accessible states in our sense is then identified
with the homogeneous coset space $G/G_o$. Such sets of states
are referred to as ``generalized coherent states" (ref
Perelomov).
\vskip.1in
Following the analogy with surface geometry we ask: What can we
learn from measurements performed using only pairs of detectors
belonging to F. Those measurements consist in observing
correlations between events occurring in a pair of detectors
$\ket{f_1}$,$\ket{f_2}$ which the rules of quantum mechanics will
relate to the function
$$p(f_1,f_2) = |\bra{f_1}f_2\rangle|^2.\eq{2}$$
The group theoretic structure of $F$ now imparts itself to
$p$ because
$$\bra{f_1}f_2\rangle = \bvac f \kvac,\; f =
f_1^{-1}f_2.\eq{3}$$ If $g_o,g_o^\prime \in G_o$ then replacing
$f$ by $g_o f g_o^\prime$ only changes the amplitude by a phase
and so
$$p(f) \equiv |\bvac f \kvac|^2 \eq{4}$$ is unchanged.
 \vskip.1in
We therefore introduce the following terminology. The
``coherence relation" between detectors $f_1,f_2$ is the double
coset $G_o\backslash f/G_o$ to which $f_1^{-1}f_2$ belongs.
Evidently quantum mechanical experiments can only obtain
information about the coherence relation between detectors.
\vskip.1in
Double cosets partition a group in the same way that left andde
right cosets do, so that $p(f)$ is a function on this partition.
We must conclude that the elementary objects to which our
experiments are addressed are not states themselves as
represented by kets in ${\cal H}$ but rather by coherence
relations represented by poins of
the double-coset space
$F^\prime = G_o\backslash G/G_o$
\vskip.1in
The form of $p$ may now be seen to have a remarkable geometric
interpretation. Let us first take note of the fact that because
$\bra{f_1}f_2\rangle$ is a scalar product, the function
$$d(f_1,f_2) \equiv \left(1 - p(f_1,f_2)\right)^{1/2}$$
is a metric on $F$, i.e.\
$$d(f_1,f_2) \geq 0 \hbox{ with equality if and only if } f_1 =
f_2 $$
$$d(f_1,f_2) = d(f_2,f_1),\; d(f_1,f_2) + d(f_2,f_3) \geq
d(f_1,f_3).\eq{5}$$
Now because  $p(f_1,f_2)$ depends only on $f = f_1^{-1}f_2$,
we find that
$$d(f_1,f_2) = d(e,f) \equiv d(f),\eq{6}$$
and the three metric properties become the following properties
of $d(f)$:
$$d(f) \geq 0 \hbox{ with equality only for the identity
relation }.$$
$$d(f) = d(f^{-1}),\; d(f) + d(g) \geq d(fg).\eq{7}$$
We can think of $d(f)$ as the ``extension" of the coherence
relation $f$ which measures its difference from the identity
relation which has zero extension.
\end

\end

 e can learn about the geometry of F
from the observation of correlations between events in pairs of
detectors 
\end

 on the history of
geometry: {\bf How do you determine the metric geometry of a
surface by intrinsic measurements, i.e.\ by operations within the
surface?} The metaphysical importance of this is that it forces
one to be a positivist, to construct models based on what is
experimentally accessible and avoid constructs that are not.
\vskip.1in
In particular one need not assign Cartesian coordinates to
points which requires the assumption that the surface is
embedded in a higher dimensional space. This is not only
unnatural, but may even be false. It makes much more sense to
characterize points by the transformations one must perform to
generate them from a reference point. The group of those
transformations provides a natural intrinsic coordinatization. 
\vskip.1in
I am going to apply this idea to quantum mechanics in the
following way: Think of the simplest experiments one can do
with polarized light.  The states to which we have access are
produced with polarizers and detected with analyzers. The
transformation group is such that the set of
accessible states can be represented by points on a sphere ---
the Poincar\'{e} sphere, and the geometry of the sphere is
manifest in the correlations we observe between a pair of points
representing a polarizer and an analyzer. Devices that
detect these correlations are in effect measuring rods for the
the inrinsic geometry of the sphere. We shall see that EPR
pairs serve this purpose, and that the group theoretic structure
is manifest in symmetry properties of the EPR states. 
\vskip.1in
All of this generalizes to every kind of quantum mechanical
measurement. We will see that when measurements are looked at
in this way rather than through the Dirac-von Neumann paradigm,
the measurement problem disappears. Moreover the reason for the
peaceful coexistence of quantum mechanics and relativity in the
presence of EPR correlations will become clear.
\end

 i.e.\  Correlations between
polarizer-analyzer events are determined by the geometry of the 
sphere. Conversely the correlations can be used to deduce the
geometry of the space inhabited by polarizers and analyzers.
\vskip.1in
I am going to show that the experiments by which one performs
metrical measurements in the space of accessible states can
be thought of as EPR experiments in which maximally entangled
states behave like measuring rods.
\end

  representing a polarizer and an analyzer. It turns
out that the correlations have a relation to the metric geometry
of the sphere, indeed the geometry can be inferred from the
experiments
\end 
\centerline{{\bf {\title{What the Algebraic Structurei
 of Entangled States Tells Us } }}}
\centerline{{\bf {\title{  About the EPR and
Measurement Problems}}}}
\vskip.1in
\centerline{{\bf Daniel I. Fivel}}
\vskip.05in
\centerline{{\bf  Department of Physics}}
\centerline{{\bf University of Maryland, College Park, MD
20742}}
\vskip.05in
\centerline{{\bf  2001}}
\noindent 
Outline:
\vskip.2in
\noindent
1. Generalized coherent state.

\noindent
2. Canonically associated Bell-EPR states.

\noindent
3. Interpretation of EPR experiments as relational measurements.

\noindent
4. Application to multimode laser systems.

\noindent
5. Implications for the measurement and EPR problems.

\vskip.2in
\centerline{1. Generalized coherent states}
\vskip.2in
\centerline{ (a) Construction.}
\vskip.1in
Let G be a locally compact group and $g \in G \to u(g)$ be an
irreducible, unitary representation of G on a Hilbert space
${\cal H}$. Let $\kvac$ be a fixed vector in ${\cal H}$ with
stability subgroup ${\cal G}_o$, i.e.\ $\kvac$ is an eigenvector of
$u(g)$ for $g \in G_o$. Select one element $g$ from each left
coset $gG_o$ and form the set F of vectors $\ket{g} =
u(g)\kvac$. The set F is in one-one correspondence with the
homogeneous space $G/G_o$ and is referred to as a system of
generalized coherent states.
\vskip.1in
Example 1 - Glauber states of n-mode lasers.
$$u(g) = e^{i\theta}U(\lambda), \;
U(\lambda) = e^{\lambda\cdot a^\dagger - \lambda^*\cdot a}$$
$$\lambda\cdot a^\dagger = \lambda_1 a_1^\dagger + \cdots +
\lambda_n a_n^\dagger,\eq{1}$$
in which the $\lambda$'s are complex numbers reprsenting
photo-currents and  $a,a^\dagger$ are the bose operators for
the modes. The group, denoted $WH_n$, is called the
Weyl-Heisenberg group. By the Stone-von Neumann theorem the Fock
representation, in which $\ket{0}$ (the Fock vacuum) is
annihilated by all
$a$'s is the only irreducible representation up to equivalence.
Note that we assume a finite number of modes, for otherwise the
group would not be locally compact. The stability subgroup of
$\kvac$ is the $U_1$ subgroup $e^i\theta I$, so the space
$G/G_o$ is labelled by the n-component complex-vectors
$\lambda$. Thus F may be represented by the states
$$\ket{\lambda} = U(\lambda)\kvac .\eq{2}$$
\vskip.1in
Example 2 - Spin 1/2 systems. Let $G = U_2, G_o = U_1$ and $u$
the two dimensional unitary representation. Then $F = U_2/U_1$
is the 2-sphere (Poincar\'{e} sphere) and the coherent states
can be written
$$\ket{\theta,\phi} = \left(\matrix{\cos(\theta/2)\cr
e^{i\phi}\sin(\theta/2)}\right).\eq{3}$$
\vskip.2in
\centerline{ (b) Completeness}
\vskip.1in
Since G is locally compact there is an invariant measure $d\mu$
and we can define the operator
$$I \equiv \int_F d\mu(g) \ket{g}\bra{g}.\eq{4}$$
Then
$$u(h)I = \int_F d\mu(g) \ket{hg}\bra{g} =
 \int_F d\mu(h^{-1}g) \ket{g}\bra{h^{-1}g} = Iu(h),\eq{5}$$
in which the last step uses the transitivity of F to change
variables and the invariance of the measure.
\vskip.1in
It follows from Schur's Lemma and the assumed irreducibility of
$u$ that by suitably scaling $\mu$ we can make I the unit
operator. Hence the coherent states are complete in ${\cal H}$.
They are not mutually orthogonal in general and the set is
actually infinitely overcomplete. The process of selecting
exactly complete subsets is of great interest but need not
concern us here.
\vskip.2in
\centerline{2. Canonically associated EPR states}
\vskip.1in
\centerline{a. Construction}
\vskip.1in
I am going to turn the above representation of the unit
operator into a peculiar type of tensor product
state. Let $u_1,u_2$ be  identical irreducible representations
of
${\cal G}$ on Hilbert spaces ${\cal H}_1,{\cal H}_2$ and let
$\ket{g,1},\ket{g_2}$ be associated generalized coherent
states with isomorphic stability subgroups. Now consider 
$$\doubleket{\beta} \equiv C \int_{\cal
F}d\mu \ket{g,1}\otimes\bra{g,2}.\eq{6}$$ Here $C$ is a constant.
Note the resemblance to I. This is a vector in ${\cal H}_1
\otimes \overline{{\cal H}_2}$. 
\vskip.1in
To intepret this object physically simply observe that
if $T$ is any anti-unitary map (e.g.\ time-reversal) then
$$\bra{g} \to T\ket{g}\eq{7}$$
is a unitary map. Thus $\doubleket{\beta}$ can be thought of as
a two-member state made of pairs in which 
member-2 is in the T-transformed state of its partner. Using
(4) one verifies that $\doubleket{\beta}$ is a unit vector with
the choice $$ C = {(Vol}({\cal{F}}))^{-1/2}
,\qquad
{Vol}({\cal{F}})= \int_{\cal{F}}d\mu =  Tr(I), \eq{8}$$
which is finite for compact groups. 
\vskip.1in
\centerline{b. The EPR property}
\vskip.1in
 Let us calculate the amplitude for
$\ket{a,1}\otimes\bra{b,2}$ in $\doubleket{\beta}$. It is
$$(\bra{a,1}\otimes\ket{b,2})\doubleket{\beta} =
\hbox{(Vol}({\cal{F}}))^{-1/2}\int_{\cal F}d\mu
\bra{a,1}g,1\rangle\langle g,2\ket{b,2} = $$ $$
\hbox{(Vol}({\cal{F}}))^{-1/2}\int_{\cal F}d\mu
\bra{a}g\rangle\langle g\ket{b} =
\hbox{(Vol}({\cal{F}}))^{-1/2}\bra{a}b\rangle.\eq{9}$$
Notice how the particle labels disappeared! {\it The state
$\doubleket{\beta}$ acts as a transfer agent by which the
scalar procuct of states in two different Hilbert spaces
becomes meaningful.} That is the secret of its ability to
produce correlations between  space-like separated events.
\vskip.1in
The state $\doubleket{\beta}$ exhibits the EPR
property: There is equal likelihood
$\hbox{(Vol}({\cal{F}}))^{-1}$ to find one member in any state
whatsoever. 
But the conditional probability for finding
particle-1 in state $\ket{a,1}$ given that its partner
is in the  state $\bra{b,2}$
is
$$p_\beta(a|b) = |\bra{a}b\rangle|^2.\eq{10}$$ 
which is unity for $a = b$. Thus it is certain to find
particle-1 in state $\ket{a,1}$ if its partner is in the
T-reversed state $\bra{a,2}$.
\vskip.1in
I will refer to these states as generalized Bell-EPR states.
It is instructive to see the calculation for the
$U_2/U_1$ coherent states:
$$\doubleket{\beta} = C \int d\Omega (\cos\theta \;
e^{-i\phi}\sin\theta)\otimes\left(\matrix{\cos\theta \cr
e^{i\phi}
\sin\theta}\right) = $$ $$
2^{-1/2}\left((1,0)\otimes\left(\matrix{1
\cr 0}\right) + (0,1)\otimes\left(\matrix{0 \cr
1}\right)\right),\eq{11}$$
which is the Bohm-Aharonov singlet --- the most familiar Bell
state.
\vskip.1in
The computation for Glauber states is trickier because although
$WH_N$ is locally compact, it is not compact. Thus the state
$\doubleket{\beta}$ is not normalizable and will have to be
defined by a limiting process. Some new physics will appear
which I will discuss at the end when I apply the ideas to
multimode laser systems.
\vskip.1in
\centerline{c. Symmetries}.
The state $\doubleket{\beta}$ transforms under a representation
of the direct product group $G \otimes G$ as follows:
Define
$$u(g_1,g_2)(\ket{a,1}\otimes\bra{b,2}) =
u_1(g_1)\ket{a,1}\otimes \bra{b,2} u_2^{\dagger}(g_2).\eq{12}$$
An argument that follows the same lines as that
leading to the representation of I above leads to:
$$u(g,g)\doubleket{\beta} = \doubleket{\beta}.\eq{13}$$s
Thus $\doubleket{\beta}$ is invariant when the transformation
representing the same group element is applied to both members.
The subgroup $G^2$ of $G\otimes G$ in which both members are
identical is thus in the stability subgroup of
$\doubleket{\beta}$. It follows that we can construct a system
of coherent states associated with $F = (G \otimes G)/G^2
\vskip.1in
Let us see that these are the states associated with quantum
cryptography.
\vskip.1in
It follows from (13) that we can produce this set by applying
 $u_1(g)\otimes I$  or $I \otimes
u_2^\dagger(g)$ to $\doubleket{\beta}$. Denote the states
$\doubleket{g}$. The scalar product of two such states is
easily computed to be
$$\doublebra{g}h\rangle\rangle = Tr(u^\dagger(g)u(h))/Tr(I) =
Tr(u(g^{-1}h))/Tr(I).\eq{14}$$
This equation can be called ``the fundamental equation of
quantum cryptography" as it reveals how the effect of a
transformation on one member (by Bob) can be detected  when
returned to a possessor of the other member (Alice). If the
representation is on an N dimensional Hilbert space and $G =
U_N$ there will be $g_j, j= 1,\cdots, N^2$ for which
$\doublebra{g_i}g_j\rangle\rangle = \delta_{ij}$. Having agreed
to only apply these, Alice can design a type of generalized
Stern-Gerlach apparatus to detect which one Bob chose.
\vskip.1in 
Observe that $G^2$ is isomorphic to G. If the coherent states
are relativistic particle states so that G includes the
Poincar\'{e} group, then $\doubleket{\beta}$ is invariant
under a group isomorphic to the Poincar\'{e} group. {\it In
this respect it is like the one-particle vacuum}. It is non-local
but transforms covariantly. 
\vskip.2in
\centerline{3. Interpretation of EPR experiments as relational
measurements}
\vskip.1in
Formula (10) takes on a particularly interesting form when we
restrict $\ket{a,1}$ and $\bra{b,2}$ to be coherent states. For
in this case $a$ and $b$ are group elements and we obtain:
$$\bra{a}b\rangle = \bvac u^{\dagger}(a)u(b)\kvac =
\bvac u(g)\kvac,\; g = a^{-1}b.\eq{15}$$ 
$$|\bra{a}b\rangle|^2 = |\bvac u(g)\kvac|^2\eq{16}$$

Now if $h = g_o g g_o^\prime $ with $g_o,g_o^\prime \in G_o$
then 
$$\bra{0}u(h)\ket{0} = e^{i\theta}\bra{0}u(g)\ket{0}.\eq{17}$$
Thus {\it because it is the modulus that appears in (10)} we
have the very important fact that  if $g = g_1^{-1}g_2$ and $h =
h_1^{-1}h_2$ belong to the same double coset $G_o\backslash x
/G_o$ then
$p_\beta(g_1,g_2) = p_\beta(h_1,h_2)$.
\vskip.1in
I shall use the term ``coherence relation" between $(g_1,g_2)$
to mean the double coset to which $g_1^{-1}g_2$ belong.
The various pairs for which the coherence relation may be
called  ``manifestations" of that relation.
We see then that it is the coherence relations between states
that we detect in EPR correlations, i.e.\ what we are probing
is the double coset space 
$F^\prime = G_o\backslash G/G_o$.
\vskip.1in
Let us next observe that the dependence of the function
$p_\beta$ on the coherence relations has a geometric meaning. 
Because $\bra{a}b\rangle$ is a scalar product, the quantity
$$\delta(a,b) \equiv \sqrt{1 - |\bra{a}b\rangle|^2}\eq{18}$$
is a metric.
It defines a function $d(g)$ on $F^\prime$ as the
common value of this function on all pairs in relation $g$. 
It can be thought of as the ``diameter" of the relation g
or as the distance of g from the identity relation.
It has  properties inherited from
the metric properties of
$\delta$. Thus,
the three metric properties
$$\delta(a,b) \geq 0 \hbox{ with equality if and only if } a =
b,$$
$$\delta(a,b) = \delta(b,a),\qquad \delta(a,b) + \delta(b,c)
\geq \delta(a,c) \eq{19}$$
imply that $d$ has the properties:
$$d(g) \geq 0 \hbox{ with equality if and only if } g = e,$$
$$d(g) = d(g^{-1}),\qquad d(a) + d(b) \geq d(ab).\eq{20}$$ 
Let us now consider experiments in which the members of an EPR
pair in the state $\doubleket{\beta}$ are detected by detectors
$\ket{g_1},\bra{g_2}$ respectively. Whenever we have a
coincidence we  say that the coherence relation $g =
g_1^{-1}g_2$ ``holds". We find that the probability of relation
$g$ holding is
$$p_\beta(g) = 1 - d(g)^2.\eq{21}$$
We can obtain this from the following hidden variable model.
Suppose that in each run a random element $h$ of G is generated
and the relation $g$ is found to hold if it is closer to the
identity than $h$, i.e.\ if $d(g) < d(h)$. Suppose that the
distribution of random group elements is such that the
probability of $h$ having $d(h) < r$ is $r^2$. Then the
probability for a coincidence is $1 - d(g)^2$ as required.
\vskip.1in
This hidden variable model is ``non-local" in that relations
are non-local. Note, however, that the model is  
``manifestly covariant", in the sense that the function $d(g)$ 
is invariant under the group $G^2$. The randomness of $h$
precludes the possibility of using coincidences to send
messages. I will refer to these models as ``relational hidden
variable models".
\vskip.1in
It is interesting to see what sort of distribution we need to
produce the relational hidden variable models. In the spin 1/2
case the distribution if found to be uniform over the
Poincar\'{e} sphere. In the $WH_N$ case the distribution is
found to be Maxwellian in the the boson number.
\vskip.1in

In what sense can
we think of detecting EPR correlations as a form of measurement:
It tests a relationship non-invasively.

\vskip.1in

\end
{\it all correlations between generalized coherent states.
\vskip.1in

Define:

\end

\end

The EPR problem and the measurementproblem are approaching
the biblical three-score and ten. So am I. I would like to see
them buried before I am. Today I will describe a new attempt at
gravedigging.
\vskip.1in 
The EPR problem arises from the observation in certain systems
of correlations between space-like separated events that cannot
be accounted for by an underlying, local hidden-variable. Thus
the mechanisim by which quantum mechanics manages to
peacefully coexist with special relativity is obscure.
\vskip.1in
The measurement problem arises from the Dirac-von Neumann
interpretation of the measurement process in which a pure state
must turn into a mixture (collapse). The former evolves
according to linear, deterministic \Schrod dynamics until
measurement, which is a non-linear, stochastic process, turns it
into a mixture.  The problem is that the theory fails to
account for how and when the transition occurs from one process
to the othr. The most popular solutions these days are (1) The
ignorance interpretation or  FAPPDO (for all practical purposes
decoherence occurs). One blames  macroscopic measuring devices
which are open to the environment causing loss of relative phase
information. (2) DRT - Dynamical reduction theories which seek
to replace the \Schrod equation with  a non-linear, stochastic
differential equation which interpolates between the two
processes.
\vskip.1in
Both problems arise from our interpretation of the model
we use for calculating observed correlations between pairs of
events. The prototype experiment involves two detectors which we
will call a polarizer and an analyzer. We observe correlations,
e.g.\ pairs of clicks indicating that something has triggered
both. The theory attempts to fit the observed correlation to a
formula in which the labels identifying the two detectors are
input. In the Dirac-von Neumann formulation the polarizer and
analyzer are represented by one-dimensional  projectors 
$\pi(a) = \ket{a}\bra{a}$ and $\pi(b) = \ket{b}\bra{b}$ in a
Hilbert space
${\cal H}$, and the observed correlations are accounted for by
the function
$$p(a,b) =  Tr(\pi(a)\pi(b)) = |\bra{b}a\rangle|^2. \eq{1}$$
Dynamical transformatiions are implemented by unitary
transformations of the projectors.
\vskip.1in
The DvN model is a way of thinking about the formula (1), a way
of thinking that leads to the measurement problem and makes
peacful coexistence seem weird. I am going to show that there is
another way to think about the formula (for closed systems)
which does not have a measurement problem and which makes
peaceful coexistence seem less weird by being manifestly
covariant. 
\vskip.1in
A striking aspect of formula (1) is that it relates the
correlation to a metric $\delta(a,b)$ on the space of states,
namely
$$ \delta(a,b) = \sqrt{1 - p(a,b)}.\eq{2}$$ 
This suggests that we apply to quantum mechanics an idea which
has an illustrious history in geometry, namely that the
metric geometry of a surface can be determined intrinsically,
i.e.\ by measurments within the surface itself without embedding
it in a linear space. The idea has a metaphysical significance
in that confining oneself to the surface means confining
oneself to what is experimentally accessible if we happen to be
residents of the surface. Since there are geometries in which
embeddability is impossible, an embeddable one might be expected
to reveal that possibility through some relations discoverable
intrinsically, and such relations are expected to be of
considerable importance.
\vskip.1in
The analogue of this idea in our search for a 
way out of the measurement problem is that we abandon the
assumption  that all of the one-dimensional projectors
correspond to physical states,  and replace it a suitable
characterization of the ``experimentally accessible" states.
\vskip.1in
I shall assume that the experimentally accessbile states are
obtainable from a reference state $x_o$ by a transformation
$x_o \to x = gx_o$, in which $g$ belongs to a group G which is
specific to the type of system that we are dealing with. We
then can use the group elements as labels for the states
provided we eliminate ambiguities resulting from the
possibility that distinct group elements
 $g_1,g_2$ produce the same effect. In such case 
$g_1^{-1}g_2$ belongs to the subgroup ${\cal G}_o$ of G which leaves
$x_o$ fixed. ${\cal G}_o$ is called the ``stability subgroup" of the
reference state. We then pick the set of distinct images of
$x_o$ by selecting one element from each coset $gG_o$. The set
of images obtained in this way are in one-one correspondence
with the points of the coset space $F = G/G_o$. This is a
homogneous space, i.e.\ it enjoys transitivity. We will call
the set of states labelled in this way a system of generalized
coherent states.
\vskip.1in 
Suppose now that the polarizer and analyzer are identified
with the cosets $g_1 G_o$ and $g_2 G_o$. The metric geometry of
$F$ determined by the observed correlations is expressed by the
way the metric $\delta(g_1,g_2)$ depends on its arguments. We
learn experimentally that this dependence is quite remarkable: 
Let $G_o\backslash g/G_o$ be the double coset containing $g$.
The double cosets partition ${\cal G}$ just as left and right cosets
do. What we find is that the value of $\delta(g_1,g_2)$ is the
same for all pairs for which $g_1^{-1}g_2$ belong to the same
double coset. In other words correlations depend on pairs only
in this combination. I will refer to this as ``the double coset
rule".
\vskip.1in
Let us pause here to see how the double coset rule result
emerges in the Dirac-von Neumann interpretation. There will be a
unitary representation $g \in G \to u(g)$ on a Hilbert space
${\cal H}$ so that the coherent states will be of the form
$$\ket{g}\equiv u(g)\ket{0},\eq{3}$$ where $\ket{0}$ is the
reference state. We then have
$$\bra{g_1}g_2\rangle = \bra{0}u^{\dagger}(g_1)u(g_2)\ket{0} =
\bra{0}u^{-1}(g_1)u(g_2)\ket{0} = 
\bra{0}u(g_1^{-1})u(g_2)\ket{0} = $$ $$
\bra{0}u(g_1^{-1}g_2)\ket{0} = \bra{0}u(g)\ket{0},\; g =
g_1^{-1}g_2.\eq{4}$$
If $g \to h = g_o g g_o^\prime$ with $g_o,g_o^\prime \in G_o$
then
$$\bra{0}u(h)\ket{0} = \bra{0}u(g_o)u(g)u(g_o^\prime)\ket{0} =
e^{i\phi}\bra{0}u(g)\ket{0},\eq{5}$$
where the phase factor appears because $\ket{0}$  is
stabilized by ${\cal G}_o$, i.e.\ it is an eigenstate of
$u(g_o)$,$u(g_o^\prime)$  and the eigenvalues are unimodular
for unitary operators. Thus when we take the squared modulus
the phase goes away, and we obtain a function which is constant
on the double cosets.
\vskip.1in
We see then that the Hilbert space formalism insures that the
double coset rule will be obeyed. The important point,
however, is that {\it the rule can be stated without reference
to the Hilbert space formalism}. It expresses the important
fact that quantum mechanical probabilities have information
about coherence relations between states not about the states
themselves. Thus for given $g$ there will be many pairs
$(g_1,g_2)$ such that
$g_1^{-1}g_2$ belongs to the g double-coset. We call these
``manifestations" of the relation.
\vskip.1in
Armed with this result I am going to show how we can
reinterpret quantum mechanics as a theory of coherence relations
between event pairs in contrast to a theory of evolving property
states. 
\vskip.1in
A clue to this comes from recalling the remarkable property of
EPR pairs: The pair carrys no information about the behavior of
each member separately but makes a strong prediction about the
state  of one member when that of the other is given.

the objects that live in the double coset space $F^\prime$ 
rather than a theory of the  coherent states that live in $F$.

The metric
$\delta(g_1,g_2)$ defines a function $d(g)$ on $F^\prime$ as the
common value of this function on all pairs in relation $g$.
Clearly it can be regarded as a distance from $g$ to the
identity $e$. It also has remarkable properties inherited from
the metric properties of
$\delta$. Thus,
the three metric properties
$$\delta(a,b) \geq 0 \hbox{ with equality if and only if } a =
b,$$
$$\delta(a,b) = \delta(b,a),\qquad \delta(a,b) + \delta(b,c)
\geq \delta(a,c) \eq{6}$$
imply that $d$ has the properties:
$$d(g) \geq 0 \hbox{ with equality if and only if } g = e,$$
$$d(g) = d(g^{-1}),\qquad d(a) + d(b) \geq d(ab).\eq{7}$$
The quantity $d(g)$ will be called
the ``diameter" of the CR and $d(g)^2$ will be called the
``cross section" of the CR.
\vskip.1in

Object is a pair in which we know nothing about each separately
but the probability of coherence relation g is p(g)
\end

\end

$served correlation will be exprssed through some function of
$g,h$ and it is from the form of this function that we construct
our mathematical model.
\vskip.1in
\end
  In order for us to use this
labelling scheme, however, we must be sure that dynamical
processes take coherent states into coherent states. It is
remarkable that the groups that we encounter have a structure
which makes this possible. To see this let us first recall the
definition of {\it semi-direct product}.  A group
${\cal G}$ is said to be the {\it semi direct product} of a subgroup
$S$ and a normal subgroup
$N$ if  every element
$g$ of ${\cal G}$ has a unique decomposition  $g =sn$
with $n \in N$, $s \in S$. We then write $G = S\odot N$.
Observe that $$(s_1 n_1)(s_2 n_2) = sn,\quad s = s_1 s_2,\;
n = n_1^\prime n_2, \;  n_1^\prime =(s_2^{-1}n_1 s_2).\eq{2}$$
Because $N$ is normal one sees that $n_1^\prime $ is in $N$ and
henc $n$ is in $N$. Also note that uniqueness means that $S$
and $N$ have only the identity $e$ as a common element. 
Now suppose that that $S$ coincides with the stability subgroup
${\cal G}_o$. Then
$$snx_o = sns^{-1}sx_o = n^\prime x_o,\; n^{\prime} =
sns^{-1}.\eq{3}$$
One sees that the distinct images of $x_o$ under ${\cal G}$ are of the
form $nx_o$ as $n$ runs over $N$. Thus the following
hypothesis insures that we can do dynamics within the set of
coherent states:
\vskip.1in
Hypothesis: (1) $G = S \odot N$. (2) Dynamical
transformations belong to $S$,\; $S \in G_o$. 
Then the coherent states can be labeled by the cosets $nG_o$ and
and evolve dynamically by automorphisms $n \to sns^{-1}$.
If $S$ is a Lie group we will have one-parameter subgroups 
$s(t)$ and the relation describing dynamics
$$n \to n(t) = s(t)ns^{-1}(t) \eq{4}$$
which is essentially the \Schrod equation.
\vskip.1in
Let us pause for an important example. For multimode laser
states the group N will be the so-called ll-Heisenberg group
which has the following form: Let $a_j,a_j^\dagger,\; j=
1,2,\cdots N$ be bose annihilation and creation operators. If
$\lambda$ is an N-component complex vector we write
$\lambda\cdot a =
\lambda_1 a_1 + \cdots + \lambda_N a_N$. Then the group $WH_N$
is generated by operators
$$f(\theta,\lambda) = e^{i\theta}U(\lambda),\qquad
U(\lambda) \equiv e^{\lambda\cdot a - \lambda^*\cdot
a^\dagger}.\eq{5}$$
The dynamical group $S$ will consist of elements of the form
$$s = e^{ia^\dagger\cdot H \cdot a},\eq{6}$$
in which $H$ is an arbitrary $N\times N$ hermitian matrix.
Then one verifies that
$$sU(\lambda)s^{-1} = U(\lambda^\prime),\; \lambda^{\prime} =
e^{iH}\lambda. \eq{7}$$
We know from the Stone-von Neumann theorem that there is only
one unitary representation of the $WH_N$ group up to
equivalence --- the Fock reperesentation. Hence we
represent the reference state by the Fock vacuum and
write $x_o = \ket{0}$ which is annihilated by $a_j$ for all $j$.
Hence $s \in G_o$ which consists of $S$ and the phase
multiplication $U_1$ subgroup of $N$. Thus the set of coherent
states can be represnted by the set $U(\lambda)\ket{0}$ as
$\lambda$ runs over all $N$-component complex vectors. These
are the states that were historically the first example of
coherent states and are often called Glauber states.
\vskip.1in
Observe that
$s$ transforms these states in such a way that
$|\lambda|^2 = |{\lambda^\prime}|^2 $, i.e.\ the expectation
value of the number operator $a^\dagger a$ is constant.
\vskip.1in
Note that Hilbert space sneaked in through the
uniqueness of the representation.

\end

Thus
elements of
$N$ are transformed into one another by similarity
transformations
$s$.

Let us make the tentative hypothesis hthat
dynamical transformations are described by elements of ${\cal G}_o$.
Then, since $g_o x_o = x_o$ we have
$$g_o g x_o = g^\prime x_o,\; g^\prime = g_o g g_o^{-1}.$$
The map $g \to g^\prime$ is an automorphism of G. 
\vskip.1in
Thus the two detectors involved in  

\end
 If I
want to tell an exerimentalist  to make a certain quantum state
I do not have to give him a set of complex numbers that are to
be its components in some Hilbert space basis. It makes his
 task clearer if I give a
recipe for {\it making} the state by applying a
specific transformation 
 to a reference state. Suppose the transformations 
form a group
${\cal G}$ and that the states produced are represented in the
Hilbert space by $$\ket{g} \equiv u(g)\kvac$$
in which $e$ is the
identity element of the group, $\kvac$ is the reference state,
and
$g
\to u(g)$ is a unitary representation of ${\cal G}$ in ${\cal H}$. 
\vskip.1in
If two elements $g,h$ produce the same image then $g^{-1}h \in
G_o$ where ${\cal G}_o$ is the subgroup of ${\cal G}$ which stabilizes
$\kvac$, i.e.\ $\kvac$ is an eigenvector of $u(g_o) \forall g_o
\in G_o$. Hence we obtain a set of distinct images of $\kvac$
by selecting one element $g$ from each coset $gG_o$. The set of
images $\ket{g}$ is in one-one correspondence with the points
of the coset space $F = G/G_o$. This is a homogeneous space,
i.e.\ it enjoys the property of transitivity.
\end

 Quantum theory deduces all correlation
predictions from expressions of the form $|\bra{a}b\rangle|^2$.
In conventional quantum mechanics we let $\ket{x}$ range over a
Hilbert sphere.  Suppose we restrict the vectors $\ket{x}$ in
the Hilbert space
${\cal H}$ to those that can be obtained by the action of a
unitary representation
$g \to u(g)$ ooup ${\cal G}$ on a reference state $\kvac$,
i.e.\ states $\ket{g} \equiv u(g)\kvac$

 I will show that there is an enormous
class of experiments to which quantum theory is applicable and
which can be

The idea is this: Suppose that we
consider a set of quantum states  to see if we can make the
measurement problem go away if we restrict the class of states
whose behavior must be accounted for

I start by asking the
following question: Does quantum theory simplify if it only
has to explain the correlations we observe in a class of
experiments 

I start by examining the
form of qu restriction of quantum theory universe in which the
only what looks like a small enough family of states that they
can be labeled by the recipe one uses for making them. I  then
show that in

 allowed
states are those that one can give a recipe for making. I

 In the Dirac-von Neumann a
formulation of quantum mechanics the analogue of a classical
system represented by a point in phase space is now represented
by 1-dimensional subspace (ray)
$\Psi$ of a Hilbert space which evolves in time via unitary
transformation. The ray is supposed to be a maximal
specification of the properties of the system at a particular
time. Properties (or observables as they are termed) are
associated with subspaces of the Hilbert space to which the ray
assigns values 1 or 0 if the ray belongs to the subspace or its
orthogonal complement respectively and otherwise assigns a
number between 0 and 1 which is interpreted as the probability
of the system having the property in question. 
The difficulty
arises from the form of the probability function: We are to write
$$\Psi = \sum_i \alpha_i \phi_i$$
where $\{\phi_i\}$ is a basis defined by the observable (an
eigenbasis for a Hermitian operator) and then compute
$|\alpha_i|^2$ for the probability that the system has the
property $\phi_i$. However it is not correct to say that the
system can be characterized as a member of an ensemble in which
there is probability $|\alpha_i|^2$ of having the property
$\phi_i$ before measurement. In technical language one cannot
replace the state vector by a mixture without producing
predictions that disagree with observation...the double slit
experiment is a standard example. Thus it appears that
``properties" only acquire determinate values upon measurement.
To account for this the theory characterizes the measurement
process through an interaction of a state with a measuring
device in which an entanglement appears and then ``collapses"
into an eigenstate of the observable. Thus there are two
distinct dynamical processes: the linear, deterministic \Schrod
evolution followed by a non-linear, stochastic collapse
process. The manner in which these two are connected is not
given. We must therefore either supply the missing connection or
revise the Dirac-von Neumann
characterization of a measurement so that it is not needed.
The first type of attack is called ``dynamic reductionism"
and, if successful, will produce new physics. There have been
a number of attempts along these lines in recent years that go
with the names Pearle, Ghirardi, Rimini, Weber, and mine.
The approach I will describe below is of the second type.

 I will first describe an algebraic
characterization of a very large class of gedanken experiments
for which the quasntum theory makes predictions. In these
experiments  correlations are observed between the behavior of a
pair of detectors which depend on the way in which the detectors
are prepared in a way that suggests an alternative
characterization of the measurement process to the way

 detected between two events. The correlations
are found to depend

event-labels
consist in an instruction for preparing the detector by the
transformation of a reference detector.

: Let G be a compact group $g\in
G \to u(g)$ an irreducible, unitary, representation of G on a
Hilbert space ${\cal H}$. Let $\kvac$ be a reference unit
vector in ${\cal H}$ and ${\cal G}_o$ its stability subgroup, i.e.\
the subgroup of ${\cal G}$ for which $\kvac$ is an eigenvector. From
the set of images $\ket{g} \equiv u(g)\kvac$ we can select a
set of distinct images by choosing one element $g$ from each
coset $gG_o$. The set of vectors obtained in this way is thus
in one-one correspondence with the homogeneous space ${\cal F} =
G/G_o$ and is referred to as a set of ``generalized coherent
states". 
\vskip.1in
The salient property of the set of coherent states is that it
is at least complete over the space ${\cal H}$. The proof is
simple, elegant, and instructive. Form the integral
$$I = \int_{\cal F}d\mu \ket{g}\bra{g},\eq{1}$$
where $d\mu$ is the invariant measure which exists for any
locally compact group. One verifies that $I$ commutes with
every $u(g)$, whence by Schur's Lemma the measure may be scaled
to make $I$ the operator.
\vskip.1in
Now let us have some theorists fun. I will use this
representation of the unit operator to construct a funny kind
of two particle state vector. I will do this by simply
inserting a tensor product sign between the ket and the bra and
labels 1,2. Thus consider the object
$$\doubleket{\beta} \equiv  \int_{\cal F}d\mu
\ket{g,1}\otimes\bra{g,2}.\eq{2}$$
To intepret this object physically simply observe that
if $T$ is any anti-unitary map (e.g.\ time-reversal) then
$$\bra{g} \to T\ket{g}\eq{3}$$
is a unitary map. Thus $\doubleket{\beta}$ can be regarded as a
two-particle state made of pairs in which the state of
particle-2 is in the T-transformed state of its partner,
particle-1. To make the interpretation as a two-particle state
complete we must normalize the state, and an easy calculation
shows that the correct formula is 
$$\doubleket{\beta} \equiv \hbox{(Vol}({\cal{F}}))^{-1/2}
\int_{\cal F}d\mu
\ket{g,1}\otimes\bra{g,2}\eq{4}$$
where
$$\hbox{Vol}({\cal{F}})= \int_{\cal{F}}d\mu \eq{5}$$
which is finite for compact groups.
\vskip.1in
Now let us observe that this state has some remarkable
properties. Let us calculate the amplitude for
$\ket{a,1}\otimes\bra{b,2}$ in $\doubleket{\beta}$. It is
$$(\bra{a,1}\otimes\ket{b,2})\doubleket{\beta} =
\hbox{(Vol}({\cal{F}}))^{-1/2}\int_{\cal F}d\mu
\bra{a,1}g,1\rangle\langle g,2\ket{b,2} = $$ $$
\hbox{(Vol}({\cal{F}}))^{-1/2}\int_{\cal F}d\mu
\bra{a}g\rangle\langle g\ket{b} =
\hbox{(Vol}({\cal{F}}))^{-1/2}\bra{a}b\rangle.\eq{6}$$
Notice how the particle labels disappeared!
\vskip.1in
This result shows that $\doubleket{\beta}$ exhibits the EPR
property: There is equal likelihood
$\hbox{(Vol}({\cal{F}}))^{-1}$ to find one member in any state
whatsoever. 
But the conditional probability for finding
particle-1 in state $\ket{a,1}$ given that its partner
is in the  state $\bra{b,2}$
is
$$p_\beta(a|b) = |\bra{a}b\rangle|^2.\eq{7}$$ 
which is unity for $a = b$. Thus it is certain to find
particle-1 in state $\ket{a,1}$ if its partner is in the
T-reversed state $\bra{a,2}$.
\vskip.1in
If ${\cal G}$ is $U_2$ and $u$ is the 2-dimensional representation of
$U_2$ with $G_o = U_1$, then $F$ is the 2-sphere and
$\doubleket{\beta}$ is the familiar Bell state in which the
members of the pairs transform like spin-1/2 particles in
time-reversal conjugate states, i.e.\ the familiar Bohm state.
It is instructive to see the calculation
$$\doubleket{\beta} = C \int d\Omega (\cos\theta \;
e^{-i\phi}\sin\theta)\left(\matrix{\cos\theta \cr e^{i\phi}
\sin\theta}\right) \propto (1,0)\left(\matrix{1 \cr 0}\right)
+ (0,1)\left(\matrix{0 \cr 1}\right).\eq{8}$$
Thus the state $\doubleket{\beta}$ generalizes the familiar
Bohm-Aharonov singlet and I will refer to it as a generalized
EPR state associated with a system of generalized coherent
states.
\vskip.1in
One sees from this construction why EPR pairs always involve an
anti-unitary mapping between the members.
\vskip.1in
Note that what $\doubleket{\beta}$ does is to compute the
scalar product between vectors in the Hilbert space of one
particle and the Hilbert space of the other.
\vskip.1in
Formula (7) takes on a particularly interesting form when we
restrict $\ket{a,1}$ and $\bra{b,2}$ to be coherent states. For
in this case $a$ and $b$ are group elements and we obtain:
$$\bra{a}b\rangle = \bvac u^{\dagger}(a)u(b)\kvac =
\bvac u(g)\kvac,\; g = a^{-1}b.\eq{9}$$
Now observe that
$$|\bra{a}b\rangle|^2 = |\bvac u(g)\kvac|^2\eq{10}$$
{\it has the same value} on all
pairs
$(a,b)$ for which $a^{-1}b$ belongs to the same double coset
$G_o\backslash g /G_o$. Double cosets partition a group just as
left and right cosets do. Thus 
{\it all correlations between generalized coherent states
detected by the associated EPR state are determined by the
double coset to which the {\bf relation} between the states
belongs.}
\vskip.1in
Let us next observe that because $\bra{a}b\rangle$ is a scalar
product, the function 
$$\delta(a,b) \equiv \sqrt{1 - |\bra{a}b\rangle|^2}\eq{11}$$ 
is a metric. In view of (10) we have
$$\delta(a,b) = d(a^{-1}b),\; d(g) \equiv \sqrt{1 - |\bvac
u(g)\kvac|^2},\eq{12}$$ 
$$p_\beta(a|b) = 1 - (d(g))^2.\eq{13}$$ 
The three properties of a metric
$$\delta(a,b) \geq 0 \hbox{ with equality if and only if } a =
b,$$
$$\delta(a,b) = \delta(b,a),\qquad \delta(a,b) + \delta(b,c)
\geq \delta(a,c) \eq{14}$$
then become
$$d(g) \geq 0 \hbox{ with equality if and only if } g = e,$$
$$d(g) = d(g^{-1}),\qquad d(a) + d(b) \geq d(ab).\eq{15}$$

We can interpret $d(g)$ as the distance of the double coset
containing $g$ from the identity double coset. The occurrence
of the square of $d$ in the formula for $p$ indicates that $p$
can be interpreted as follows: 
\vskip.1in
Suppose that in each run of an experiment involving
$\doubleket{\beta}$ a random group element $h$ is generated, and
suppose we observe a correlation between $a,b$ if and only if the
relation $g = a^{-1}b$ between them is closer to the identity
than $h$, i.e.\ if and only if $d(g) < d(h)$. If the
distribution of the random $h$ is such that the probability of
obtaining an $h$ with $d(h) < r$ is proportional to $r^2$, then
the probability that a correlation between two coherent states
in relation $g$ will be observed will be $1 - (d(g))^2$. Thus
the quantum mechanical prediction will be reproduced by this 
hidden variable theory. Note that because it depends on the
relation between the states this is a {\it non-local} hidden
variable theory. Thus its ability to reproduce quantum
mechanics is not a violation of Bell's Theorem.
\vskip.1in
It is natural to inquire as to what distribution of group
elements with respect to the invariant measure $d\mu$ is
required to make this rule work. I will return to this later,
and for the moment examine the consequences of having a hidden
variable theory of this kind.
\vskip.1in
Let us first observe that the theory is ``covariant" with
respect to the group ${\cal G}$ in the following sense:
\end

variable $0 \leq \epsilon \leq 1$
is generated. Suppose further that if $a,b$ are such that $g =
a^{-1}b$ is closer to the identity in the $d$ metric than
$\epsilon$, a correlation is observed and otherwise not.  If the
random
$\epsilon$ is so distributed that the probability of finding it
in the range $\epsilon > d$ is $1 - d^2$, we will reproduce the
quantum mechanical prediction.

\end

 The behavior of entangled states lies at
the heart of both problems, and my purpose in this paper is to
show that the algebraic structure of such states  provides some
new insights. Let us first review what the role of entanglement
is for the two problems.
\vskip.1in
(1) The EPR problem: It is possible to do delayed choice
experiments with  entangled pairs demonstrating the existence of
correlations between space-like separated events. Bell showed
us that the form of these correlations defies a local hidden
variable explanation, i.e.\ they cannot be accounted for by
assigning values to the members of the pair separately. The
question then is how nature manages to produce such
correlations without super-luminal signalling that would
contradict special relativity. These days the problem is often
referred to as the ``peaceful coexistence" problem.
\vskip.1in
(2) The measurement problem: In the Dirac-von Neumann (DvN)
interpretation of quantum mechanics one extracts information
about a system from its state vector $\psi$ by asking about the
subspaces of the Hilbert space to which $\psi$ belongs. 
 the information coded into
the state vector is expressed by the subspaces of Hilbert
space to which the vector belong, i.e.\ the information one can
extract from knowing the coefficients relative to a basis.

 This arises from the
Dirac-von Neumann notion of ``observable". It is postulated
that the quantum state (wave-function) carries information
about properties of a system expressed by the eigenvalues of
Hermitian operators associated with the observables. The information,
however, is only statistical unless the state of the stystem is
in an eigenstate of the observable. The identification of
determinate properties of a system with membeship in subspaces
of the Hilbert space is called the ```eigenvalue-eigenstate
link".
\vskip.1in
 If a state is a 
 superposition of
eigenstates of an observable, the linear dynamics implies that
its inteaction
 with a measuring device for that observable produces
an entangled state that is not an eigenstate.   This is then
supposed to ``collapse" into various constituents with a
probability equal to the squared modulus of the coefficient of
the state relative to an eigenbasis. The theory gives no account
of how and when the linear deterministic \Schrod evolution
chnges into the non-linear, stochastic collapse process. (Ref
Leggett)
\vskip.1in
The identification of ``properties" of a state with its
membership in eigen-subspaces is referred to as the
``eigenvalue-eigenstate link".  The DvN notion of an observable
is predicated on the linear kinematics, i.e.\ the  Hilbert sphere
(unit rays) which represents the physical states is embedded in
a linear space. Questions of the form: ``does the state belong to
such-and-such a subspace of the Hilbert space then make sense.

There is, however, a way to formulate the kinematics such that
we never have to leave the space of physical states. This will
lead us to the notion of ``generalized coherent state". I will
first show that when we restrict the allowed states to be states
of this kind there is no measurement problem.
\end

 Let us begin by examining the information content of
a quantum state. In the DvN formulation the measurement of an
observable ``asks" a state ``to which one of a set of
orthogonal subspaces of the Hilbert space do you belong?"  A
definite answer is gotten if and only if the state vector  is
an eigenvector. 

information one gets from knowing the coefficients of the state
in an eigen-basis of the hermitian operator corresponding to
the observable. But if we want to tell an experimentalist to
make a state we don't hand him a list of complex numbers. We 
give instructions for making the state by the transformation of a
reference state. The transformations form a group G in general
so that we can represent the states by
$$\ket{g} = u(g)\kvac,  \eq{1}$$
where $g \in G$, 
$e$ is the identity of G,  $\ket{0}$ is the reference state in a
Hilbert space ${\cal H}$, and
$u(g)$ is a unitary representation of G in ${\cal H}$. It may
be that two group elements $g_1$ and $g_2$ give the same
result which means that $g_1^{-1}g_2$ leaves $\ket{0}$ invariant.
Thus we can achieve a unique labelling by selecting one element
from each coset $gG_o$ where ${\cal G}_o$ is the subgroup of ${\cal G}$ that
leaves the reference state invariant --- its so-called
``stability subgroup". The set of distinct labels is then the
homogeneous coset space $F = G/G_o$, and the corresponding states
are referred to as a set of ``generalized coherent states". 
\vskip.1in 
For
experiments with Stern-Gerlach magnets or polarizers the group
G will be the three dimensional rotation group, and ${\cal G}_o$ will
be the subgroup of rotations about a fixed axis. The space F 
= G/G_o$ of coherent states may then be identified with a 
sphere (the Poincar\'{e} sphere).
\vskip.1in
 For laser states we take the
Fock vacuum $\ket{0}$ as the reference state and form the states
$$\ket{\lambda} = U(\lambda)\ket{0},\; U(\lambda) = e^{\lambda
a^\dagger - \lambda^* a}, \eq{2}$$
in which $a,a^\dagger$ are the bose annihilation and creation
operators for a mode, and $\lambda$ is a complex number
characterizing the photocurrent. The group ${\cal G}$, known as the 
Weyl-Heisenberg (WH) group, is represented by operators
$e^{i\theta}U(\lambda)$, and ${\cal G}_o$ is the subgroup of operators
$e^{i\theta}I$. Thus the space $ F = G/G_o$ of coherent states
may be identified with the complex
$\lambda$-plane.
\vskip.1in
Although the coherent states turn out to be infinitely
overcomplete (see below) they do not form a linear space, i.e.\
linear combinations of coherent states are not in general
coherent states. Nonetheless we can describe what we observe
without forming such combinations. To see why note that when
laser beams are superposed it is the currents that we add not
the state vectors. Thus the superposition of $\ket{\lambda}$
and $\ket{\mu}$ will be described by $U(\lambda + \mu)\ket{0}$
which is (up to phase) $U(\lambda)U(\mu)\ket{0}$. 
\vskip.1in
Now consider the family of
dynamical processes that transform $N$-mode laser states into
laser states with constant expectation value for the total
photon number. We describe
an N-mode laser  by simply letting
$\lambda$ be an
$N$-component complex number and $a = (a_1,\cdots,a_N)$ and
understand $\lambda \cdot a$ to mean $\lambda_1 a_1 + \cdots +
\lambda_N a_N.$ The dynamical processes will
be described by transformations of the form $V(t) = e^{-i{\bf
H}t}$ where
$${\bf H} = \sum_{j,k = 1}^N a_j^\dagger H_{jk} a_k,\eq{3}$$
and $H_{jk}$ is an $N\times N$ hermitian matrix.
Observing that $$VU(\lambda)V^{-1} = U(\lambda(t)),\quad
\lambda(t) = e^{-iHt}\lambda,\; V\ket{0} = \ket{0},\eq{4}$$ we see
that
$$V(t)\ket{\lambda} = \ket{\lambda(t)}\eq{5}$$
i.e.\ the coherent states transform into one another.
\vskip.1in
\end
We see then that if we make the additional hypothesis that only
dynamical processes are permitted that take coherent states
into coherent states, it is neither necessary nor sensible to
characterize observables by projectors on linear subspaces  as
Dirac and von Neumann do. I will call these $CS-to-CS$ systems. 
I will confine myself to such systems and 
As the last example shows there is an
enormous class of quantum mechanical systems to which this
hypothesis applies.

 v the fact that $N$ is an arbitrary integer,
and
$H$ is an arbitrary
$N\times N$ Hemitian matrix means that

How then should we characterize
measurements within a set of coherent states? To answer this
let us examine the form that amplitudes take within such a
set.
\vskip.1in
For $g_1,g_2 \in G$
$$\bra{g_1}g_2\rangle = \bra{0} u^{\dagger}(g_1)u(g_2)\ket{0} =
\bra{0} u(g)\ket{0},\; g = g_1^{-1}g_2.\eq{6}$$
Thus amplitudes are constant on elements of the form
$G_o g G_o$, which is to say they are functions on the
double-cosets $G_o\backslash g/G_o$. These partition ${\cal G}$ in the
same way that left and right cosets do, and we shall 
adopt the term ``coherence relation" to mean a member of the
set of double cosets.
\vskip.1in
Now observe that because $\bra{g_1}g_2\rangle$ is a scalar
product, the quantity
$$d(g_1,g_2) \equiv \sqrt{1 - |\bra{g_1}g_2\rangle |^2} \equiv
d(g),\; g = g_1^{-1}g_2
\eq{7}$$ is a metric. Since it depends only on the relation
$g_1^{-1}g_2$, the metric properties (positivity,symmetry,
triangle ineuality) become:
$$ d(g) \geq 0, \; d(g) = 0 \Longleftrightarrow\;  g = e
$$
$$d(g) = d(g^{-1}),\qquad d(g) + d(h) \geq d(gh). \eq{8}$$
\vskip.1in
The the quantum mechanical probabillity function $p(g_1,g_2)$
is computed from the distance $d(g)$ of the double coset $g =
g_1^{-1}g_2$ to the identity in the $d$-metric by:
$$p(g_1,g_2) = |\bra{g_1}g_2\rangle |^2 = 1 - d(g)^2.\eq{9}$$
It follows that {\it all we can learn from experiments about the
states
$g_1,g_2$ is the ``cross-section" of the coherence
relation between them defined as the square of its distance from
the identity in the d-metric .}
\vskip.1in
Now let us see that experiments with EPR pairs are well-adapted
to measuring the cross section of coherence relations. In fact
we shall see that for each coherence group G there is a
canonically defined maximally entangled EPR pair whose
correlations  probe 
the cross-sections of coherence relations. 
\vskip.1in
We can assume with no loss in generality that G is
locally compact so that there is an invariant measure $d\mu$
by which we can integrate over F. The ``volume" of F is
defined by 
$$vol(F) \equiv \int_F d\mu .\eq{10}$$
For locally compact groups that are also compact this will be
finite. 
Now if $u$ is irreducible there is a
representation of the unit operator adapted to the coherent
states. For if we define
$$I = \int_F d\mu(g) \ket{g}\bra{g},\eq{11}$$
then
$$u(h)I = \int_F d\mu(g) u(h)\ket{g}\bra{g} =
\int_F d\mu(g) u(hg)\ket{0}\bra{0} u(g^{-1}) = Iu(h) \eq{12}$$
where the last step uses the transitivity of F  to change
variables $g \to h^{-1}g$ and then uses the invariance of the
measure. Thus I
commutes with
$u(h)$ for all
$h \in G$, and hence by the assumed irreducibility of $u$ and
Schur's Lemma we can scale the measure such that $I$ is the
unit operator on ${\cal H}$. This proves that the coherent
states are at least complete. (They are actually infinitely
overcomplete in general).
\vskip.1in
We are now going to use (11)
 to construct generalized Bell
states. To do this we first recall one of the curious
features of Bell states: They are  are linear combinations of
terms
$\ket{a}\otimes
\ket{b^T}$ in which T is an {\it anti-unitary} map, i.e
$$\bra{x^T}y^T\rangle = \bra{y}x\rangle.\eq{13}$$ (In the case
of the celebrated Bohm-Aharonov singlet T is time-reversal.) Now
if T is anti-uitary the map
$$\ket{x^T} \to \bra{x} \eq{14}$$
is unitary.  We can then simplify calculations by supposing
that there are two kinds of states: R-states described by kets
in ${\cal H}$
and $L$-States described by bras in the dual space
$\overline{{\cal H}}$. Thus our pair states will be linear
combinations of $\ket{a,1}\otimes\bra{b,2}$ that ``live" in
${\cal H}_1\otimes \overline{{\cal H}}_2$. The index 1,2 is
present to take account of the possibility that the $R$
and
$L$ states are not necessarily duals of one another. For
example they may be photons with different frequencies. We
shall use the notation $|\, \cdot\,\rangle\rangle$ and
$\langle\langle
\,\cdot\,|$ to indicate the pair states and their duals.
\vskip.1in
We can now turn the representation (11) of the unit operator
into such a pair state, namely let$$\doubleket{\beta} \equiv
C\int_F d\mu
\ket{g,1}\otimes\bra{g,2},\eq{15}$$
 where $C$ is a normalization constant. From (11) one verifies
that the choice $$C = (vol(F))^{-1/2}\eq{16}$$ makes $\beta$ a
unit vector. It is non-zero for compact groups. For locally
compact but non-compact groups one must do the analysis through a
limiting process.
\vskip.1in
Now let $p(g_1,1; g_2,2)$ be the probability
in the pair state $\beta$
that the R-member is in the state $\ket{g_1,1}$ {\it and} the
$L$-member is in the state $\bra{g_2,2}$, i.e.\ that the pair
is in the state $\doubleket{\gamma} \equiv
\ket{g_1,1}\otimes\bra{g_2,2}$. We compute the amplitude
$$\doublebra{\gamma}\beta\rangle\rangle =
C\int_F d\mu \bra{g_1,1}g,1\rangle\bra{g,2}g_2,2\rangle =
$$ $$
C\int_F d\mu \bra{g_1}g\rangle\bra{g}g_2\rangle =
C\bra{g_1}I\ket{g_2} = vol(F)^{-1/2}
\bra{g_1}g_2\rangle.\eq{17}$$
It follows that the probability of finding one member in some
state when nothing is given about the other is $1/vol(F)$. Thus
from Bayes Law
$$p(g_1,1; g_2,2) = (1/vol F)p(g_1,1|g_2,2)\eq{18}$$
where
$$p(g_1,1|g_2,2) = |\bra{g_1}g_2\rangle|^2 = 1 -
d(g)^2\eq{19}$$
is the conditional probabillity for the R-member to be found in
state $\ket{g_1,1}$ given that its L-partner is found in
$\bra{g_2,2}$.
 Notice that the normalization has disappeared
so that a limiting process will give the same result even for
non-compact groups.
\vskip.1in The
disappearance of the labels 1,2 in (18) is remarkable.
 For it means that if the state
$\doubleket{\beta}$ exits, it will exhibit correlations between
states in two different Hilbert spaces, as e.g.\ the two
particles of an EPR pair. 
\vskip.1in
Now let us look more closely at $\doubleket{\beta}$ to see that
a symmetry of the state is what is responsible for its
wonderful properties:
\vskip.1in
First observe that pair states are themselves coherent states
associated with a group that is canonically related to the
coherence group G of the two members, namely the direct
product group $G \otimes G$. The members are pairs 
$(g_1,g_2)$,
the components of which are elements of G. Composition is
component wise, i.e. $(g_1,g_2)(h_1,h_2) = (g_1h_1,g_2h_2)$.
Let the unitary representation $u_1(g)$ act on the R space
and the unitary representation  $u_2(g)$ act on the L-space.
Take the reference state to be
$$\doubleket{e} = \ket{e,1}\otimes\bra{e,2},\eq{20}$$ and define
$$ \doubleket{g_1,g_2} = u(g_1,g_2)\doubleket{e} =
u(g_1,1)\ket{e,1}\otimes\bra{e,2}u^\dagger(g_2,2)
= \ket{g_1,1}\otimes\bra{g_2,2},\eq{21}$$
One readily verifies that $u(g_1,g_2)$ defines a representation
of $G \otimes G$. The stability subgroup of $\doubleket{e}$ is
$G_o \otimes G_o$, so the coherent pair states are identified
with the homogeneous space $(G \otimes G)/(G_o \otimes G_o)$.
\vskip.1in
The group $G\otimes G$ has a subgroup $G^2$ consisting of pairs
$(g_1,g_2)$ with $g_1 = g_2$. One sees from (18) that
this is a stability group for $\doubleket{\beta}$ i.e.\
$$u(g,g)\doubleket{\beta} = \doubleket{\beta}, g \in G,
\eq{22}$$  or equivaently
$$(u(g) \otimes 1)\doubleket{\beta} = (1 \otimes
u(g^{-1}))\doubleket{\beta}.\eq{23} $$ 
This property of $\doubleket{\beta}$  is the essential one for
quantum cryptography, for it says that the effect of a
transformation on one member of the pair by A can be
detected on receipt by B but not by an interceptor. It is also
clear from this property that the scalar product 
$$\doublebra{g_1,g_2}\beta\rangle\rangle \eq{24}$$ can only
depend on
$g_1,g_2$ in the combination $g_1^{-1}g_2$ as we saw above.
\vskip.1in
The family of coherent states $G\otimes G/G^2$ are the
generalized Bell states. They are the distinct images of
$\doubleket{\beta}$ under the action of $G \otimes G$ and
because of the $G^2$ invariance of $\doubleket{\beta}$ they are
in one-one correspondence with the elements of $G/G_o$ itself.
In a sense they act like a single ``non-local" system under
transformations of G, somwhat like a measuring rod under
translations.
\end

 \end

 is intrinsically non-local
correlations in the state
$\doubleket{\beta}$ can test coherence relations between states
belonging to different Hilbert spaces. In particular we can 

Let us take note of two
consequences of this:  One  readily checks that for any $h \in
G$ 
$$(u(h) \otimes
u(h^{-1})\doubleket{\beta}=\doubleket{\beta}\eq{20}$$ or
equivalently
$$(u(h) \otimes 1)\doubleket{\beta} = (1 \otimes
u(h)\doubleket{\beta}.\eq{21}$$
Equation (21) might be called the fundamental equation of
quantum cryptography because it states that a receiver (Bob)
acting on one member can reproduce the effect on
$\doubleket{\beta}$ caused by a sender (Alice) acting on the
other member. 
\vskip.1in
Another consequence is the EPR property of $\doubleket{
\beta}$, i.e.\ that while it is equally likely to find one
member of a pair in any state if nothing is known about the
partner, it is certain that the R-member will be found in state
$\ket{s,1}$ if its L-partner is found in $\bra{s,2}$. 
The EPR assumptions (realism, locality, completeness) lead to
the conclusion from this property that we should be able to
construct sets $\Lambda_1(s)$ and $\Lambda_2(s)$ with a measure
$\nu$ such that
$$p(g_1,1; g_2,2) = \nu(\Lambda_1(g_1)\cap
\Lambda_2(g_2)).\eq{22}$$ Bell's No-Hidden Variable Theorem
tells us that  (18) cannot be reproduced by an expression of the
form (22).
However,
since for coherent states $p(g_1,1; g_2,2)$ depends only on
its arguments in the combination $g = g_1^{-1}g_2$, we can
reproduce (18) in the form
$$1 - d(g)^2 = \nu(\Lambda(g)).$$
Indeed all this equation says is that the square of a metric
can be used as a measure of sets 
 by making
$\Lambda(g)$ the complement of the  repr If we give up locality,
however, there  is a simple and natural way to reproduce (1 It
is this that makes quantum cryptography possible (correlations
between what Bob does and what Alice does). It also leads to the
existence of correlations between space-like separated events. 
\vskip.1ine 
The fact t

\end 

 But suppose that what we know about a state
is more  state is characterized in an experimentally
more useful way, namely by giving a recipe whereby it may be
prepare.these transformations form a group G and the states are
denoted:

 It has long been
realized (ref Fivel) that maximal EPR correlations always  occur
between members of  an entangled pair that are related by some
anti-unitary transformation $\ket{s} \to \ket{s^T}$. In the case
of the familiar Bohm-Aharonv singlet, for example, T is
time-reversal. I will begin by examining why this is so, and
from the examination show that the states which exhibit EPR
correlations --- let us call them ``generalized Bohm states"
(GBS) --- have a remarkable algebraic structure. A by-product
will be a new perspective on the EPR and measurement problems
in which such states play an essential role. 
\vskip.1in
In dealing with states that are combinations
of tensor products $\ket{a}\otimes\ket{b^T}$ we obtain both a
notational and conceptual simplification by observing that if
$T$ is anti-unitary the map
$\ket{b^T} \to
\bra{b}$  from Hilbert space ${\cal H}$ to its dual
$\overline{{\cal H}}$ is  {\it unitary}. Thus we can imagine
that there are ``conjugate" states:  R-states and L-states
described respectively by kets and bras. The pairs are then
described by objects of the form
$q =
\ket{a}\otimes\bra{b}$ which ``live" in the tensor product space 
${\cal H}\otimes\overline{{\cal H}}$. Observe that there is a
natural one-one linear map $q \leftrightarrow \hat{q} =
\ket{a}\bra{b}$ which associates
 operators in ${\cal H}$ with pair states in ${\cal
H}\otimes\overline{{\cal H}}$ by simply dropping or inserting  
the
$\otimes$. One then verifies that the scalar product
of elements in the pair space is expressed in terms of
operators in ${\cal H}$ by the formula
$$\bra{q}r\rangle = Tr(\hat{q}^\dagger \hat{r}).\eq{1}$$
I will refer to the map $q \to \hat{q}$ as the ``state to
operator (SO)" map.
\vskip.1in
A state $\beta$ in ${\cal H}\otimes\overline{{\cal H}}$ will
be called a generalized Bohm state (GBS) if the conditional
probability for finding the R-member in
$\ket{a}$ if the L-member is found in $\bra{b}$ is given by
$$p(a|b) = |\bra{b}a\rangle|^2.\eq{2}$$
The EPR property follows, i.e.\  it is equally
likely to find a member in any state if no conditions are
imposed on its partner, but it is certain that the R-member
will be found in state $\ket{a}$ if its partner is found in
state $\bra{a}$.
\vskip.1in
{\bf Theorem} A state is a generalized Bohm state if and only
if its image under the SO map is a multiple of the identity.
\vskip.1in
Proof: The amplitude in the pair state $\gamma$ for the R
member to be found in $\ket{a}$ and the L member to be found in
\bra{b}$ is the scalar product of $\ket{a}\otimes\bra{b}$ with
$\gamma$ i.e.\ $ Tr((\ket{b}\bra{a})\hat{\gamma}) =
\bra{a}\hat{\gamma}\ket{b}$. The required formula for $p(a|b)$
for all $a,b$ requires that $\hat{\gamma}$ be a multiple of the
identity. QED
\vskip.1in
Thus, remarkably, the generalized Bohm states correspond to the
density operator of an unpolarized mixture under the SO map.
\vskip.1in
It follows that if $\beta$ is a GBS and $u$ is any unitary
opertor on ${\cal H}$ then under the SO map
$$(u \otimes u^{-1})\beta \to u\hat{\beta}u^{-1} =
\hat{\beta},\eq{3}$$ whence
$$(u \otimes u^{-1})\beta = \beta.\eq{4}$$
It is upon this identity that quantum cryptography depends for
when rewritten
$$(u \otimes I)\beta = (I \otimes u)\beta,\eq{5} $$ it shows tht
a transformation on one member (by Bob) is equivalent to a
transformation on the other (by Alice). 
\vskip.1in
Before going further I must briefly digress to review the
notion of generalized coherent state.
\vskip.1in
Generalized coherent states are states which can be defined by
the recipe one uses to make them by the transformation of a
reference state. Thus suppose ${\cal G}$ is a group, $g \to u(g)$ a
representation of ${\cal G}$ on ${\cal H}$, ${\cal G}_o$ a subgroup, and
$\ket{0}$ a vector in ${\cal H}$ which is stabilized by ${\cal G}_o$.
Then we obtain a set of distinct images $\ket{g} \equiv
u(g)\ket{0}$ by selecting one element from each coset $gG_o$.
This set of states is thus in one-one correspondence with the
points of the coset space $ F = G/G_o$. This is a homogeneous
space, i.e.\ it has the trnsitivity property. We refer to the
set of states F constructed in this way as a set of generalized
coherent states.
\vskip.1in
Now suppose that G is a compact topological group with
invariant measure $d\mu$ and that the representation $g \to
u(g)$ is irreducible. It is then easy to demonstrate that a set
of generalized coherent states is complete by showing that the
unit operator has the representation: 
\vskip.1in 
$$I = \int_F d\mu \ket{g}\bra{g}.\eq{6}$$
To verify this one needs only verify that $I$
 commutes with
$u(h)$ for all $h \in G$, so that by Schur's Lemma the measure
can be scaled to make $I$ the unit operator. Proving the
commutation makes use of transitivity to permit a change of
variables and the invariance of the measure.
\vskip.1in
In view of our observation above about the SO image of
generalized Bohm states, we see from (6) that there is a
representation of the Bohm state associated with every set of
generalized coherent states namely:
$$\beta = C\int_F d\mu \ket{g}\otimes\bra{g},\eq{7}$$
where $C$ is a normalization constant. Indeed from (6) a simple
computation shows that to make $\beta $ a unit vector we must
choose
$$ |C| = (Vol(F))^{-1/2},\quad Vol(F) = \int_F d\mu,
\eq{8}$ which is non-zero for compact groups.
\end

et F of distinct images with the coset space
$G/G_o$, and we refer to the set of vectors $\ket{g} =
u(g)\ket{0}$ in which we select one memeber from each coset, as a
set of generalized coherent states. 
\vskip.1in
Now suppose that G is compact so that there is an invariant
measure $d\mu$ and we can define a pair state
$$\beta = C\int_F d\mu\ket{g}\otimes\bra{g}\eq{5}$$
which contracts to
$$\hat{\beta} = C\int_F d\mu\ket{g}\bra{g}.\eq{6}$$
From the transitivity of F and invariance of $d\mu$ we
readily verify $\hat{\beta}$ commutes with all $u(h)$ and
so by Schur's Lemma we establish that $\beta$ is a generalized
Bell state. Notice that this also establishes that the coherent
states are at least complete (in fact they turn out to be
infinitely overcomplete).
\vskip.1in
{\bf Corollary:} $$(u(h) \otimes u^{-1}(h))\beta = \beta
\eq{7}$$ i.e.\ $\beta$ is an eigenstate with eigenvalue unity of
all operators of the form $u(h)\otimes u(h^{-1})$. 
 It follows that
in any dynamical process where the R-member is transformed by 
$\ket{a} \to u\ket{a}$ and the L-member by $\bra{a} \to
\bra{a}u^\dagger = \bra{a}u^{-1}$ the Bell state is {\it
invariant}. This is the secret of ``peaceful coexistence", for
if the two members are particle states that are subjected to
the same Poincar\'{e} group transformation, the Bell state will
not change. 
\vskip.1in
Now consider transformations of pair coherent states under
the direct product group $G \otimes G$ which has elements
$(g,h)$ with $g,h \in G$ and componentwise composition.
Since 
$$(u(g_1)\otimes u(h_1^{-1})(u(g_2)\otimes u(h_2^{-1}) =
u(g_1)u(g_2) \otimes u(h_2^{-1})u(h_1^{-1}) = $$ $$u(g_1
g_2)\otimes u((h_1 h_2)^{-1}),\eq{8}$$
we see that $(g,h) \to u(g) \otimes u(h^{-1})$ is a
representation of $G \otimes G$. This group has a subgroup
$G^2$ consisting of pairs with identical entries, and we thus
conclude that since a Bell state is stabilized by any 
element of $G^2$  we can characterize the images of a Bell
state under $G \otimes G$ as the set of coherent states
$(G\otimes G)/G^2$. 
\vskip.1in
Let us see that quantum cryptography relies on this property:
For the invariance under $G^2$ means that 
$$(u(h) \otimes 1)\beta = (1 \otimes u(h))\beta,\eq{9}$$
so that the effect of a transformation on one member is
equivalent to the same transformation applied to its partner.
Thus Alice can learn about what Bob did to an R-member from a
transformation performed on an L member.
\end

  We can now give a simple characterization of
generalized Bell states by the following simple theorem:  A
state is a generalized Bell state if and only if it contracts to
a multiple of the identity.
\vskip.1in

The pair state 
$\ket{a}\otimes\bra{a}$ in ${\cal H}\otimes\overline{{\cal H}}$
resembles the projection operator $\ket{a}\bra{a}$ in ${\cal
H}$ and this resemblance will have many consequences as we
shall see. Note how scalar products are computed:
$$(\ket{a}\otimes\bra{b})\cdot (\ket{c}\otimes \bra{d}) =
\bra{a}c\rangle \bra{d}b\rangle.$$ Also note how operators act
on pair states: $$(A \otimes B)(\ket{a}\otimes\bra{b}) =
A\ket{a}\otimes\bra{b}B.$$
\vskip.1in
Suppose now that ${\cal G}$ is an arbitrary compact group and $g \to
u(g)$ is an irreducible, unitary, representation on ${\cal H}$.
Let ${\cal G}_o$ be a subgroup and $\ket{0}$ a vector in ${\cal H}$
which is stabilized by ${\cal G}_o$. Then we obtain a set of distinct
images $\ket{g} = u(g)\ket{0}$ by selecting one element from each
coset $gG_o$. This set of vectors $F$ which is in one-one
correspondence with the set of cosets $G/G_o$ is called a
system of coherent states associated with G and ${\cal G}_o$. F is a
homogeneous space, i.e.\ it enjoys an obvious transitivity
property. This set of states is at least complete, i.e.\ it may
be overcomplete. The proof of this is elegant and very
instructive. Let $d\mu$ be the invariant measure on G (which
exists for any locally compact group and consider the operator
$$I = \int_F d\mu(g) \ket{g}\bra{g}$$
Then 
$$u(h)I = \int_F d\mu(g)u(h)u(g)\ket{0}\bra{0} u^\dagger(g).$$
Since $u(h)u(g) = u(hg)$, a change of variables $g \to h^{-1}$
making use of transitivity and invariance of the measure
transforms the right side into $Iu(h)$. Thus I commutes with
every operator
$u(h)$, whence the irreducibility and Schur's Lemma implies
that I becomes the unit operator if when the measure $d\mu$ is
properly normalized.  This proves the at least completeness
of the coherent states. In general they are infinitely over
complete.
\vskip.1in
Now consider the pair state which resembles I, i.e.\
$${\cal B} \equiv C\int_F d\mu(g)\ket{g}\otimes\bra{g}.$$
Following the same steps used in verifying that I is the unit
operator one readily finds that
$$(u(h)\otimes u(h^{-1}){\cal B} = {\cal B}$$ 
i.e.\ ${\cal B}$ is an eigenstate with eigenvalue unity of all
operators of the form $u(h)\otimes u^\dagger(h) = u(h^{-1$. Now
observe that 
$$(u(g_1)\otimes u(h_1^{-1})(u(g_2)\otimes u(h_2^{-1})=
u(g)\otimes u(h^{-1}),\;\; g = g_1 g_2,\;\; h = h_1 h_2.$$
Thus $u(g)\otimes u(h^{-1})$ is a representation of $G \otimes
G$ and the operators $u(g) \otimes u^(g^{-1})$ represent the
subgroup $G^2$ consisting of pairs with identical elements. 
Thus the state ${\cal B}$ is stabilized by $G^2$ and the set
of disinct images of ${\cal B}$ under $G \otimes G$ is
identified with the homogeneous space $(G\otimes G)/G^2$. 
\end

There is a group ${\cal G}$ which transforms the states of each member
of the pair into one another, so that pair states are
transformed into one another by elements of the direct produc
group $G \otimes G$. Generalized Bell states are invariant under
the subgroup $G^2$ of $G \otimes G$, i.e.\ the pairs of the
form $(g,g)$.
\vskip.1in
\end
  doing will show that there is a general method for
constructing states exhibiting maximal EPR correlations. The
construction reveals that these states, which I will call
generalized Bell states (GBS), have a natural symmetry group
G, and  conversely there is such a state associated with every
group within a very large class that embraces all of the
groups we ever encounter in physics (locally compact
topological groups). 
\end

 We are accustomed to characterize quantum
mechanical predictions in the following language: A state
$\ket{0}$ is  prepared at time $t = 0$ (perhaps by passing a
beam through a polarizer or Stern-Gerlach filter); it evolves
for time $t$ under some unitary transformation $\ket{0} \to
\ket{a} = U(t)\ket{0}$ with e.g.
$U(t) = e^{-itH}$; and when passed through an analyzer for the
state
$\ket{b}$ we find that 
$$p = |\bra{b}a\rangle|^2$$
gives the probability of passage.
\vskip.1in
We know from Bell's theorem that this prediction cannot be
reproduced by a local hidden variable theory, i.e.\ it is not
possible to associate a point set $\Lambda(s)$ with every state
$\ket{s}$ with a measure $\mu$ on sets such that
$$p =
\mu(\Lambda(b)\cap\Lambda(a)).$$

\end
is the fraction that passes.

 If you ask someone to
give you an example of an entangled state they are likely to
show you the Bohm-Aharonov  pair, the spin zero state of two
spin-1/2 particles
$$\doubleket{B} =
2^{-1/2}(\ket{\uparrow}\otimes\ket{\downarrow} -
 \ket{\downarrow}\otimes\ket{\uparrow}).\eq{1}$$
The remarkable property of this state is this: It is equally
likely for one member to be in any spin state $\ket{a}$ if
nothing is known about the partner, but if it is given that one
member is in state $\ket{b}$, the probability $p(a|b)$ of
finding its partner in state $\ket{a}$ is given by
$$p(a|b) = |\bra{a}{b^T}\rangle|^2, \eq{2}$$
in which the superscript T indicates time reversal.
I will call (1) ``the EPR property".  For
spin-1/2 particles $T$ may be defined by $T\ket{\uparrow} =
\ket{\downarrow},\; T\ket{\downarrow} = -\ket{\uparrow}$. In
other words it is certain that one member will be found in a
state if its partner is found in the T-conjugate state. 
\vskip.1in
It is important to understand the role of T-conjugation in
producing this remarkable property. T is an {\it anti-unitary}
operator, i.e.\
$$\bra{a^T}b^T\rangle = \bra{a}b\rangle^* = \bra{b}a\rangle
\eq{3}.$$ Let $\ket{j},\; j = 1,2$
be any spin basis. 
Then the probability of finding the two particle state
$\doubleket{a,b} \equiv \ket{a}\otimes\ket{b}$ when the
system is in the state 
$$\doubleket{B} =
2^{-1/2}\sum_j\ket{j}\otimes\ket{j^T}\eq{4}$$
is
$$p(a,b) = |\doublebra{a,b}B\rangle\rangle|^2. $$
Now 
$$\doublebra{a,b}B\rangle\rangle = 
2^{-1/2}\sum_j\bra{a}j\rangle\bra{b}j^T\rangle
=2^{-1/2}\sum_j\bra{a}j\rangle\bra{j}b^T\rangle = $$
$$ 2^{-1/2}\bra{a}I \ket{b^T}.\qquad
I = \sum_j\ket{j}\bra{j}.\eq{5}$$
Recognizing $I$ as the unit operator we have
$$p(a,b) = (1/2)|\bra{a}b^T\rangle|^2.\eq{6}$$
It follows from this that the probability for finding one
particle in state $\ket{b}$ if no restriction is placed on the
state of its partner is $1/2$. Thus from Bayes law
$p(a,b) = (1/2)p(a|b)$, and (2) follows.
\vskip.1in
Now let us look at this calculation in a slightly different
way: Observe that since $T$ is anti-unitary, the map
$\ket{a^T} \to \bra{a}$ is a {\it unitary} map. Thus time
reversed states behave like members of the {\it dual } space
$\overline{{\cal H}}$ of the the Hilbert space ${\cal H}$. In
other words one can think of bra vectors as T-reversed kets.
Thus we shall imagine that there are two kinds of ``particles"
r-particles described by $| \rangle$ vectors and l-partcles
described by $\langle |$. The two partcle states will then be
sums of tensor procucts of the form $\ket{a}\otimes\bra{b}$
We represent $\doubleket{B}$ by
$$\doubleket{B} = 2^{-1/2}\sum_j\ket{j}\otimes\bra{j}\eq{7}$$
and $\doubleket{a,b}$ by $\ket{a}\otimes\bra{b}$. . Computing the scalar
product of $\doubleket{a,b}$ and $\doubleket{B}$ is now
trivially performed to give the result above, i.e.\ by
describing T-reversed kets as bras the role of
anti-unitarity in producing the EPR correlation becomes
transparent.
\vskip.1in
 Note the
similarity of the form of the two particle state $\doubleket{B}$
to the one particle unit operator. A consequence of this form
is that for any unitary operator $U$ on ${\cal H}$ we obtain 
$$(U\otimes U^{-1})\doubleket{B} = \doubleket{B}.\eq{8}$$
This symmetry property suggests the following  general method
for constructing maximally entangled states.
\vskip.1in
Let ${\cal G}$ be a compact topological group (the reader may
simply think of something like the rotation group whose
elements can be parametrized by Euler angles which vary
continuously over finite intervals). On such groups there is
an invariant measure $d\mu(g)$ by means of which one can
integrate over the group. Now let $g \to u(g)$ be an
irreducible, unitary representation of G on ${\cal H}$ and let
$\ket{0}$ be a reference vector in ${\cal H}$. Finally let
${\cal G}_o$ be the subgroup of ${\cal G}$ for which $\ket{0}$ is an
eigenvector. This is called the stability subgroup of $\ket{0}$.
Thus  elements of the form $gg_o$ for given $g$ and arbitrary 
$g_o
\in G$ have the same effect on $\ket{0}$. We therefore
 select one element from each left coset
$gG_o$ to form the set $F = G/G_o$. These are the set of
distinct images of $\ket{0}$ and are referred to as a set of
coherent states associated with G and $\ket{0}$. The set F forms
what is known as a ``homogeneous space" because it enjoys a
transitivity property --- any element is obtained from any
other element by a transformation in G. 
\vskip.1in
Now define
$$\doubleket{B} = C\int_F d\mu \ket{g}\otimes\bra{g},\eq{9}$$
where $C$ is a normalization constant to be determined.
If $h$ is any element of ${\cal G}$ we have
$$(u(h) \otimes u^{-1}(h))\doubleket{B} =
C\int_F d\mu u(h)u(g)\ket{0}\bra{0} u^{-1}(g)u^{-1}(h) = $$
$$
C\int_F d\mu u(hg)\ket{0}\bra{0} u^{-1}(hg) =
\doubleket{B},\eq{10}$$ where in the last step one made a change
of variables justified by the transitivity of F and then used
the invariance of the measure.
\vskip.1in
Observe that the same steps used to establish (10) show that
the one particle operator
$$I = \int_F d\mu\ket{g}\bra{g}\eq{11}$$
commutes with all $u(h)$ and hence, by the assumed
irreducibility and Schur's Lemma, $I$ is a multiple of the
identity. We may therefore normalize the measure  so that $I$ is
itself the unit operator and we will then find
that the proper choice for $C$ to normalize $\doubleket{B}$ is
$$C = \hbox{vol(F)}^{-1/2},\quad \hbox{vol F} =
 \int_F d\mu.\eq{12}$$
\vskip.1in
Let us see how Bohm-Aharonov pairs appear in this general
context: Let $G = S0_3$ and $u$ be the spin-1/2
representation. The states $\ket{g}$ can be identified with
vectors 
$$\ket{\theta,\phi} \to \left(\matrix{\cos(\theta/2)\cr
e^{i\phi}\sin(\theta/2)}\right)$$
i.e.\ F is just the Poincar\'{e} sphere with invariant measure
proportional to the solid angle $d\Omega = \sin\theta d\theta
d\phi$. Thus when we integrate over $d\Omega$ we will find
$$\int d\Omega \ket{\theta,\phi}\otimes\bra{\theta,\phi}
\propto
\left(\matrix{1 \cr
0}\right)\otimes\left(\matrix{1 \cr
0}\right) 
 + \left(\matrix{0 \cr
1}\right)\otimes \left(\matrix{0 \cr
1}\right) $$
with the cross terms having vanished upon integration over
$\phi$. 
\end
Calculating as in the Bohm-Aharonov example one
verifies that in the state $\doubleket{B}$ one member of a pair
may be found with equal likelihood in any state $\ket{g}$, but
the conditional probability of finding one member in state
$\ket{g}$ if its partner is  in state $\bra{h}$ will be
$$p(g|h) = |\bra{g}h\rangle|^2 = |\bra{0} u(g^{-1})u(h)\ket{0}|^2
= |\bra{0} u(w)\ket{0}|^2,$$ $$ w = g^{-1}h.\eq{13}$$

\end
   If a theorist wishes to instruct an experimentalist to
prepare a specific quantum state, it will be of little use to
the experimentalist if he is given a list of complex numbers
that are to be its components in some Hilbert space basis.
 What the experimentalist  requires is a {\it recipe} for
preparing the state, e.g.\ to turn a Stern-Gerlach magnet in a
prescribed way or to turn on
 a prescribed laser-generating current. In general the recipes
specify a transformation  to be applied to a reference state,
and the transformations form a group that is characteristic of
the type of device used to prepare the state. For spin states
prepared by Stern-Gerlach magnets this is the rotation group 
$SO_3$. For laser states generated by photocurrents $j$ it is
the Weyl-Heisenberg (WH) group represented by
exponentials of bose operators in which j
 appears.
\vskip.1in
States described by such recipes are called {\it
generalized coherent states}. The WH states were first
introduced into optics by Glauber$^{\bf 1}$, and the concept
was subsequently generalized by Perelomov$^{\bf 2}$ who, with
his co-workers, discovered most of what we know about them.
\vskip.1in
Although the class of generalized coherent states 
 embraces every kind of state that one can give a recipe for
making, it is not linearly closed, i.e.\ linear combinations
of coherent states are not necessarily coherent states. Thus
if only generalized coherent states were allowed to exist in
the world, it would make no sense to identify observable
physical properties  with membership in subspaces of Hilbert
space. This identification, 
known as the ``eigenvalue-eigenstate link", is a 
fundamental precept of the  orthodox Dirac-von Neumann (DvN)
formulation of quantum mechanics, and, when combined with the
linearity of
\Schrod dynamics, is known to be the cause of the measurement
problem. In a world with only coherent states, i.e.\ states
that are completely prescribed by the prescription for making
them, it is neither sensible nor necessary to introduce
Dirac-von Neumann observables, and one can expect the
measurement problem to disappear. 
\vskip.1in
This motivates the task I will address below, namely
to replace the DvN formalism with what I shall
call a ``coherent state model" (CSM) of quantum
mechanics. In this model only coherent states will be
permitted, \Schrod dynamics will be unchanged, and all
predictions of the DvN model which are applicable to coherent
states will continue to hold.
\vskip.1in
Let us first examine the form of quantum mechanical
amplitudes between coherent states. Suppose that ${\cal G}$ is a
group and $g \in G \to u(g)$ is a unitary representation of G
acting in a Hilbert space ${\cal H}$. If the unit vector
$\ket{0}$ is the reference state, the coherent states associated
with G will be of the form $\ket{g} = u(g)\ket{0}$. We would
like to label the states by $g$, but must first deal with the
possibility that two different group elements $g_1$ and $g_2$
may produce the same state. In such case $g = g_1^{-1}g_2$ 
belongs to the ``stability subgroup" ${\cal G}_o$ of $\ket{0}$,
i.e.\
$u(g)$ leaves $\ket{0}$ invariant. To  obtain a unique
labeling we select one element $g$ from each
coset $gG_o$. The set $F = G/G_o$ of cosets is a homogeneous
space (i.e.\ enjoys a transitivity property), and its points
will be in one-one correspondence with a set of unit vectors to
which we shall refer as the set of coherent states associated
with ${\cal G}$ and the reference state
$\ket{0}$.
\vskip.1in
Suppose now that $\ket{0}$ and $\ket{0^\prime}$ are reference
states. Amplitudes betwee coherent states $\ket{g_1} =
u(g_1)\ket{0}$ and $\ket{g_2} = u(g_2)\ket{0^\prime}$ then
have the form 
$$\bra{g_2}g_1\rangle = \bra{0^\prime}u(g)\ket{0} \equiv f(g),\; 
g = g_1^{-1}g_2.$$
Observe that if $g_o$ and $g_o^\prime$ are  arbitrary members
respectively of the stability subgroups ${\cal G}_o$ of $\ket{0}$ and
$G_o^\prime$ of $\ket{0^\prime}$ then $f(g) = f(g_o^\prime g
g_o)$. Thus amplitudes are constant on {\it double cosets}
$G_o^\prime\backslash g/G_o$. Double cosets partition a group
in the same way that left and right cosets do, and we see that
amplitudes are functions on that partition. 
\vskip.1in
The basic rule of quantum mechanics asserts that $p(g_2,g_1) =
|\bra{g_2}g_1\rangle|^2$ is the probability that a system in
state $\ket{g_1}$ will be detected in state $\ket{g_2}$.
Becasue of its constancy on double cosets we can now interpret
this as the probability that a {\it relation} $g$ between the
states ``holds", it being understood that the set of possible
relations is the set of double cosets $G_o^\prime\backslash
G/G_o$.
\vskip.1in
Now recall that when we do experiments with EPR pairs, i.e.\
maximally entangled two-particle states,  we  test relations
between states rather than properties of the states
individually. Let us see that such pairs provide a natural
way to probe relations between states.
\vskip.1in
 Recall that EPR pairs
are states formed from tensor products of states $\ket{a}$ and
$\ket{a^T}$ where the superscript $T$ indicates an
anti-unitary transformation of $\ket{a}$. For the familiar
Bohm-Ahronov pair $T$ is time-reversal. Now suppose that G is
a compact group with invariant measure $d\mu(g)

 Let us now show that there is
associated with any compact group G a state that acts like Bell
statesdesign a type of gedanken experiment that tests the basic
rule as it applies to relations between states. For this we
want to treat the two states on the same footing. Suppose the
states are particle states. We construct a special two-particle
state
$\doubleket{B}$ the memebers of which will interact with
particle detectors. This two-particle state has the

\end

associated with these
two reference states  have the form
$$\bra{0
\end

Axiom (1) There is a group ${\cal G}$ associated
with every quantum mechanical system. On physical grounds
we can assume that G
is locally compact, so that t phere exists an invariant
measure
$d\mu(g)$ by means of which we can integrate over the group. 
\vskip.1in
Axiom (2) The states are identified with points $x$ in a
topological space $X$ on which G acts as a group of
transformations. The set of allowed states are the images $x =
gx_o$ of a reference point $x_o$ under the action of $g$.
 Once we have defined a group G that characterizes
the production recipe,  we can use its elements
$g$ as labels for the states. However, it is possible that two
different group elements $g_1,g_2$ produce the same state.
This means that
$g_1^{-1}g_2$ belongs to a group $S$ that leaves the
reference state invariant called the {\it stability
subgroup} of the reference state. To obtain an unambiguous
labelling of the states we have only to select one element
$g$ from each of the cosets of G over S. The set of labels
thus forms the space
${\cal F} = G/S$. It is a homogeneous space,
i.e.\ it enjoys the transitivity property whereby any point
is accessible from any other by action of an element of ${\cal G}$.
\vskip.1in
We shall suppose that the states of a system can be
represented by points in some sort of topological space $X$ and
that the elements $g$ of ${\cal G}$ are bi-continuous transformations
$x \to gx$. Let $x_o$ denote the point corresponding to the
reference state, so that for $g \in S$ we have $gx_o = x_o$.
The homogeneous space ${\cal F}$ will then be represented by
the distinct images of $x_o$.
\vskip.1in
In order for this description to be adequate for quantum
kinematics it is necessary that dynamical transformations carry
elements of ${\cal F}$ into elements of ${\cal F}$ in a manner
that preserves the group theoretic and topological structure.
It suffices for this that dynamical transformations $s$ be
members of $S$, for we will then have:
$$s(gx_o) = (sgs^{-1})sx_o = (sgs^{-1})x_o  = g^\prime x_o,\;
g^\prime = sgs^{-1}.\eq{1}$$
The map $g \to g^\prime$ is  an {\it automorphism} of ${\cal G}$
and $gx_o \to g^\prime x_o$ is a homeomorphism of $X$.
\vskip.1in
It is instructive to see how this description of physical
processes looks in the usual quantum mechanical formalism for
the important case of multimode bose systems: The space $X$
will be a Hilbert space ${\cal H}$ referred to as {\it Fock
space} in which the bose operators  $a
=(a_1,\cdots,a_\nu)$ and their adjoints act in the well-known
way on a reference state $x_o = \ket{0}$ called the Fock
vacuum.  Let
$\lambda = (\lambda_1,\cdots,\lambda_\nu)$ be a vector with
complex elements. The usual (Glauber) coherent states are
$$\ket{\lambda} = n(\lambda)\ket{0},\; n(\lambda) =
e^{\lambda\cdot a^\dagger - \lambda^*\cdot a}\ket{0} ,\eq{2}$$
in which $\ket{0}$ is the Fock vacuum annihilated by all $a_i$.
Let
$N$ be the Weyl-Heisenberg group which can be represented by
operators of the form $e^{i\theta}n(\lambda)$. Indeed by the
Stone-von Neumann theorem the Fock representation is unique
(up to equivalence). Now let $S$ be the group of operators of
the form
$$s(A) = e^{ia^\dagger\cdot A \cdot a},\eq{3}$$
in which $A$ ranges over the $\nu\times \nu$ hermitian
matrices. Observe that 
$$s(A)n(\lambda)s^{-1}(A) =
n(\lambda^\prime),\; \lambda^\prime = e^{iA}\lambda,\eq{4}$$
and since
$s(A)$ stabilizes the Fock vacuum
 we have
$$s(A)n(\lambda)\ket{0} = n(\lambda^\prime)\ket{0} ,\eq{5}$$
which means that
dynamical transformations of the form $s(A)$ take
coherent states into coherent states. Since $S$ is a Lie
group it has one parameter subgroups of the form 
$e^{-i\tau {\bf H}}$ where ${\bf H}$ is a hermitian operator
of the form $a^\dagger\cdot a$. Thus the orbit of
coherent states for one parameter subgroups of $S$ will obey
the \Schrod equation with Hamiltonian {\bf H}. 
The space ${\cal F}$ resembles a classical phase space. It
consists of all complex $\nu$ dimensional vectors $\lambda$
which are transformed dynamically by elements $e^{iA}$ of the
the unitary group $U_\nu$. This means that the expectation
value $|\lambda|^2$ of the boson number is constant. What we
have here is a subgroup of the linear canonical
transformations  which apply to all kinds of optical systems. 
Indeed since $A$ ranges over all hermitian matrices the family
of dynamical systems included is, in a sense, as large as that
embraced by conventional quantum mechanics. 
\vskip.1in
We can 
 abstract from this example to characterize the general class
of systems in which the dynamics is consistent with the
maintenance of coherence as follows: Let $N$ and $S$ be given
groups and let there be a homomorphism $s \to \alpha_s$  of $S$
into the automorphism group of $N$.  Then  we can construct
(ref Mackey) a group
$G \equiv 
S\odot N$ called the {\it semi direct product} of $S$ and $N$
such that every element
$g$ of ${\cal G}$ has a unique decomposition into the form $g =sn$
with $n \in N$, $s \in S$, and sns^{-1} =
\alpha_s(n)$. Thus elements of $N$ are transformed into one
another by similarity transformations $s$.
\vskip.1in
 Any 
unitary representation of ${\cal G}$ on a Hilbert space ${\cal H}$
defines (by restriction) a representation
$u(n)$ of
$N$ and $u(s)$ of $S$ on ${\cal H}$. Any vector
$\ket{0}$ which has $S$ as a stability group  can then serve
as a reference for coherent states $\ket{n} = u(n)\ket{0}$ which
will transform into one another by
$$u(s)u(n)\ket{0} = u(\alpha_s(n))\ket{0}.\eq{6}$$ 
The
\Schrod ``equation" for the coherent states is the map 
$$n \to \alpha_{s(\tau)}(n)\eq{7}$$
as $s(\tau)$ varies over a one-parameter subgroup of $S$.
\vskip.1in
By analogy with the Glauber states we will refer to
$\ket{0}$ as a ``vacuum" state and $S$ as the {\it stability group
of the vacuum}. Each of the coherent states $\ket{n}$ has a
stability group $nSn^{-1}$ which is isomorphic to $S$.
\vskip.1in 
The group $S$ determines the type of physical object
associated with
 the coherent states. Thus if $S$ is the
Poincar\'{e} group (space-time translations and Lorentz
transformations), the associated objects can be thought of as
relativistic particles.
\vskip.1in
   We next make an important observation:  
 {\it A given group may have different (non-isomorphic)
decompositions into semi-direct products}, {\it and there will
be  qualitative physical differences between the corresponding
coherent states.} We shall be particularly concerned with the
following situation in which two different types of
coherent state appear: Let
$G = S\odot N$ define a system of coherent states. We construct
{\it pair coherent states} by applying the direct product group
$\Gamma \equiv G
\otimes G$ to a pair reference state. The elements of 
$\Gamma$ are pairs 
$\gamma = (g_1,g_2)$ with components $g_1,g_2$ in ${\cal G}$, and 
the group composition is component-wise. The obvious way to
write this as a semi-direct product is
$$\Gamma = S_I \odot N_I,\;\; S_I = S\otimes S,\;\;
N_I = N
\otimes N. \eq{8}$$
 If $S$ contains the Poincar\'{e} group
${\cal P}$ then
$S_I$ will contain ${\cal P}\otimes {\cal P}$. Hence  the pair
coherent states associated with (8) behave like independent
pairs of relativistic particles. 
 We shall refer to such states
as type-I or {\it factored}  coherent states.
\vskip.1in
But now observe that $\Gamma$ has another semi-direct
product decomposition:
$$\Gamma = S_{II} \odot N_{II},,\; S_{II} = G^2,\; N_{II} =  (G
\otimes E)
\hbox{ or } 
 (E\otimes G),\eq{9}$$
in which $G^2$ is the subgroup of pairs $(g_1,g_2)$ with
$g_1 = g_2$, and E is the identity subgroup of G.
With this decomposition the reference state is no longer
stabilized by $S \otimes S$, but is stabilized by its subgroup
$S^2$ which belongs to $S_{II}$. Thus if $S$ contains ${\cal
P}$, the reference state is stabilized by $P^2$ which is
isomorphic to
$P$. Thus the pair coherent states associated with the
decomposition (9) have a relativistic {\it single
particle-like} behavior under a group isomorphic to the
Poincar\'{e} group. But $G^2$ also contains such
transformations as $N^2$ in which both memebers are subject to
the same transformation $n \in N$. This will not resemble a
space-time symmetry in general. We shall call these type-II
or, for reasons that will become clear below, {\it entangled
coherent states}. They behave like {\it non-local} but
relativistically covariant objects, in effect like ``
measuring rods" or ``clocks" that determine intervals between
events.   
\vskip.1in
Amplitudes between type-I and type-II coherent states now have
an interesting physical interpretation. One can think of
type-I states as describing a pair of detectors placed at two
different space-time points and set to detect particular
one-particle coherent states. I will call these detections
``events". One can now think of the type-II reference state as 
a measuring rod. The rules of quantum mechanics assert that if
a system is in the state $\doubleket{II}$ the probability
that the two detectors 

sone states produced by the application of
$\gamma = (g_1,g_2)$ to the independent pair 
\end

 The question now naturally
arises as to whether quantum mechanical correlations occur
between type-I and type-II pairs. If so it must be possible to
represent the states in the {\it same} Hilbert space. Thus
there must be a representation $\gamma \to U(\gamma)$ of
$\Gamma$ in a Hilbert space ${\cal H}_\Gamma$, the elements of
which will be indicated by ``double-kets" $\doubleket{}$. There
 must be vectors $\doubleket{I},\doubleket{II}$
representing the corresponding reference states which are
stabilized respectively by $S_I,S_{II}$. There will then be
amplitudes between states of the form $\ket{\gamma,I} =
U(\gamma)\ket{I}$ and $\ket{\gamma,II} = 
U(\gamma)\ket{II}.$
\vskip.1in
Amplitudes betwee type-I and type-II coherent states have a
remarkable group theoretic structure. For observe that:
$$\doublebra{\gamma_2, II}\gamma_1,I\rangle\rangle =
\doublebra{II}U(\gamma)\doubleket{I},\;\; \gamma =
\gamma_2^{-1}\gamma_1. \eq{11}$$
Thus {\it amplitudes are constant on  double cosets of
the relation $\gamma$ connecting the two group elements of
$\Gamma$.} 
\vskip.1in
It follows that the amplitude is unchanged if $\gamma =
(g_1,g_2)$ is replaced by $(h,h)(g_1,g_2)(g_o,g_o^\prime)$
where $h$ is any element of ${\cal G}$ and $(g_o,g_o^\prime) \in
S\otimes S$. Choosing $h = (g_1 g_o)^{-1}$ it follows that the
amplitude depends only on $g_o^{-1}g
g_o^{\prime}$ where $g = g_2^{-1}g_1$. Thus it is constant on
double cosets
$S\backslash G/S$.
\vskip.1in
Let us now derive a formula for the amplitude that explicitly
reveals this constancy. We must first have a Hilbert space
${\cal H}_\Gamma$ on which we can represent  $\Gamma$. To do
this suppose that we have a representation $g \to u(g)$ of ${\cal G}$
on a Hilbert space ${\cal H}$ with reference state $\ket{0}$
stabilized by $u(g)$. Then we can construct the dual
representation $g \to u^\dagger(g)$ on the dual space
$\overline{\cal H}$. Thus double-kets will belong to the
space
${\cal H}\otimes \overline{{\cal H}}$, and if $\gamma =
(g_1,g_2)$ we can define  
$$U(\gamma)(\ket{a}\otimes\bra{b}) = u(g_1)\ket{a}\otimes
\bra{b}u^\dagger(g_2).\eq{12}$$ 
Clearly the vector
$$\doubleket{I} = \ket{0}\otimes \bra{0}\eq{13}$$ is stabilized
by
 $S \otimes S$. Thus we can represent type-I coherent states
produced by the action of $\gamma = (g_1,g_2)$ on the IP
vacuum $\doubleket{I}$ as
$$\doubleket{(g_1,g_2),I} = \ket{g_1}\otimes\bra{g_2}.\eq{14}$$
\vskip.1in
Our next task is to obtain a representation of
$\doubleket{II}$ in ${\cal H}_\Gamma$. A clue to the form of
this state comes from the demonstration of the completeness of
the coherent states. Suppose ${\cal G}$ is locally compact,
so that there is an invariant measure $d\mu$ by which we can
integrate over the group and over the space ${\cal F} = G/S$.
Consider the operator then observe that the operator
$$I = \int_{\cal F}d\mu\ket{g}\bra{g}\eq{15}$$
commutes with  $u(h)$ for all $h \in G$. If $u$ is irreducible
it follows from Schur's Lemma that (with proper normalization
of the measure) $I$ is the unit operator on ${\cal H}$. The
(at least) completeness of the coherent states follows.
They are infinitely overcomplete in general. Now let us turn
this operator into a vector in ${\cal H}\otimes\overline{{\cal
H}$ by inserting $\otimes$ between the bra and the ket and
introduce a normalization constant $C$. Thus let
$$\doubleket{II} = C\int_{\cal F}d\mu\ket{g}\otimes\bra{g}.
\eq{16}$$ The argument used to show that $I$ commutes with all
$u(h)$ now shows that $\doubleket{0}$ is invariant under
of $u(h,h) = u(h)\otimes u^{\dagger}(h)$ for all $h$, i.e.\ it
is stabilized by $S_{II}$. One then verifies that the choice
$$C = V_{\cal F}^{-1/2}, \; V_{\cal F} \equiv \int_{\cal F}
d\mu \eq{17}$$
normalizes $\doubleket{II}$. The state is normalizable if and
only if $V_{\cal F}$ is finite. This will be the case if ${\cal G}$
is compact. If it is only locally compact the state is
improper and must be handled with care. We now compute the
amplitude above and obtain:
$$\doublebra{II}U(\gamma)\doubleket{I} =
C\int_{\cal F}d\mu \bra{0}
u^{\dagger}(g_2)\ket{g}\bra{g}u(g_1)\ket{0} =
C\bra{0} u(g) \ket{0},\; g = g_2^{-1}g_1.\eq{18}$$
This is a function on the double cosets $S\backslash G/S$ as
noted above.
\vskip.1in

\end

 ${\cal
H}_\Gamma$ type-II reference state. We use the notation
$\doubleket{}$ (double-ket) to indicate vectors in in ${\cal
H}_\Gamma$ and write
$\doubleket{I}$ and
$\doubleket{II}$ respectively for the

Since states of type-I and
type-II are labelled by cosets $\Gamma/S_I$ and
$\Gamma/S_{II}$ respectively it is natural to guess (an we
shall indeed see) that such correlations

 whether quantum mechanical correlations between
pair-coherent states belonging 
${\cal G}$ but possibly different semi-direct product decompositions
associated with non-isomorphic sub-groups $S$ and $S^\prime$.

 In this
discussion the notation ${\cal G}$ can  also be  for $G\otimes G$.
Since one type will be labelled by cosets $G/S$ and the other
by cosets $G/S^\prime$ it is clear that correlations can only
depend on the {\it double coset} $S^\prime\backslash G/S$ to
which an element $g \in G$ belongs.

We shall now make the hypothesis that quantum mechanical
correlatons only occur between states that 

Now let us recall that the experiments to which we apply
quantum theory detect correlations between pairs of events
e.g.\ the correlation between the passage of light
through a polarizer and an analyzer. In the conventional
formulation one says that the polarizer produces a state, and
the analyzer tests the state. We can, however, treat the
polarizer and analyzer more symmetrically, i.e.\ we can regard
the polarizer-analyzer as a single state produced by the
action of a group element $\gamma \in \Gamma$ on the 
 with a pair $\gamma \in \Gamma$ acting on a pair
reference state. In setting up the experiment the polarizer
and analyzer are uncorrelated so that our description of 
\end

 Suppose then we think of $IP$ coherent states
labelled by
$\gamma = (g_1,g_2)$ as states of a pair consisting of a
polarizer and an analyzer which is set up  We shall write
$u(\gamma)$ for a unitary representation of $\Gamma$ on a
Hilbert space
${\cal H}_\Gamma$  and indicate
 IP and EP  respectively by $u(\gamma)\doubleket{0}$
and $u(\gamma)\doubleket{B}$ where $\doubleket{0}$ is
stabilized by $S_I \equiv S \otimes S$ and $\doubleket{B}$ is
stabilized by $S_{II} \equiv G^2$. A quantum mechanical 
amplitude between
$u(\gamma_1)\doubleket{0}$ and $u(\gamma_2)\doubleket{B}$
then takes the following form:
$$\doublebra{B}u^\dagger(\gamma_2)u(\gamma_1)\doubleket{0} =
\doublebra{B}u(\gamma)\doubleket{0},\; \gamma =
\gamma_2^{-1}\gamma_1 .\eq{10}$$

 Observe that this function is
constant on {\it double cosets} $S_{II}\backslash \Gamma /S_I$.
Double cosets partition a group into subsets just as left and
right cosets do, and this partition is dictated by our choice of
coherent state {\it types}. Amplitudes are therefore determined
by the subset in the partition to which the {\it relation}
$\gamma$ between the coherent
states belongs.
\end

\doubleket{B}\gamma_2^\dagger \gamma_1\doubleket{0}
\otimes G$ and write  $\doubleket{0}$ and $\doubleket{B}$ for 
IPCS and EPCS reference states. If we ha

 and that $\ket{0_1},\ket{0_2}$ are the 
$G
\otimes G$ In the following we will use ${\cal G}$ to mean either ${\cal G}$
or $G
\otimes G$, or indeed any group for which we have various
semi-direct product decompositions which we denote $S_j$
indexed by $j$, and we refer to the  corresponding coherent
states as $S_j$-coherent states. 
\vskip.1in
We make the following hypothesis: Whenever quantum mechanical
correlations are observed between  $S_j$-coherent states for
various $j$,  the reference states can be represented by points
$\ket{0_j}$ in the same Hilbert space ${\cal H}$, with the
same correspondence $g \to u(g)$ for all $j$. Thus we write
$$\ket{g,j} = u(g)\ket{0_j}\eq{10}$$
 to represent $S_j$-coherent
states. Then amplitudes between $S_1$ and $S_2$ coherent states
have the form:
$$\bra{g_2,2}g_1,1\rangle =
\bra{0_2}u^\dagger(g_2)u(g_1)\ket{0_1} =
\bra{0_2}u(g)\ket{0_1}, \; g = g_2^{-1}g_1.\eq{11}$$
Thus amplitudes are determined by the group theoretic
relation $g$ between $g_1$ and $g_2$ and are unchanged if
$g_j$ is replaced by any other member of the coset $g_jS_j$.
In other words {\it amplitudes are functions on the double
cosets} $S_2\backslash G /S_1$. 
\vskip.1in
Double cosets partition a group in the same way that left and
right cosets do. Thus {\it when we choose reference states we
are choosing the way that the transformation group G is to be
partitioned into sets on which an amplitude function will be
defined.}
\vskip.1in
Let us now examine the form that the amplitudes take
for the group $G \otimes G$ for $S_1$, $S_2$ corresponding
respectively to $IPCS$ states and $EPCS$. We must first
find a suitable Hilbert space in which to represent $G
\otimes  G$ and construct states $\doubleket{0_j}$ to
represent the vacuua assoicated with

\end

Quantum mechanical amplitudes
will then have the form:
$$\ bra{g

 The
possibility of quantum mechanical correlations between the two
different types of coherent states suggests that we must be
able to represent them both in the same topological space $X$,
i.e.\ that the same representation $g \to u(g)$ characterize
the action of ${\cal G}$ on reference states and that the reference
states stabilized by $S$ and $S^\prime$  be representable by
vectors $\ket{0}$ and $\ket{0^\prime}$ in thw same Hilbert
space ${\cal H}$. With this hypothesis it makes sense to
compute ampliudes connecting an $S$-coherent state $\ket{g}

\vskip.1in
Represent S-coh  

(g,g)(g_1,e)(g_o,g_o^\prime) = (e,

 Since either $G/S$ or $G/S^\prime$ will be our
coherent states we 
\end

 But since $G^2$
also contains $(g,g)$ for arbitrary $g \in G$, the coh 

\end

\Schrod dynamics for a large class of systems can be implemented
within the coherent state framework, let us now show that this
framework provides  new insight into the ``philosophical"
problems that arise from the intrinsically statistical aspect of
quantum mechanics, i.e.\ the measurement problem and the EPR
problem. To do this we first show that the phenomenon of {\it
entanglement} which is the source of both of these problems has
a remarkable group theoretic meaning that is revealed by the
coherent state formalism.
\vskip.1in
We begin by observing that the experiments addressed by quantum
mechanics always involve {\it two} devices which will be called
the {\it polarizer} and {\it analyzer}. These devices
have the same sort of transformation properties and so will be
associated with the same group ${\cal G}$. The experimental
set up can then be described by the action of the direct
product group $G\otimes G$ in which the elements are pairs
$(g_1,g_2)$ of elements of ${\cal G}$ and the composition law is
component-wise.
\vskip.1in
In order to achieve unambiguous labelling of the {\it pair
states}, as we shall call them, we must factor out the
stability subgroup of the pair reference state. The first
question we must ask, therefore, is what sort of choices do we
have in for the pair reference state based on what we know
about the separate reference states for the polarizer and
analyzer?
\vskip.1in
We can get some insight from examining the bose system in which
the reference state is the Fock vacuum $\ket{0}$. We use the term
``vacuum" because of its invariance to a group $S$ of
transformations which (if we extend to infinitely many modes)
can characterize all of the transformations of the Poincar\'{e}
group, i.e.\ space-time translations and Lorentz
transformations. If we now form a composite of two bose systems
the obvious choice of reference state is a state
$\doubleket{0}$ which is invariant under $S \otimes S$, a
state which physicists would call the ``two particle vacuum"
. Since this group allows us to select elements
$(s_1,s_2)$ with no particular relation between the
components, no correlation between polarizer and analyzer will
be introduced by such a choice. Moreover if ${\cal G}$ is a semi-direct
product
$S\odot N$, then $G\otimes G$ is a semi-direct product
$(S\otimes S)\odot (N\otimes N)$ and the space of coherent
states obtained in this way is just the cartesian product space
$({\cal F},{\cal F})$. 
\vskip.1in
At this point we make an observation which is crucial for our
discussion below: There is another way to decompose $G\otimes
G$ into a semi-direct product  $S^*\odot N^*$ such that $S^*$
is the stability subgroup for a state which can be interpreted
as the ``pair vacuum" in contrast to the ``two-particle vacuum".

\end
\to gx
\vskip.1in
It is important to keep in mind that there may be coherent
states with the same G but different, even
non-isomorphic, subgroups
$S$ and $S^\prime$, so that the corresponding spaces ${\cal
F}$ and ${\cal F}^\prime$ 
 characterize  physically quite different classes of
coherent states. Indeed we will see below that it is precisely
this kind of difference that distinguishes simple tensor
product coherent states from entangled ones.
\vskip.1in
In conventional quantum mechanics the reference state is 
described by a unit vector $\ket{0}$ in a Hilbert space
${\cal H}$, and G will be represented by unitary operators $g
\to u(g)$ in ${\cal H}$. Thus we represent the
points of
${\cal F}$ by unit vectors $\ket{g} = u(g)\ket{0}$ in which one
$g$ is taken from each coset of $G/S$, and the term  {\it set of
coherent states} will refer to this  set of unit vectors. 
Since the elements of
$S$ stabilize $\ket{0}$ it must be an eigenvector of $u(g)$ for
all $g \in S$. 
\vskip.1in
For the purposes of physics there is no loss in generality in
assuming that ${\cal G}$ is locally compact, so that there is an
invariant measure $d\mu$ by which we can integrate over
${\cal F}$. One then observes from the transitivity of ${\cal
F}$ and the invariance of the measure that
$$ I = \int_{\cal F}d\mu \ket{g}\bra{g} \eq{1}$$
commutes with $u(g)$ for all $g \in G$. Hence if we assume
that $u$ is irreducible, it follows from Schur's lemma that
 $I$ is the identity
operator in ${\cal H}$ once we normalize the measure.
Thus the set of coherent states  is at least complete and in
general is found to be infinitely overcomplete (ref Perelomov).
\vskip.1in
For later reference we observe that the ``volume" of ${\cal F}$
is
$$V_{\cal F} \equiv Tr(I) = \int_{\cal F}d\mu .\eq{2}$$
If ${\cal G}$ is compact this is finite. Otherwise one must handle
normalizations with care.
\vskip.1in
Although the coherent states are complete, linear
combinations of coherent states are not in general coherent
states. Hence the coherent state labelling scheme will only
be useful in dynamical processes that leave the reference
state invariant and map states $u(g)\ket{0}$ into states
$u(g^\prime)\ket{0}$ in such a way that the correspondence $g
\to g^\prime$ preserves the structure of the group G, i.e.\
is an {\it automorphism} of G. This is the case for a vast
array of physical systems and is possibly a universal property
of Hamiltonian systems that occur in
nature.
\vskip.1in
It is useful to see how these conditions are met in multimode
bose systems: Let $a =(a_1,\cdots,a_\nu)$ be   bose
annihilation operators, and let $\lambda =
(\lambda_1,\cdots,\lambda_\nu)$ be a vector with complex
elements. The usual (Glauber) coherent states are
$$\ket{\lambda} = n(\lambda)\ket{0},\; n(\lambda) =
e^{\lambda\cdot a^\dagger - \lambda^*\cdot a}\ket{0} ,\eq{3}$$
in which $\ket{0}$ is the Fock vacuum annihilated by all $a_i$.
Let
$N$ be the Weyl-Heisenberg group which can be represented by
operators of the form $e^{i\theta}n(\lambda)$. Indeed by the
Stone-von Neumann theorem the Fock representation is unique
(up to equivalence). Now let $S$ be the group of operators of
the form
$$s(A) = e^{ia^\dagger\cdot A \cdot a},\eq{4}$$
in which $A$ ranges over the $\nu\times \nu$ hermitian
matrices. Observe that 
$$s(A)n(\lambda)s^{-1}(A) =
n(\lambda^\prime),\; \lambda^\prime = e^{iA}\lambda,\eq{5}$$
and since
$s(A)$ stabilizes the Fock vacuum
 we have
$$s(A)n(\lambda)\ket{0} = n(\lambda^\prime)\ket{0} ,\eq{6}$$
which means that
dynamical transformations of the form $s(A)$ take
coherent states into coherent states. Since $S$ is a Lie
group it has one parameter subgroups of the form 
$e^{-i\tau {\bf H}}$ where ${\bf H}$ is
hermitian. Thus the orbit of coherent states for one
parameter subgroups of $S$ will obey the \Schrod equation
with Hamiltonian {\bf H}. 
The space ${\cal F}$ resembles a classical phase space. It
consists of all complex $\nu$ dimensional vectors $\lambda$
which are transformed dynamically by elements $e^{iA}$ of the
the unitary group $U_\nu$. This means that the expectation
value $|\lambda|^2$ of the boson number is constant. What we
have here is a subgroup of the linear canonical
transformations  which apply to all kinds of optical systems. 
\vskip.1in
We can 
 abstract from this example to characterize the general class
of systems in which the dynamics is consistent with the
maintenance of coherence as follows: Let $N$ and $S$ be given
groups and
$\alpha_s(n)$ a homomorphism of $S$ into the automorphisms of
$N$. Then  we can construct (ref Mackey) a group ${\cal G}$ called
the semi-direct product of $S$ and $N$ and denoted $S\odot N$
such that every element
$g$ of ${\cal G}$ has a unique decomposition into the form $g =sn$
with $n \in N$, $s \in S$, and sns^{-1} =
\alpha_s(n)$. Thus elements of $N$ are transformed into one
another by similarity transformations $s$.
 Any irreducible
unitary representation of ${\cal G}$ on a Hilbert space ${\cal H}$
defines (by restriction) a representation
$u(n)$ of
$N$ and $u(s)$ of $S$ on ${\cal H}$. Any vector
$\ket{0}$ which has $S$ as a stability group  can then serve
as a reference for coherent states $\ket{n} = u(n)\ket{0}$ which
will transform into one another by
$$u(s)u(n)\ket{0} = u(\alpha_s(n))\ket{0}.\eq{7}$$ 
By analogy with the Glauber states we will refer to $\ket{0}$ as
a ``vacuum" state.  The \Schrod
``equation" in general is the map 
$$n \to \alpha_{s(\tau)}(n)\eq{8}$$
as $s(\tau)$ varies over a one-parameter subgroup of $S$.
\vskip.1in
Once we restrict states to those that are obtainable by the
action of a given group G, we find that the amplitudes, by
which correlations between stats are computed under the rules
of quantum mechanics, take on a remarkable group theoretic
structure. Since, as noted above, there may be different
kinds of coherent states associated with the same G, let us
consider amplitudes between states $u(g_1)\ket{0}$ and
$u(g_2)\ket{0^\prime}$ in which $g_1,g_2 \in G$ and
$\ket{0},\ket{0^\prime}$ are stabilized by possibly
non-isomorphic subgroups $S$ and $S^\prime$. Thus amplitudes
have the form
$$f_{S^\prime,S} (g_2,g_1) = 
\bra{0^\prime}u^\dagger(g_2)u(g_1)\ket{0} =
 \bra{0^\prime}u^{-1}(g_2)u(g_1)\ket{0} = $$ $$
\bra{0^\prime}u(g_2^{-1})u(g_1)\ket{0} =
\bra{0^\prime}u(g_2^{-1}g_1)\ket{0} =
\bra{0^\prime}u(g)\ket{0} , \; g = g_2^{-1}g_1.\eq{9}$$
Thus {\it the amplitude is the same for all $g$ belonging to
the same double coset} $S^\prime\backslash g/S$. 
\vskip.1in
Let us examine this result in the light of the usual way in
which we characterize quantum mechanical predictions: 
\vskip.1in
In the
conventional interpretation we prepare a system in a state
$\ket{g_1}$ with a device, let us call it a ``polarizer", and
test the state by passing it through another device called
an ``analyzer", which is also described by a state vector
$\ket{g_2}$. The theory predicts that the probability
$p(g_1,g_2)$ that a system prepared by the polarizer
$\ket{g_1}$ will pass an analyzer $\ket{g_2}$ is given by
$$p(g_1,g_2) = |\bra{g_2}g_1\rangle|^2.\eq{10}$$\vskip.1in
Hidden variable attempts to explain the lack of determinacy in
the outcome do so by attaching an additional labels
$\gamma_1,\gamma_2$  to
$\ket{g_1}$ and $\ket{g_2}$ respectively which, were they known
in each experimental run, would determine whether a system that
passes the polarizer will pass  any given analyzer. The hidden
variable $\gamma_1$ does not ``know"
in advance what the analyzer setting
$\ket{g_2}$ is going to be, so that the theory  must be able to
accommodate ``delayed choice".} The impossibility of
reproducing the quantum mechanical probability formula in such
a model is expressed by Bell and related  ``no-go"
theorems.
\vskip.1in
But now we see that, because our restriction to coherent
states makes
 amplitudes (and hence probabilities) depend only on
 {\it relations}
$g = g_2^{-1}g_1$ between  polarizer and analyzer labels,
  the theory needs only predict {\it whether
the relation $g$ ``holds"}. That means in particular that
 the following simple-minded rule gives a hidden
variable theory that reproduces the quantum mechanical
prediction: In each run of the experiment there is a random
variable
$\epsilon$ which assumes values in the interval
$[0,1]$, and the relation $g$ holds in a given run if
and only if
$|\bra{0^\prime}u(g)\ket{0}|^2 < \epsilon$. {\it The hidden
variable expresses an uncontrollable property of the universe
in which the experiment takes place not an uncontrollable
property of the polarizer and analyzer.} Clearly there is no
problem with delayed choice here --- changing the analyzer
setting merely changes the nature of the relation being tested.
\vskip.1in
Of course the reason why this rule doesn't violate the no-go
theorems is that it is {\it drastically non-local}. To make
this clear suppose $\ket{g_1},\ket{g_2}$ are particle states,
and $S$ is the Poincar\'e group. Since this is itself a
semi-direct product of the space-time translations and the
Lorentz group, we can write $g = ns$ with $s$ restricted to
space-time translations. Thus if we represent the elements of
$S$ by $s(x) = e^{-ix\cdot {\bf P}}$ where ${\bf P}$ is the
4-momentum operator, it follows that 
 $$g = g_2^{-1}g_1 = n_2^{-1}e^{-i(x_1 - x_2)\cdot {\bf
P}}n_1.\eq{11}$$
Thus $g$ depends on the space-time {\it interval} between the
events describing the passage of a particle through the
polarizer and analyzer. Note, however, that as long as the
polarizer and analyzer states transform properly under Lorentz
transformation, the relation $g$ will be Poincar\'{e}
covariant because of the form of (11). Thus the notion of a
quantum mechanical relation is non-local but Poincar\'{e}
covariant in precisely the same sense that a space-time
interval is non-local but Poincar\'{e} covariant. This is
why there is ``peaceful coexistence" between quantum
mechanical correlations and special relativity.
\vskip.1in
We can obtain a more intuitive feeling for the non-locality by
formulating the theory in a way that treats the polarizer and
analyzer on the same footing rather than making the distinction
between the observer and the observed.  Let us represent
polarizer states by vectors
$\ket{g} = u(g)\ket{0}$ in ${\cal H}$ as before, but now we
represent analyzer states by vectors $\bra{g} = \bra{0}
u^{\dagger}(g)$ in the dual space $\overline{{\cal H}}$. The
combination of a polarizer and analyzer state is then
represented by  vectors in the space ${\cal
H}\otimes\overline{{\cal H}}$ which are transformed by
$$u(g_1,g_2)(\ket{h_1}\otimes\bra{h_2}) =
u(g_1)\ket{h_1}\otimes\bra{h_2}u^{\dagger}(g_2) =
\ket{g_1 h_1}\otimes\bra{g_2 h_2}.\eq{12}$$
Observe that $u(g_1,g_2)$ is a representation in ${\cal
H}\otimes\overline{{\cal H}}$ of the direct product group $G
\otimes G$ in which the members are pairs $(g_1,g_2)$ and the
group composition is component-wise.
\vskip.1in
To  coherent pair states we must have a 
subgroup $S^*$ of $G \otimes G$ which stabilizes a
reference vector in ${\cal H}\otimes
\overline{{\cal H}}$ and is such that $G \otimes G$ is the
semi-direct product of $S^*$ and a normal subgroup $N^*$ of $G
\otimes G$.
\vskip.1in
Remarkably $G \otimes G$ has two distinct types of semi-direct
product decompositions:
\vskip.1in
Type 1: Choose the reference vector 
$$ \doubleket{0} \equiv \ket{0} \otimes \bra{0}, \eq{13a}$$
which is stabilized by $S^* = S \otimes S$. The corresponding
semi-direct product decomposition is $G \otimes G} = S^* \odot
N^*$ with $N^* = N \otimes N$. One then verifies that the
coherent states are the factorized states $\ket{g_1}\otimes
{\bra{g_2}$ in which $g_1$ and $g_2$ range independently over
${\cal F}$, and the homogeneous space of pair coherent states
associated with this type is just ${\cal F}\otimes {\cal F}$.
\vskip.1in
Type 2: Choose the reference vector
$$ \doubleket{B} \equiv V_{\cal F}^{-1/2}\int_{\cal F}d\mu
\ket{g}\otimes 
\bra{g}.
\eq{13b}$$
This is stabilized by the subgroup $S^* = G^2$ of $G \otimes G$
consisting of pairs $(g_1,g_2)$ with $g_1 = g_2$. One verifies
that $G \otimes G$ is the semi-direct product of $G^2$ with
either of the normal subgroups $G \otimes E$ or $E \otimes G$,
in which $E$ is the identity subgroup. The intrinsic
``non-locality" of $\doubleket{B}$ is revealed by the identity
$$u(g,e)\doubleket{B} = u(e,g^{-1})\doubleket{B},\eq{14}$$
where $e$ is the identity of ${\cal G}$. Here a transformation
affecting the polarizer only is seen to have the same effect as
the inverse transformation applied to the analyzer only. We
will see that the Type 2 coherent states are the maximally
entangled states, i.e.\ they are generalized Bell states, and
it is the identity (13) that makes quantum cryptography
possible. 
\vskip.1in
The amplitude in state $\doubleket{B}$ for finding the pair
in state $u(g_1,g_2)\doubleket{0}$ is:
$$\doublebra{B}u(g_1,g_2)\doubleket{0} =
V_{F}^{-1/2}\bra{0}u(g)\ket{0},\;\; g = g_2^{-1}g_1. \eq{15}
$$
It follows that the probability that the relation $g$ holds,
i.e.\ that the analyzer is in state $\ket{g_2}$ given that
the polarizer is in state $\ket{g_1}$ with $g_1, g_2$ such
that $g = g_2^{-1}g_1$ is 
$$p(g) = |\bra{0} u(g) \ket{0}|^2.\eq{15}$$
Here we see the perfect EPR correlation when $g_1 = g_2$
i.e.\ when $g$ is the identity  $e$.
\vskip.1in
Thus the non-locality of the function $\bra{0} u(g) \ket{0}$ can be
seen to be a consequence of the non-locality of $\doubleket{B}$.
If we replace $\doubleket{B}$ by the vacuum $\doubleket{0}$ the
amplitude will factorize and there will be no correlation. 

\end

\end
\vskip.1in

\end
 
 we can say that each experimental

 Experimentally we cannot
distinguish pairs
$(g_1,g_2)$ from pairs $(h_1,h_2)$ which are in the same
{\it relation}, i.e.\ for which
$g_1^{-1}g_2 = h_1^{-1}h_2$. That $g$ expresses a {\it
directed} relation between reference states  is seen from
the fact  that predictions are unchanged when the reference
states are interchanged and
$g$ is replaced by its inverse. 
\vskip.1in
By restricting to coherent states we can now
reinterpret the theory in the following way: In
the conventional interpretation the theory predicts the
probability that a system in a prescribed state will be
found in another prescribed state. The reinterpreted theory
 {\it predicts the probability that a
prescribed (directed) relation holds between prescribed
reference states.} The impact of this reinterpretation on the
measurement problem will be examined below.
\vskip.1in
Let us next observe that coherent states have a natual duality
structure. Corresponding to states that are transformed by
the action of $g \in G$ there are states that are transformed by
 the inverse $g^{-1}$. If the former are represented in a
Hilbert space ${\cal H}$ by vectors $\ket{g}$, we can represent
the latter by vectors $\bra{g}$ in the dual space
$\overline{{\cal H}}$. Alternatively we can introduce an
anti-unitary operator $T$ and observe that the map
$T\ket{g} \to \bra{g}$ is unitary. Thus the states in
$\overline{\cal H}$ behave like $T$-reversed states in ${\cal
H}$. It is more natural to use the $\bra{g}$ characterization.
Thus we may imagine that there are two kinds of particles:
R-particles represented by kets and L-particles represented by
bras.
\vskip.1in 
We can represent a pair of coherent states consisting of an R
and an L particle by vectors
$\ket{g_1}\otimes\bra{g_2}$ in ${\cal H}\otimes \overline{{\cal
H}}$. The group
$G\otimes G$ acts on them. The members of
this group are pairs
$(g_1,g_2)$ and composition is component-wise. We construct a
representation $u(g_1,g_2)$ of $G \otimes G$ in ${\cal H}\otimes
\overline{{\cal H}}$ from the representation $u(g)$ of ${\cal G}$ by
the rule
$$u(g_1,g_2)(\ket{h_1}\otimes\bra{h_2}) = 
u(g_1)\ket{h_1}\otimes \bra{h_2}u(g_2^{-1}) = \eq{9} $$ $$
(u(g_1)u(h_1)\ket{0}\otimes\bra{0} u^{\dagger}(h_2)u^\dagger(g_2) =
 u(g_1 h_1}\ket{0}\otimes\bra{0} u^{\dagger}(g_2 h_2) =
\ket{g_1 h_1}\otimes \bra{g_2 h_2}.$$
\vskip.1in
Coherent states with respect to the group $G \otimes G$ will be
called {\it pair coherent states}. To construct such states
following the prescription above we must first obtain a
semi-direct product decomposition  $${\cal G}\otimes {\cal G} =
{\cal S}\odot {\cal N},\eq{10}$$
where ${\cal S}$ is a subgroup that stabilizes some reference
vector in ${\cal H}\otimes \overline{{\cal H}}$.
\vskip.1in
 Let us
consider two candidates for ${\cal S}$. The first which leads
to a trivial generalization of single coherent states is 
${\cal S} = {\cal S}_o \equiv S \otimes S$. One sees that
$G\otimes G}/{\cal S}_o$ is just the tensor product ${\cal F}
\otimes {\cal F}$, i.e.\ these coherent states are {\it
factorizable} in
${\cal H}\otimes \overline{{\cal H}}.$ We obtain a more
interesting type of pair coherent state by choosing ${\cal S} =
G^2$, where $G^2$ is the subgroup of $G\otimes G$
consisting of those pairs $(g_1,g_2)$ for which $g_1 = g_2$.
The set of coherent states
$${\cal B} \equiv (G\otimes G)/G^2 \eq{11}$$
will be called, for reasons that will appear below, {\it
generalized Bell states}. Observe that both the factorizable
states and the generalized Bell states correspond 
to  semi-direct product decompositions of $G \otimes G$. In the
former ${\cal N} = N \otimes N$. In the later ${\cal N}$ can be
taken to be $G \otimes E$ or $E \otimes G$ in which $E$ is the
identity subgroup of ${\cal G}$.
\vskip.1in
The reference state for the factorizble states and the Bell
states are, respectively
$$\doubleket{0} \equiv \ket{0}\otimes \bra{0}\;\hbox{ and }\;
\doubleket{B} \equiv \int_{\cal
F}d\mu\ket{g}\otimes\bra{g}.\eq{1}$$ Note the resemblance of the
vector $\doubleket{B}$ in ${\cal H}
\otimes \overline{{\cal H}}$ to the operator I (see eq. 1).
Indeed, when one checks that $\doubleket{B}$ is stabilized by
$G^2$ the steps are identical to those used in showing that $I$
commutes with all $u(g)$.
\vskip.1in
Let us now  examine experiments with pair coherent states in
the light of our reinterpretation of quantum mechanics. Thus
exeriments test whether relations $(g_1,g_2)$ hold between
reference states. There are then three types of
experiment in which the amplitudes are readily
computed to be:
$$ f_{00}(g_1,g_2) \equiv
\quad\doublebra{0}u(g_1,g_2)\doubleket{0} = \bra{0} u(g_1)\ket{0}
\bra{0} u(g_2)\ket{0}^*,
\eq{13a}$$
$$
f_{B0}(g_1,g_2)\equiv\quad\doublebra{B}u(g_1,g_2)\doubleket{0} =
\bra{0} u(g_2^{-1}g_1)\ket{0},
\eq{13b}$$
$$ f_{BB}(g_1,g_2)\equiv
Tr(u(g_2^{-1}g_1)).
\eq{13c}$$
In (a) both reference states are the pair vacuum and there is
no correlation. In (b) one reference state is the Bell state,
the other is the pair vacuum, and we observe EPR correlations
--- note that the correlation is perfect when $g_1 = g_2$. In (c)
we have the type of experiment upon which quantum cryptography
is based, i.e.\ one in which both reference states are  Bell
states.  Note that in the latter it is possible to choose
discrete sets of $g$'s that exhibit trace orthogonality even
though the states themselves are not orthogonal.
\vskip.1in
Let us examine case (b) which provides some insight into the
EPR problem:
\vskip.1in
The Bell state $\doubleket{B}$ is invariant under 
 $G^2$ which is isomorphic to ${\cal G}$. In particular it is
invariant under $S^2$ which is isomorphic to $S$. Thus it is
more symmetric than the one-particle vacuum (if we use this
term for the reference state $\ket{0}$.) It is, however,
non-local in the sense that it is not invariant under $S\otimes
S$, i.e.\ under independent  transformations of the particles
by the symmetry group $S$. If we use the semi-direct product
factorization  $G = S\odot N$ in computing the amplitude in
(13b) we see that
$$f_{B0}(g_1,g_2)\equiv =
\bra{n_2}u(s_2^{-1}s_1)\ket{n_1} \eq{14}$$
so that if $u(s_1)  = e^{-i\tau_1 {\bf H}},\;  u(s_2) =
e^{-i\tau_2{\bf  H}}$ we will obtain the amplitude
$$f_{B0} = \bra{n_2}e^{-i(\tau_1 - \tau_2){\bf
H}}\ket{n_1}.\eq{15}$$ Observe that this quantity not only
transforms properly under the one particle symmetry group $S
\otimes S$, but depends on
$\tau$  only through the difference $\tau_1 - \tau_2$. Thus
 if the individual members transform covariantly under the
Lorentz group, the correlations will be Poincar\'{e} invariant.
Thus {\it the ingredients needed for the so-called peaceful
coexistence of EPR correlations and special relaivity are
exhibited explicitly by this decomposition.}

\end

 The invariance of
$\doubleket{B}$ under
$G^2$ but not under
$S\otimes S$ means that although it is not like the two
particle vacuum $\doubleket{0}$, it has a symmetry g
\end

 ${\cal G}\otimes {\cal
G}$ can be written 
 
\end

 reference
states in
${\cal H}\otimes\overline{{\cal H}}$ which are stabilized by 
 
\end

Suppose now that we have a system of coherent states defined
by a group ${\cal G}$ and a stability subgroup $S$. Consider a
composite system in which a pair of elements $(g_1,g_2)$
produce states by acting on a reference pair. The group G is
now replaced by $G \otimes G$, the direct product of G with
itself. The group multiplication is component-wise. We now
look for ``canonically" defined  pair coherent states (PCS) by
looking for factorizations of $G \otimes G$ into semi-direct
products.
\vskip.1in
We refer to the factorization
$$G \otimes G = (S \otimes S)\odot (N \otimes N) \eq{9}$$
as the ``trivial" factorization. It will be seen that the
corresponding coherent states $(G \otimes G)/(S \otimes S)$
belong to the homogeneous space ${\cal F}\otimes {\cal F}$,
i.e.\ they are simple tensor
products of single particle coherent states.
\vskip.1in
Remarkably, however, there is a non-trivial factorization that
is independent of the one-particle dynamics, namely
$$G \otimes G = G^2 \odot (G\otimes E) \hbox{ or } G^2 \odot
(E \otimes G), \eq{10}$$ in which
$G^2$ is the subgroup consisting of pairs $(g_1,g_2)$ with $g_1
= g_2$, and  $E$ is the subgroup of ${\cal G}$ consisting only of the
identity. We will see that the coherent states associated with
this factorization, i.e.\
$${\cal B} \equiv (G\otimes G)/G^2 \eq{11}$$
 are maximally
entangled, i.e.\ we may call them {\it generalized Bell
states}. 
\vskip.1in
Let us examine the behavior of the coherent states ${\cal B}$
under the subgroup of one-particle dynamical transformations $S
\otimes S$. Noting that $(s_1,s_2)(g,g)$ produce the same state
for all $g$ when acting on ${\cal B}$ states, it follows that
the effect of $(s,e)$ is the same as $(e,s^{-1})$ where $e$ is
the identity of ${\cal G}$.

transforming one member by $s$ is the same as
transforming its partner by $s^{-1}$. This suggests an elegant
way to represent generalized Bell states.
\end

 the orbit depends only on
$s_1 s_2^{-1}$. Thus the states
${\cal B}$ recognize only the {\it relative} displacement of the
constituents under dynamical transformations. Thus the
transformation of a ${\cal B}$ state of one particle in a pair by
$s$ cannot be distinguished from the transformation of its
partner by
$s^{-1}
\vskip.1in
Let us think of the pair as consisting of two parcticles which
we shall call R and L particles. One sees that the state
obtained from a  in which the R particle has been transformed by
$s$ is identical to the state in which the L particle has  been
transformed by $s^{-1}$. If the transformations $s(\tau)$ form
a one-parameter subgroup of $S$ one sees that the presense of
an $R$ particle at $\tau$ is indistinguishable from the
presence of an $L$ particle at $-\tau$. If $\tau$ is
interpreted as ``time" we would say that an R particle moving
forward in time is indistinguishable from an L particle moving
backwrd in time.
\vskip.1in
A convenient way to incorporate this conjugacy of $R$ particles
and L particles is to represent $R$ particles by vectors
$\ket(g) = u(g)\ket{0}$ in ${\cal H}$ and $L$ particles by
vectors $\bra{g} = \bra{0} u(g^{-1} = \bra{0} u^{\dagger}(g)$ in
the dual space $\overline{{\cal H}}$. We then represent
$G\otimes G$ by unitary operators on
${\cal H} \otimes \overline{{\cal H}}$ namely
$$u(g_1,g_2)(\ket{h_1}\otimes \bra{h_2}) =
u(g_1)\ket{h_1}\otimes \bra{h_2}u^{\dagger}(g_2) =
$$ $$
u(g_1 h_1)\ket{0}\otimes \bra{0} u^{\dagger}(g_2 h_2)=
\ket{g_1 h_1}\otimes \bra{g_2 h_2}.\eq{12}$$
The reference states with stability subgroups $
\end

particles by
$\bra{g Having identified two kinds of dynamically interesting
choices of stability groups namely
$S_o = S\otimes S$ and $S_B = G^2$ there will be three types of
experiment testing relations $(g_1,g_2)$ between the
corresponding reference states. The results are constant
on the following three double cosets: 
$$S_o\backslash G \otimes G/S_o,\quad 
S_o\backslash G \otimes G/S_B,\quad S_B\backslash G \otimes
G/S_B\eq{12}$$

The amplitudes corresponding to testing the relation
$(g_1,g_2)$ in these three experimental
\end

 from which one can produce
coherent states of
$G\otimes G$, we now apply our new interpretation of quantum
mechanics to 

\end

 choice suggested by the resolution
(1) of the identity:
Suppose we replace $\ket{g}\bra{g}$ in the integrand of (1) by
$\ket{g}\otimes \bra{g}$ so that the object we obtain is no
longer an operator in ${\cal H}$ but  rather a vector in
 the tensor product
space of
${\cal H}$ with its dual $\overline{{\cal H}}$. We use a
$\doubleket{}$ symbol  for such objects and define
$$\doubleket{B} = \int_{{\cal F}}d\mu
\ket{g}\otimes\bra{g}.\eq{8}$$ One may think of ket vectors as
representing one kind of particle called R-particles and bra
vectors as representing a different kind called L-particles.
 We represent the group
$G \otimes G$ by operators $u(g_1,g_2) \equiv u(g_1) \otimes
u^\dagger(g_2)$ so that
$$(u(h_1) \otimes u^\dagger(h_2))(\ket{g_1}\otimes\bra{g_2}) =
u(h_1)\ket{g_1}\otimes \bra{g_2}u^\dagger(h_2) =
\ket{h_1 g_1}\bra{h_2 g_2}.\eq{9}$$
It follows from (1) that
$$(u(h) \otimes u^\dagger(h))\doubleket{B} =
\doubleket{B},\;\hbox{ for all } h\in G.\eq{10}$$
Thus $\doubleket{B}$ is stabilized by the subgroup $G^2$ of $G
\otimes G$ consisting of elements $(g_1,g_2)$ with $g_1 = g_2$.
\vskip.1in
We are thus led to examine a type of pair coherent state which
is very different than the factorable ones. The stability
subgroup $G^2$ has a the subgroup $S^2$ which is isomorphic to
the group
$S$ which stabilizes the one particle reference $\ket{0}$. In
this sense $\doubleket{B}$ is at least as symmetric as the
one-particle vacuum $\ket{0}$. Indeed it is more symmetric in
general because it is invariant under $(g,g)$ even when $g
\notin S$. Note, however, that the reference state in
${\cal H}\otimes \overline{{\cal H}$ for coherent states based
on $S \otimes S$, namely the two particle vacuum $\doubleket{0}
\equiv
\ket{0}\otimes
\bra{0}$, is invariant under $(s_1,s_2)$ even when $s_1
\neq s_2$. 
\vskip.1in
In our
reinterpretation of of quantum mechanics  experiments test
prescribed relations between prescribed reference states. For
pair systems we have examined two types of reference states 

 Thus a possible class of experiments tests the relation
$(g_1,g_2)$ between the pair ``vacuum"  $\doubleket{0}$ and the
state $\doubleket{B}$. Let us compute it:
$$\doublebra{B}(u(g_1,g_2))\doubleket{0} =
\int_{{\cal F}}d\mu\bra{0}
u^{\dagger}(g_2)\ket{g}\bra{g}u(g_1)\ket{0}
 =
\bra{0} u(g_2^{-1}g_1)\ket{0}.\eq{11}$$
This means that one observes perfect correlation in
$\doubleket{B}$ when $g_1 = g_2$. We see then that the states
$\doubleket{B}$ are maximally entangled states. We call them
generalized Bell states. 
\vskip.1in
Formula (11) may  be stated in the
language of our reinterpretation as a general quantum
mechanical theorem: {\it The
probability that relation
$(g_1,g_2)$ holds between a generalized Bell state and the
vacuum coincides with the probability that the relation
 $g_2^{-1}g_1$ holds between the one-particle vacuum and
itself.}
\vskip.1in
It is interesting to contrast the relations between
$\doubleket{B}$ and $\doubleket{0}$ with
those between these states and themselves. Thus
$$\doublebra{0}u(g_1,g_2)\doubleket{0} = \bra{0} u(g_2)\ket{0}^*
\bra{0} u(g_1)\ket{0}^,\eq{12a}$$
$$\doublebra{B}u(g_1,g_2)\doubleket{B} = Tr(u(g_2^{-1}g_1)).
\eq{12b}$$
\vskip.1in
One verifies that
$$\doublebra{B}B\rangle\rangle = \hbox{ vol}({\cal F}) =
\int_{\cal F} d\mu,\eq{13}$$
which is finite if and only if $G/S$ is compact. In the
non-compact case one must make sense of $\doubleket{B}$ through
a limiting process.
\vskip.1in
Because $\doulbeket{B}$ is invariant under $G^2$, i.e.\ under
any transformation such that both memebers are transformed in
the same way, we are motivated to consider the notion that
generalized Bell states are a kind of vacuum.
\end